\documentclass[12pt]{article}

\usepackage{scicite}
\usepackage{graphicx}
\usepackage{times}
\usepackage{color}
\usepackage[caption=false]{subfig}
\usepackage{csquotes}
\usepackage{lineno}
\usepackage[normalem]{ulem}
\usepackage{sectsty}

\usepackage{xr}
\makeatletter
\newcommand*{\addFileDependency}[1]{
  \typeout{(#1)}
  \@addtofilelist{#1}
  \IfFileExists{#1}{}{\typeout{No file #1.}}
}
\makeatother
 
\newcommand*{\myexternaldocument}[1]{%
    \externaldocument{#1}%
    \addFileDependency{#1.tex}%
    \addFileDependency{#1.aux}%
}
\myexternaldocument{si}

\sectionfont{\fontsize{15}{10}\selectfont}
\subsectionfont{\fontsize{13}{10}\selectfont}

\newenvironment{sciabstract}{%
\begin{quote} \bf}
{\end{quote}}

\title{Deeply nested structure of mythological traditions worldwide}

\author
{Hyunuk Kim,$^{1,2}$ Marcus J. Hamilton,$^{3,4\ast}$ Woo-Sung Jung,$^{5,6}$\\ and Hyejin Youn$^{2,7, 8\ast}$\\
\\
\normalsize{$^{1}$Department of Management and Entrepreneurship,}\\
\normalsize{Martha and Spencer Love School of Business, Elon University, Elon, NC 27244, USA}\\
\normalsize{$^{2}$Northwestern Institute on Complex Systems, Evanston, IL 60208, USA}\\
\normalsize{$^{3}$Department of Anthropology, University of Texas at San Antonio, San Antonio, TX 78249, USA}\\
\normalsize{$^{4}$School of Data Science, University of Texas at San Antonio, San Antonio, TX 78207, USA}\\
\normalsize{$^{5}$Department of Industrial and Management Engineering,}\\
\normalsize{Pohang University of Science and Technology, Pohang 37673, Korea}\\
\normalsize{$^{6}$Department of Physics,}\\
\normalsize{Pohang University of Science and Technology, Pohang 37673, Korea}\\
\normalsize{$^{7}$Department of Management \& Organizations,}\\
\normalsize{Kellogg School of Management, Northwestern University, Evanston, IL 60208, USA}\\
\normalsize{$^{8}$Santa Fe Institute, Santa Fe, NM 87501}\\
\normalsize{$\ast$ E-mail: marcus.hamilton@utsa.edu, hyejin.youn@kellogg.northwestern.edu}
}

\date{}

\begin{document} 

\baselineskip 24pt

\maketitle
\vspace{-0.5cm}

\begin{sciabstract}
All human societies present unique narratives that shape their customs and beliefs. Despite cultural differences, some symbolic elements (e.g., heroes and tricksters) are common across many cultures. Here, we reconcile these seemingly contradictory aspects by analyzing mythological themes and traditions at various scales. Our analysis revealed that global mythologies exhibit both geographic and thematic nesting across different scales, manifesting in a layered structure. The largest geographic clusters correspond to the New and Old Worlds, which further divide into smaller bioregions. This hierarchical manifestation closely aligns with historical human migration patterns at a large scale, suggesting that narrative themes were carried through deep history. At smaller scales, the correspondence with bioregions indicates that these themes are locally adapted and diffused into variations across cultures over time. Our approach, which treats myths and traditions as random variables without considering factors like geography, history, or story lineage, suggests that the manifestation of mythology has been well-preserved over time and thus opens exciting research avenues to reconstruct historical patterns and provide insight into human cultural narratives.
\end{sciabstract}

All human societies present unique narratives that establish and sanctify their customs, institutions, and taboos. Nevertheless, there are recurring symbolic motifs such as heroes and tricksters widely recognized across many different cultures. 
All human cultures have rich tapestries of stories about their origins, norms, and values \cite{levi1983raw, descola2013,harari2014sapiens, bascom1954four, toelken1996dynamics}. 
Examples abound: A legendary hero's journey in search of immortality; a mythical dispute between the sun and moon to illuminate the sky; a tiger and a bear striving to become human. These narratives not only captivate but also function as essential societal frameworks \cite{smith2017cooperation}. Gilgamesh's quest for immortality tells us about the vainglorious drive for fame and immortality and the importance of wisdom, friendship, and the acceptance of life's impermanence \cite{kramer1963sumerians}. The Sun and Moon's mutual agreement after the dispute to share the duty reinforces Agta's cultural values of cooperation and gender equality \cite{smith2017cooperation}. The tiger and bear represent two tribes, and the bear's metamorphosis into human form earned legitimacy and kingship as the founder of the Korean kingdom. Intended or not, such stories serve as powerful social technologies binding groups together while differentiating them from others \cite{harari2014sapiens, smith2017cooperation}.

Indeed, myths, folktales, and parables are narrative traditions established to sanctify customs, taboos, and institutions and encode systems of beliefs, values, and norms ~\cite{wiessner2014embers,eliade1959sacred, csapo2005theories, dissanayake2015art, mehr2019universality, tangherlini2013}.
While many of these powerful social technologies have been lost to time, others remain as relevant today as they were to our ancestors, upholding cultural values, social interactions, and shared identities that both unite and distinguish societies ~\cite{sims2011living, tomasello2010origins,smith2017cooperation, dunbar2014conversations, michalopoulos2021folklore, white2021, cohen2009}. For example, while many of us may not believe in mythical creatures, the hero's quest still resonates as we accept adult responsibilities, and the enduring principles of rewarding virtue and punishing vice still shape our moral compass \cite{campbell1949hero, michalopoulos2021folklore}.  
Whether consciously or subconsciously, whether by design or happenstance, this rich cultural heritage continues to weave itself into the fabric of modern society, explaining cultural differences in social conformity, tolerance, morality, trust, and gender equality \cite{michalopoulos2021folklore, bloom2012religion, jackson2019good, jackson2023supernatural, gelfand2011differences}.

As such, mythology is a timeless human universal \cite{brown1991human}. 
To fully appreciate these complex cultural constructs, one must examine them from diverse perspectives in their roles both as universal social technologies and unique cultural expressions \cite{harris2001rise, boas1920methods}. 
On the one hand, as in general purpose technologies, certain elemental building blocks within these cultural constructs are sufficiently general to appear frequently and combine with more specialized niche elements, creating a nested structure among these building blocks \cite{rosenberg1982inside, lipsey2005economic, arthur2009nature, mariani2019nestedness, hosseinioun2023nested}. 
On the other hand, similar to all other forms of creative expression, such as artwork and architecture, mythology too has its own complexities that resist simple comparison. Rich in information, these manifestations of unique cultural expression may not seem immediately comparable at a superficial level. 
And yet, remarkably similar themes (e.g., creation myths, trickster tales, great floods, virgin births, and tales of the afterlife) and narrative structures (e.g., a hero's journey and other emotional arcs) are found across cultures widely dispersed in time and space, reminding us of shared human universal expressions \cite{campbell1949hero, boyd2020narrative, segal1998jung, thuillard2018large, Ross20123065, witzel2012origins, jackson2019emotion, reagan2016emotional, Youn16022016}. 

These common building blocks have been identified as \textit{motifs}, \textit{indexes} or \textit{tropes} to illustrate the underlying threads that connect diverse narratives and cultures \cite{berezkin2015folklore, garcia2021simpsons, daSilva150645, dhuy2017studying, uther2004types}. 
Notably, widely recognized motifs and indexes in Eurasia folktales, such as `Little Red Riding Hood', `The Wolf and the Kids', and `The Kind and Unkind Girls' serve as indispensable keystones for analyses of species-like cultural evolution \cite{bortolini2017inferring, pagel2016anthropology}.
Similarly, the pervasive `Cosmic Hunt' narrative, which portrays star constellations as hunters pursuing game, manifests in sufficient variation around the world to establish a mythological phylogeny over tens of thousands of years \cite{berezkin2005cosmic, berezkin2015folklore, dhuy2013phylogenetic, dhuy2013cosmic,tehrani2017phylogenetics, tehrani2013phylogeny}. Phylogenetic work has provided deep insights into cultural evolution and has proven to be a rich source of hypotheses for subsequent research \cite{berezkin2005cosmic, berezkin2015folklore, dhuy2013cosmic, dhuy2013phylogenetic, Ross20123065, tehrani2013phylogeny, tehrani2017phylogenetics}.

\begin{figure*}
\label{fig1}
\centerline{\includegraphics[width=\textwidth]{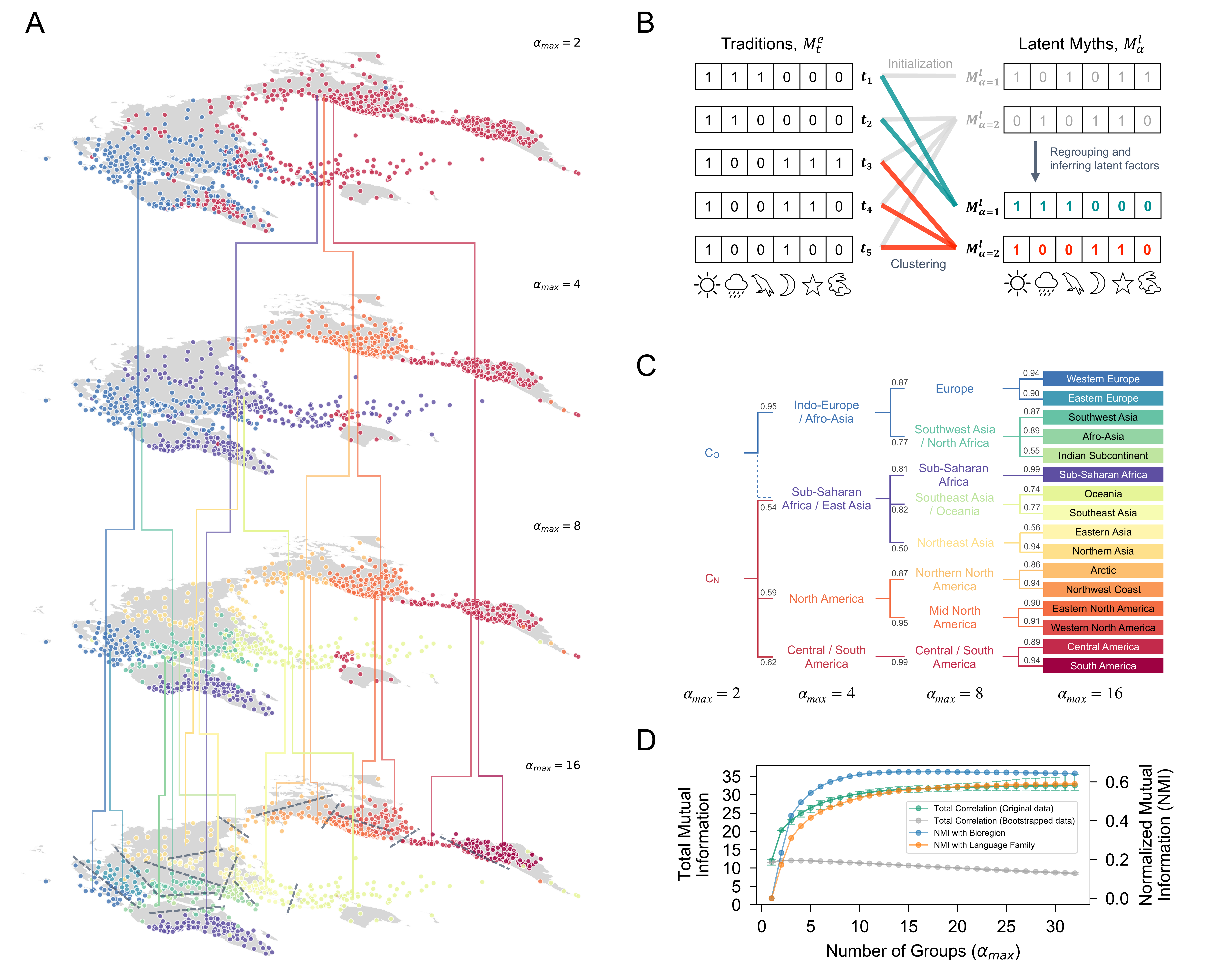}}
\noindent {\bf Figure 1.} The nested structure of mythological traditions and motif communities in space. (A) Individual traditions (circles) are embedded and nested into broader geographic communities at larger scales clustered at different scales in decreasing order. The color scheme of these geographic motif communities is followed in Fig. 1C. The marker size indicates the number of motifs that appear in a given tradition. (B) A schematic of the clustering algorithm \cite{NIPS2014_5580, gallagher2017anchored}. Empirically observed motif co-occurrence for tradition $i$ is expressed as a binary vector $t_i$. At each iteration, traditions are successively clustered under a putative latent factor, $M_\alpha$, for a cluster $\alpha$ where $i \in \alpha$. Through successive iterations, the latent factors are updated to $M^l_\alpha$ to maximize their explanatory power, and their final configuration is referred to as \textit{latent myth}. (C) 
The overarching structure of clustered traditions across scales exhibits both nested geographic clusters and nested latent myths. 
Embedded in a branching tree, motifs that comprise latent myth at higher levels are most likely retained in low levels (80\% on average). (D) On the left axis, the average total multivariate mutual information (explanatory power) of inferred clusters (green) and bootstrapped data (gray) with increasing scales. 
On the right axis, normalized mutual information between inferred clusters and bioregions (blue) and language families (orange). 
\end{figure*}

In this study, we assess whether a more relaxed set of assumptions can still uncover the expected structural variations in global mythology, typically identified through more targeted studies.
Drawing from principles of scaling theory and the renormalization group theory, our analysis probes the global and scale-dependent characteristics of mythological motifs, both as stand-alone elements or as part of larger clusters \cite{barenblatt2003scaling,kadanoff2000statistical,levi1955structural, west2018scale}.
As such, our approach is intentionally unsupervised and model-free, devoid of any particular guiding model (i.e., evolutionary models or cultural mechanisms) that typically examines the semantic nature or historical context of specific narratives.
Instead, we aim for a broad and quantitative exploration of the overall motif structure. 
Sidestepping these traditional factors isn't to dismiss their relevance; rather, we posit that our methodology offers a complementary structure for examining a wider context for the cultural evolution of mythologies by capturing the granular variation and complex combinations that appear across different scales in the global landscape of mythical motifs.

Our coarse-graining framework reveals that global mythologies are nested both geographically and thematically across broad scales (Fig. 1). We first cluster mythology traditions at different scales and overlap those clusters. We find that these clusters are hierarchically nested, with smaller clusters embedded within larger ones. Notably, the largest clusters encompass the old and new worlds, while the smallest clusters correspond to bioregions (Fig. 2). The bioregions suggest local variations and diffusion of narrative themes, while the larger patterns indicate that motifs were carried through human migration.

Our phylogenetic tree, derived from these variations, supports this inference on cultural dynamics (Fig. 3). Additionally, the nested geographic structure aligns with the hierarchical structure of the motifs themselves (Fig. 4 and Fig. 5). This integration of nested structures in thematic and geographic spaces indicates that mythologies are preserved over history and are deeply embedded in human culture.

\section*{Coarse-graining Mythological Traditions}

Our study is grounded in an extensive database, compiled by Yuri Berezkin, which spans nearly three thousand motifs across a thousand mythological traditions, extending beyond WEIRD (Western, Educated, Industrialized, Rich, Democratic) societies  \cite{henrich2010beyond, henrich2010most, berezkin2015folklore} (SI Section \ref{si:data}). 
While the large volume of observations generally ensures strong statistical power, its high dimensionality combined with a limited number of motifs per tradition requires particular attention to inference methods resistant to potential bias and noisy incomplete observations ~\cite{berezkin2015folklore, dhuy2017studying, thuillard2018large, tangherlini2013, tangherlini2016} (SI Section~\ref{si:data}).  

The Correlation Explanation (CorEx) \cite{NIPS2014_5580} is our choice of method for coarse-graining the global mythological traditions across different scales. As opposed to an evolutionary model, this method utilizes a multivariate mutual information framework by treating the presence/absence of motifs in a mythological tradition as random variables at multiple scales, and thus offers a model-free and unsupervised approach. Additionally, its flexibility allows us to adjust the granularity by controlling the group size to be coarse-grained, $|\alpha|$. Finally, it is well-suited for analyzing high-dimensional yet sparse binary matrices, providing invaluable insights into the latent structures within our dataset ~\cite{NIPS2014_5580, gallagher2017anchored} (SI Section~\ref{si:corex}). 

Fig. 1 illustrates how we coarse-grain information in mythological traditions and infer the corresponding latent factors that best \textit{explain} the information contained in the observed motif distributions within each group $\alpha_i$. Both grouping traditions and inferring the latent factors that explain the distribution of motifs within the group undergo iterative updates (from gray to red/green lines in Fig. 1B). In each iteration, motif vectors -- indicating the presence or absence of motifs within a tradition, as shown in the left-hand vectors of Fig. 1B -- are regrouped such that 
each group's total mutual information with respect to the inferred latent factor across the entire structure reaches a maximally explainable collection of motifs (on the right in Fig. 1B) (SI Section~\ref{si:corex}). 

We define the final collection of motifs that encapsulates the coarse-grained information of group $\alpha$ as a \textit{latent corpus of myth} (a hidden body of mythological motifs), which we denote as $M^l_\alpha = [m_1, m_2, ..., m_{\max}]$ where $m_i$ indicates presence/absence of motif $i$. The superscript $l$ serves to differentiate the latent corpus, $M^l_\alpha$, from the directly observed data $M^e_t$ in each individual tradition $t$. 
Note that our latent myth does not simply equate to a centroid in data clouds or an average of motif frequencies in observed traditions, formally put, $M^l_\alpha \neq <M^e_t>_{t \in \alpha}$. 
This distinction is crucial for understanding our choice of mutual information as a coarse-graining method, especially as we cannot assure unbiased sampling of observed traditions in the presence of random noise, which is often assumed in statistical sampling. Thus, we use $M^l_\alpha$ as a latent myth unless otherwise indicated.
For clarity, we will use the term `group' to describe these coarse-grained traditions and motifs, while the term `cluster' will specifically refer to geographically associated structures.

\section*{Hierarchically Nested Geographic Clusters}

Fig. 1A illustrates coarse-grained mythological traditions at various scales across multiple layers. 
In each layer, traditions $t$ are geographically embedded and color-coded by their final group membership $\alpha$, denoted as $t \in \alpha_i$, where $\alpha_i \in \alpha$ whose total number, $|\alpha|$, indicates the level of coarseness. 
The granularity of traditions increases as we move downward from the top layer, representing the coarsest scale, to the lower layers of the figure.
Even at this most abstract scale $|\alpha| = 2$, the geographical structure is evident: Traditions of one group (blue) are found throughout Old World Afro-Eurasia ($\alpha_{old\_world}$) while traditions of the other group (red) span across Sub-Saharan Africa and Asia to the New World and Oceania ($\alpha_{new\_world}$). Note that the New World is entirely red, while the Old World is a blend of both groups.

Increasing granularity subsequently discovers four smaller groups, $|\alpha'|=4$ (the prime differentiates clusters at a subsequent scale). 
Again, mythological traditions are geographically clustered into reasonably well-defined groups, including North America (orange), Central/South America (red), Indo-Europe/Afro-Asia (blue), and Sub-Saharan Africa/East Asia (purple), which extends across Sahul (Australia and New Guinea). 

These newly identified smaller-scale clusters are nested within the previously identified communities from the preceding layer. We denote these 
$\alpha'$(orange) and $\alpha'$(red) are mostly subsets of $\alpha_{new\_world}$, while $\alpha'$(blue) is subsumed within $\alpha_{old\_world}$. 
Traditions in $\alpha'$(purple) are interesting as they share information from both $\alpha_{new\_world}$ and $\alpha_{new\_world}$, but the additional analysis of shared latent myth places them in $\alpha_{new\_world}$  (See Fig. 1C). However, it is also interesting that $\alpha'$(purple) is the region where $\alpha_{new\_world}$ and $\alpha_{new\_world}$ overlap in the layer above, indicating that increasing the resolution seems to largely resolve the geographic clustering between the Old and New Worlds. We will explain this analysis further in the following section.

Further increasing the resolution uncovers the deeply 
nested structure, finding independently identified smaller groups as sub-groups of the previous larger-scale clusters across subsequent scales. For example, on the third layer, we discover a New World Northern group, $M^l_{Northern}$, spanning the North American Arctic and Subarctic, which is similar to the geographic range of the Athabaskan and Inuit-Yupik language families. In the Old World, the Indo-European/Afro-Asiatic cluster (blue) divides into northern and southern clusters, as does the Asian component of the purple Afro-Eurasian cluster from the layer above. Interestingly, the Sub-Saharan Africa cluster (purple) resembles the geographic distribution of the Niger-Congo language family. Sahul remains isolated (red).  

Potentially, we can increase the resolution all the way to $|\alpha| = |\{t\}|$, the finest possible scale where a group has only one member,  $M^e_{t} = M^l_\alpha$. 
Indeed, as we increase the scale of resolution, each latent myth $M^l_\alpha$ better captures the observed motifs therein, $M^e_{t \in \alpha}$, but we are losing generality instead, presenting a trade-off we make as a theorist (see Methods). 
At the finest possible scale, for example, the latent myth would perfectly describe each unique tradition, representing an extreme particularism where individual mythologies are entirely unique and thus cannot be coarse-grained. 
At the coarsest possible scale, one latent corpus of myths would universally account for all observed traditions, offering the most parsimonious but least informative model. 

Fig. 1D (green line) captures this trade-off well as a non-linear increase in explanatory power with $|\alpha|$, sharply increasing but eventually plateauing around 15 (SI Section~\ref{si:bootstrap}). This trade-off structure is unique to the data.
When the empirically observed $M^e_t$ is randomized to destroy their nested structure, increasing resolution no longer provides additional information (grey curve in Fig. 1D), and thus no benefit to increasing resolution~\cite{strona2014fast} (SI Section~\ref{si:bootstrap}).  This result holds even when randomizing $M^e_t$ with fixed row and column totals. The bootstrapped analysis contrasts with the empirical observation (blue curve) that increasing resolution rapidly gains information initially but diminishes beyond the resolution of $\sim15-20$ clusters. 

Given the diminishing returns of information gain while losing generality, we set the optimal scale around $|\alpha| = 16$, thus avoiding over-fitting while ensuring sufficient statistical power. Coincidentally, but perhaps not surprisingly, this optimal scale falls within the spatial range of global bioregions and linguistic families. Fig. 1D (blue line) and Fig. 2 show that these coarse-grained traditions at the optimal scale align well with global bioregions.

The bioregions suggest local variations and diffusion of narrative themes, indicating that certain mythological themes are specific to particular regions. Larger patterns, on the other hand, show that these motifs spread through human migration, implying that as humans moved, they carried their mythologies with them. These cultural dynamics are further supported by the phylogeny tree in Fig. 3, constructed using the nearest neighbor network method, which shows that large-scale geographic clusters are consistent with early human migrations.

\section*{Biogeography and Cultural Diversity of Mythological Clusters} 

Fig. 2 uncovers remarkable agreement of the spatial distribution of clusters at their optimal scale (colored by cluster) with terrestrial zoogeographic bioregions (19 subregions of amphibians, 19 of birds, and 34 of mammals) \cite{Holt74}. For further details, refer to Fig. S\ref{si:fig:bioregion} and SI Section~\ref{si:bioregion}.
To quantify this alignment, we compute the normalized mutual information (NMI) between the membership of mythological traditions and each of the three bioregions (see Methods and SI Section~\ref{si:bioregion}).
This metric measures the amount of information gained about one membership by observing another, treating them as random variables, and thus quantifying how much two memberships are aligned. 
Fig. 1D, indicated by a blue line, along with the numbers enclosed in Fig. 2, demonstrates a high degree of mutual information between the clusters of mythological traditions and the aforementioned bioregions.

\begin{figure*}
\label{fig_main_bioregion}
\centerline{\includegraphics[width=0.7\textwidth]{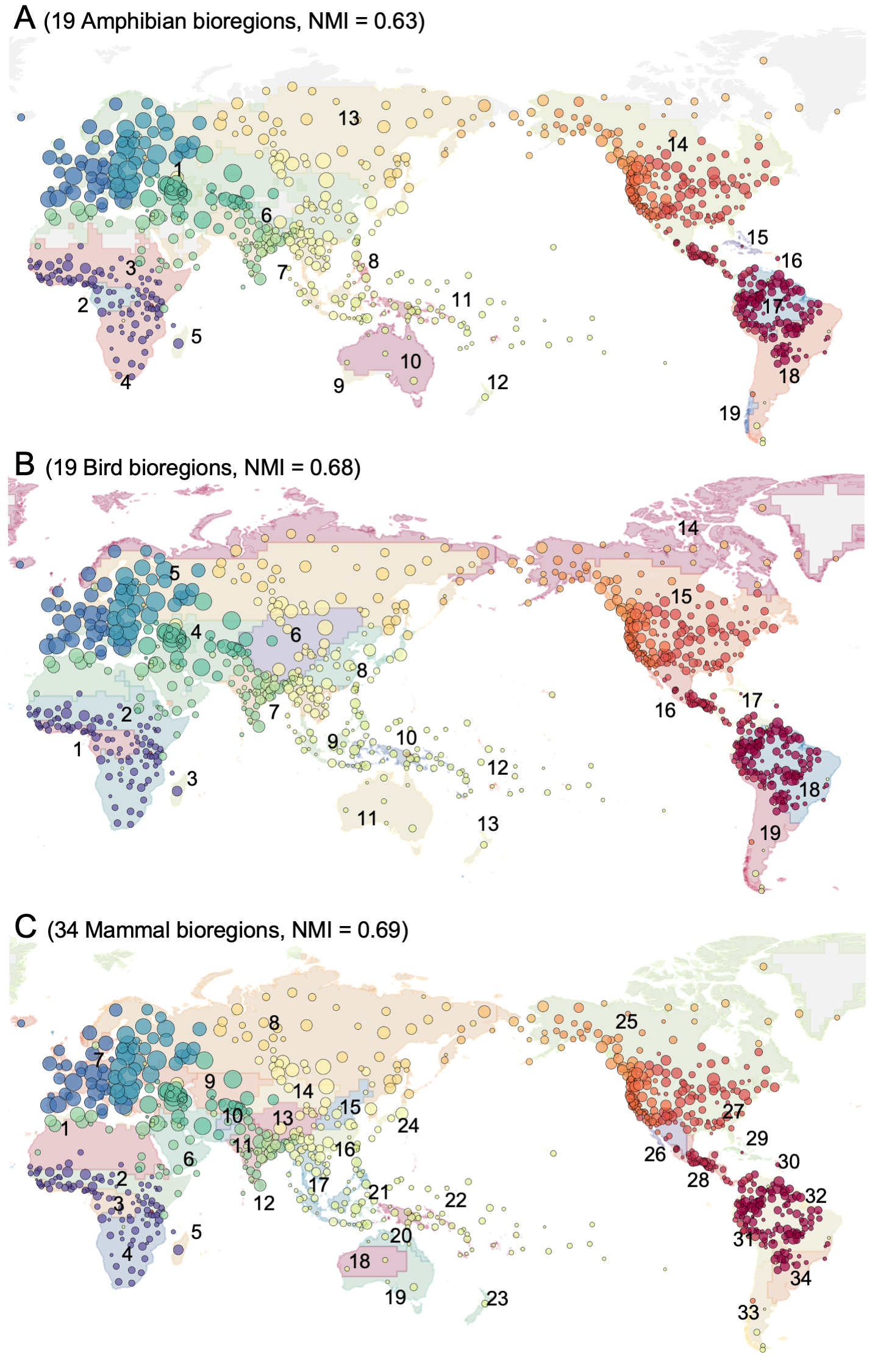}}
\noindent {\bf Figure 2.} The mythological clusters at the scale of $|\alpha| = 16$ overlain on bioregions, including (A) amphibians, (B) birds, and (C) mammals whose regions are colored ~\cite{Holt74}. We quantify their overlap memberships by calculating the normalized mutual information (NMI) between mythological clusters and these bioregions in Fig. 1D (blue line), which is exceptionally higher than what is expected from the bootstrapped groups (gray line) (SI Section~\ref{si:bioregion}).
\end{figure*}

Biogeography and language affiliation are both widely recognized to influence the dispersion of cultural traits in space ~\cite{harcourt2012human, lomolino2017biogeography,bortolini2017inferring}. Biogeography restricts demic movements (the dispersal and interaction of people) in space, either through environmental discontinuities or physical barriers to movement, such as oceans, mountain ranges, major rivers, or deserts, thus creating natural scale limits \cite{lomolino2017biogeography}. Likewise, shared language facilitates information transmission within a linguistic group, while language discontinuities act as boundaries, imposing limited scales of cultural variation. Importantly, local traditions are likely to share evolutionary histories, a phenomenon known as ``Galton's Problem'' in anthropology \cite{mace1994comparative}. 

To disentangle the marginal effects of biogeography and language family on the geographic boundaries of the clusters, we conducted an AMOVA (analysis of molecular variance) to decompose the motif variation into three hierarchical levels: 1) among biogeographic regions; 2) among language families within biogeographic regions; and 3) within language families \cite{excoffier1992analysis, bortolini2017inferring} (SI Section~\ref{si:amova}). 
We find that biogeographic boundaries explain about 4\% of the variance ($p$-value$<$0.001), similar to the variance explained by language families within biogeographic regions.
Both the mutual information analysis shown in Fig. 1D and the results of the AMOVA indicate that biogeography is more of a boundary to motif dispersal than language family affiliation. As a result, neighboring traditions within biogeographic boundaries are likely to share similar motifs, often reflecting specific features of local environments~\cite{michalopoulos2021folklore}, contributing to the forming of motif clusters as identified in Fig. 2 (SI Section~\ref{si:bioregion}). 

\begin{figure}[h]
\centerline{\includegraphics[width=0.6\textwidth]{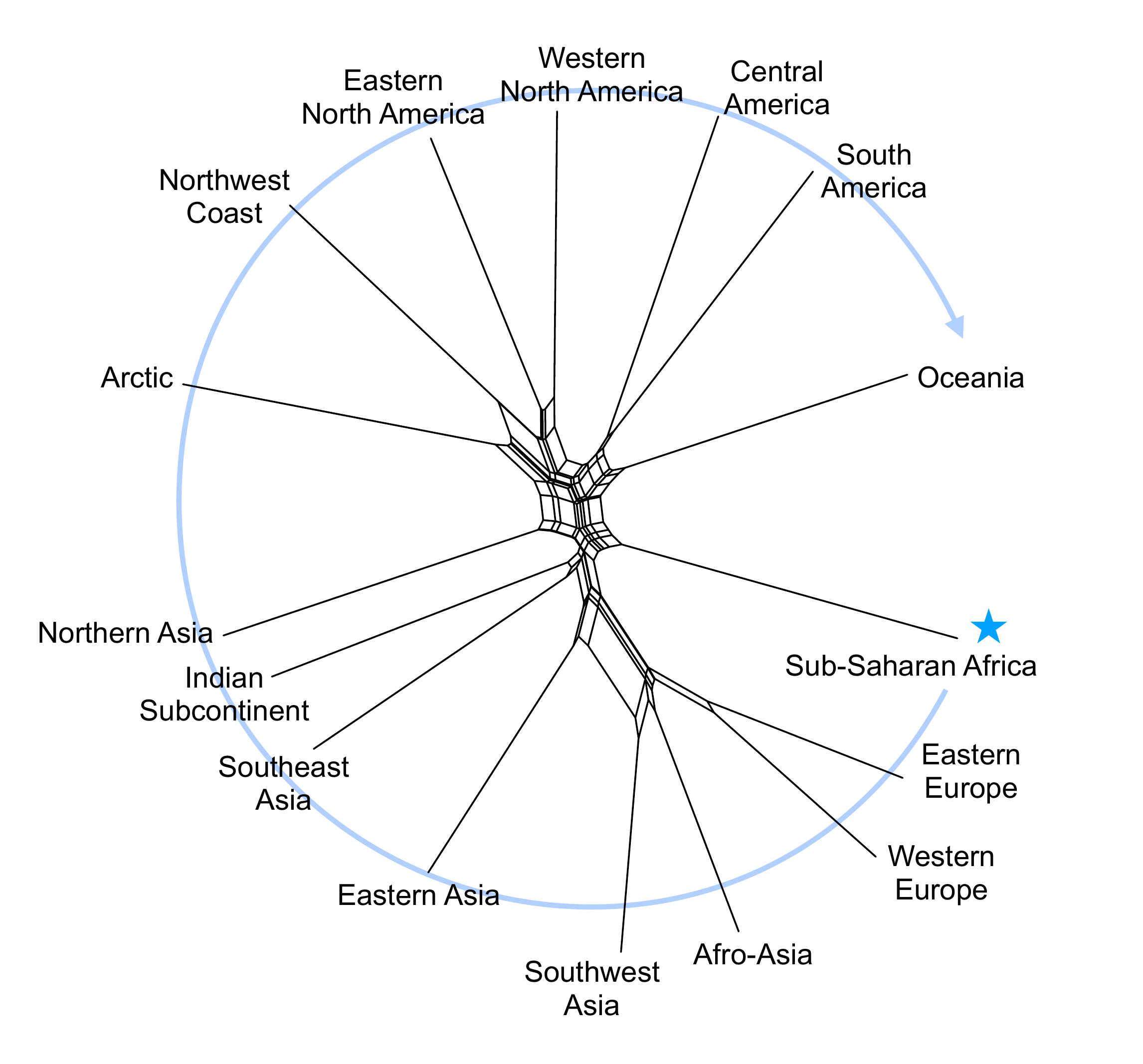}}
\noindent {\bf Figure 3.} A phylogenetic network of latent myths constructed using the Neighbor-Net algorithm~\cite{neighbornet}, connecting each leaf to one or more parent nodes based on similarities. The Jaccard distance is used as the metric for this network. The sequence of clusters in the network aligns well with historical human migration patterns.
\end{figure}

The hierarchical manifestation of clusters (Fig. 1) and their spatial distribution associated with biogeography and language (Fig. 2) suggest complex relationships between clusters. To visualize these relationships, we built a phylogenetic network based on the Jaccard distance between latent myths using the Neighbor-Net algorithm \cite{neighbornet}, an agglomerative method that allows more than one parent for each leaf (Fig. 3). In this network, edge length represents the degree of difference between two leaves. Interestingly, although we did not impose any geographical information during network construction, the detected clusters are grouped into the Old and New Worlds. The sequence of clusters aligns well with the prehistoric dispersal of humans out of Africa. This alignment is significant compared to 1,000 random permutations.

\section*{Latent Myths Embedded in Nested Hierarchy}

Multi-scale groups $M^l_\alpha$ have two dimensions: aggregated traditions $t \in \alpha$ and latent myths $m_i \in M^l_\alpha$ representing the coarse-graining of each group's information.  
Unlike mythological traditions, whose hierarchically nested structure is easily observable through geographical clusters, the latent structure of motifs is less evident. Therefore, we analyzed motifs distributed across latent myths, including both their frequency across scales and their geographical and hierarchical distributions. 

The retention rate, $r_{\alpha_i, \alpha'_k}$ measures the proportion of motifs that remain when aggregated into a larger scale, from $\alpha'$ to $\alpha$ (see Methods). 
As Fig.~1C illustrates, high retention rates (indicated by numbers at each branching point between varying scales) are maintained across scales, suggesting a hierarchical nesting of latent myths themselves \cite{mariani2019nestedness}. 
For example, latent myths of Western and Eastern Europe ($M^l_{West_Europe}$ and $M^l_{East_Europe}$) preserve 90\% of their motifs when aggregated into the broader European cluster  $\alpha_{Europe}$, as displayed in the blue branches at the upper-right of Fig. 1C.

This coarse-graining structure continues. 
Mythological traditions are further aggregated within Southwest Asia/North Africa, retaining nearly 80\% of their motifs in the larger-scale latent myth, that is, $M^l_{blue}$. 
Take Netsilik as another example, a local mythological tradition that belongs to an Arctic motif collection. Almost 90\% of the Arctic latent myth is also affiliated with the Northern North American cluster, which geographically branches off a more general North American motif community that itself emerged from the original separation of the Old and New Worlds. The majority of motifs that are found in the Netsilik latent myth are retained all the way from the initial Old World/New World split (Fig. 1A), demonstrating a remarkably deep nested structure.
It should be noted that the observed nesting isn't a byproduct of methodological choices; rather, each cluster level is identified independently of any preconceived nested structure.

\begin{figure}[!]
\centerline{\includegraphics[width=0.8\textwidth]{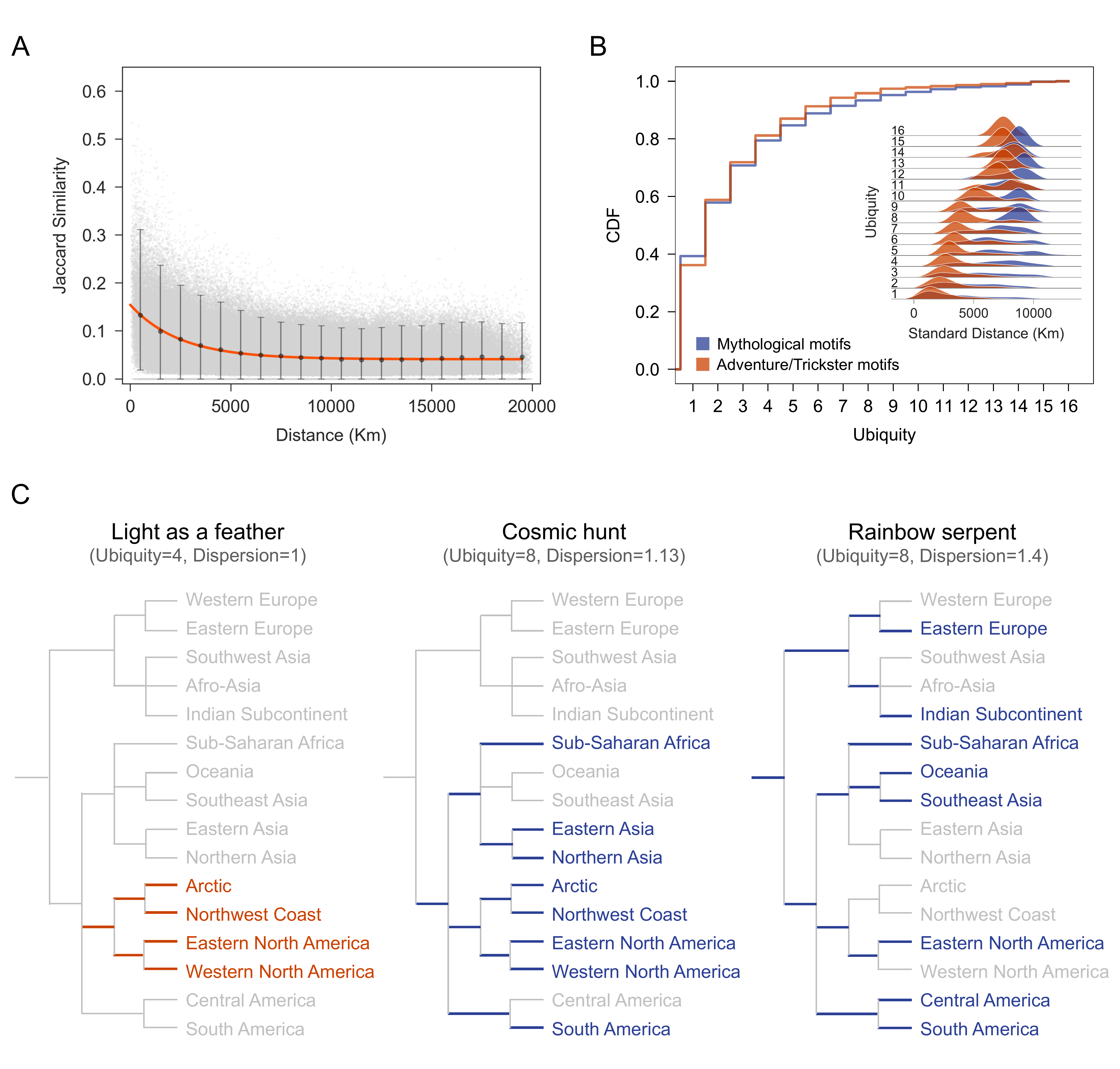}}
\noindent {\bf Figure 4.} (A) Spatial auto-correlation of mythological traditions, measured by Jaccard distance between their motif incidences (gray) as a function of geographic distance $x$. 
Binned averages (black points with 95\% error bars) are well-fit with an exponential function, $e^{-x/x_0}$ (red) where the estimated characteristic distance $x_0$ is $2,725$km, suggesting radial diffusion plays a role in transmission within a scale somewhat smaller than that of an average continent. 
(B) The distributions of motif ubiquity (latent frequency): Most motifs exhibit more than three mythological clusters at the scale of bioregions, and adventure and trickster motifs tend to be more localized than other mythological motifs (as shown in the inset).
(C) Motif dispersion: a motif's distribution across nested structures by the total branch lengths among its mythological clusters. For instance, `Light as a Feather' has minimal dispersion with a ubiquity score of four, while `Rainbow Serpent' and `Cosmic Hunt' have similar ubiquity scores to each other but differ in their distribution. `Rainbow Serpent' is more widely dispersed, contributing significantly to the nested structure. We normalize the total lengths with the minimum length for a given ubiquity. 
\end{figure}

High retention rates across different scales indicate that persistent conservative motifs serve as underlying threads that connect diverse narratives and cultures at different scales.  
What precisely are these long-lasting and stable motifs, and to what extent are they ingrained in the hierarchical tree? 
We use two metrics to characterize these motifs: ubiquity $U$, which quantifies the extent to which motifs are repeatedly found at the tree's leaves, and dispersion $D$, which gauges the depth at which these recurring motifs are spread throughout the tree as shown in Fig. 4C. 
With these two quantities, we identify motifs that contribute to the hierarchical nestedness of clusters, as well as those that induce variations within them. 

Fig. 4B shows a cumulative distribution of ubiquity where we find nearly half of all motifs are limited to one or two terminal clusters, that is, $\{ m_i | U (m_i) \leq 2 \}$ where $U (m_i) = |\{\alpha | m_i \in M^l_\alpha\}|$ (see Methods and SI Section~\ref{si:ubiquity}). 
These motifs, by their relative uniqueness, contribute to variation in mythologies, preventing certain groups from merging and thus forming distinct tree branches.
Conversely, ultraconserved motifs, retained across all latent myths, play a crucial role in maintaining the tree's overall depth, persisting across all latent myths and upholding the global structure. 
However, these motifs alone don't entirely explain the consistent hierarchical nesting observed in Fig. 1. 
Rather, motifs with a moderate level of ubiquity, accounting for roughly 10\% of the total, play a significant role in this structure. Examples include widely recognized motifs like the `Cosmic Hunt' and `Rainbow Serpent.'

Fig. 4B inset demonstrates that the geographic reach of these motifs is not simple but shows an almost bimodal distribution for those motifs.
We quantify a motif's dispersion as the average radius across all traditions in which it appears (SI Section~\ref{si:standard_distance}). 
Generally, dispersion increases with a motif's ubiquity.  
For example, `Taste of Blood' (ubiquity=2) is localized to specific regions, including the Indian Subcontinent and Western North America, and `Singing Bird of the Hero' only occurs in Eastern North America and Western Europe. Both show a narrow radius of their geographic distributions.  
Conversely, the average radius grows with the number of traditions in which a motif appears. 
Intriguingly, motifs with medium ubiquity show a bimodal dispersion, indicating that different categories of motifs serve various functions in sustaining the overall narrative architecture.

Consequently, we distinguish between motifs of cosmological myths and folklore (i.e., motifs related to adventure and tricks). 
Cosmological motifs, often tied to elemental concepts like the Sun, Moon, Earth, and Water, enjoy a wider geographic reach. These motifs resonate with ancient Paleolithic narratives and appear to be deeply ingrained in human culture ~\cite{dhuy2013cosmic, dhuy2013phylogenetic, Youn16022016}. 
On the other hand, folklore motifs are more localized, particularly in regions of low ubiquity, which is consistent with existing knowledge  \cite{berezkin2005continental}. 
This contrast in distribution is evident in the inset of Fig. 4B: cosmological motifs (blue) are geographically more widespread than their folklore counterparts (orange).
These differences create a rich tapestry of cultural histories that simultaneously occupy the same narrative ecosystems \cite{thompson1955myths}. Such observations underscore the need for further research to delve into the functional distinctions between these motif categories.

While ubiquity quantifies the prevalence of motifs across latent myths, the dispersion $D (m_i)$ provides an additional layer of complexity by accounting for the depth of the tree. 
The dispersion measures the total length of branches connecting the clusters where a motif occurs, $\{ \alpha_i | m_i \in M^l_{\alpha_i}\}$ (highlighted in Fig. 4C). 
To control for the influence of ubiquity on dispersion, we normalize this total branch length by the minimum total branch length of motifs of the same ubiquity (see Methods and SI Section~\ref{si:dispersion}).
For example, $D=1$ indicates the motif appears in neighboring clusters, while $D>1$ indicates the motif's presence across latent myths of non-adjacent clusters.

Indeed, comparing ubiquity and dispersion for each motif further uncovers the motif's contribution to the nuanced structures. 
Fig.~4C illustrates how dispersion captures the variation in the distribution of motifs for a given level of ubiquity.
Take the motif `Light as a Feather,' which is confined to North America but appears in all four of its latent myths -- Arctic, Northwest Coast, Eastern, and Western -- yielding a $U$ = 4, and its minimal dispersion ($D=1$) indicates its presence in adjacent clusters. Therefore, this motif is important at a small scale. 
`Cosmic Hunt' is twice as ubiquitous ($U$ = 8), linking by 17 branch lengths in a total of eight clusters spanning multiple continents but only slightly more dispersed ($D$ = 1.13), in comparison to `Rainbow Serpent' ($U$ = 8 and $D$ = 1.4). The wide distribution of `Rainbow Serpent' results in a dispersion that diverges significantly from the minimum, suggesting a pattern that does not favor geographic clustering in Fig.~1A. 

Low ubiquity motifs with high dispersion are of particular interest. These motifs, as listed in Table~S\ref{si:tab:max_dispersion_motifs} in the SI, occur only in a few clusters but span the deeply nested tree structure (SI Section~\ref{si:dispersion}). 
This dispersion suggests two possibilities: either these motifs were once prevalent but have disappeared in adjacent clusters, or they have been independently reinvented multiple times through convergence (i.e., suggestive of the human collective unconscious) \cite{jung1959archetypes}.
While our analysis does not address these two generative models directly, given the motifs' high retention rates across a branching tree, we lean towards the former explanation. This would suggest that such widely dispersed but infrequently occurring motifs likely stem from a shared ancestry rather than repeated convergent innovation, reinforcing the idea that mythologies are generally more conservative in their evolution compared to the more locally and recently innovated folktales.

\section*{High-order Structure in Mythological Motifs}

In the rich tapestry of global mythologies, motifs are not isolated threads but essential building blocks to be bundled with others, resulting in a high-order structure.
Such complex relations can be represented in a motif network, where two motifs  $m_i$ and $m_j$ are connected when they co-appear in an underlying latent myth $M^l_{\alpha}$. 
Using a `multiscale backbone' then allows us to sidestep incidental ties and identify statistically significant connections \cite{Serrano6483} (SI Section~\ref{si:backbone}). 

Fig. S\ref{fig:backbone_entire} illustrates a resulting network with tie strength weighted by the co-occurrence frequency. 
At the core of the backbone network structure (Fig. S\ref{fig:backbone_entire}) are the four ultraconserved motifs universally present across all latent myths, contributing thus to high retention rates in Fig. 1C. 
These motifs include `Female Sun', `The Mysterious Housekeeper', `Tasks of the In-Laws', and `The Obstacle Flight' which are recurrent with various thematic tales of tricksters, supernatural women, celestial bodies, and adventure throughout the world. 
While the ubiquity of these particular motifs is well-documented, they represent merely the surface of a more complex, nested structure that spans the globe \cite{green1997folklore}. 

\begin{figure}[!]
\label{fig_hierarchy_nested}
\centerline{\includegraphics[width=0.65\textwidth]{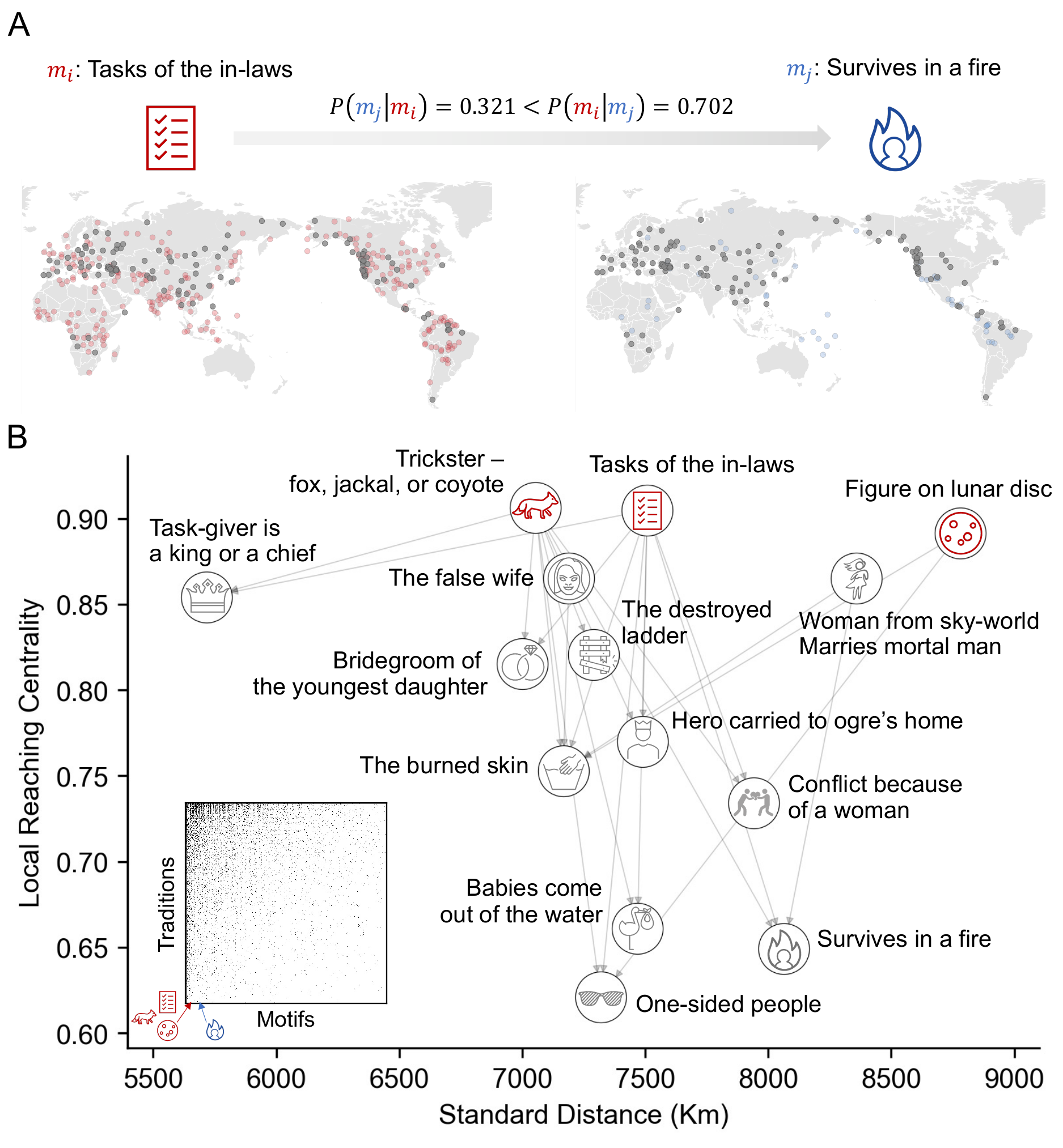}}
\noindent {\bf Figure 5.} Motif hierarchy and conditional probabilities. (A) Schematic for inferred directionality from the asymmetric conditional probability of a motif contingent on another motif. The `Survives in a fire' motif (blue) is more likely to appear if `Tasks of the in-laws' (red) exist than the opposite, as is a conditional probability as well as in the map where their coexistent are marked in black (see Methods and SI) \cite{jo2020extracting}. 
(B) The hierarchy is constructed from the aggregated weighted directions of all possible motif pairs. A node’s horizontal and vertical positions are, respectively, its standard distance (as in Fig. 4B) and local reaching centrality, which is the proportion of other nodes reachable from that node \cite{mones2012hierarchy}. Only highly ubiquitous motifs are shown to ensure statistical power and for visualization. 
The inset shows $M^e_t$. 
\end{figure}

The last piece in the puzzle of decoding the hidden nested structure is to examine conditional probabilities -- essentially, how one motif begets another. 
This goes a step beyond the mere co-occurrence illustrated in Fig. S\ref{fig:backbone_entire}, where directional relationships were notably absent. 
The expectation of asymmetry is confirmed by examining upper-triangle fills in the empirically observed $M^e_t$ (inset of Fig. 5B). Note that this upper-triangle shape isn't perfect due to the modular, rather than mono-nested, nature of the structure, as already shown in Fig. 1 \cite{mariani2019nestedness}. 

Instead of tracking which motifs frequently appear together, we examine which motifs actually set the stage for others, acting as foundations in the empirical mythological data, that is, $M^e_t$, not $M^l_\alpha$.  
Essentially, we calculate the conditional probability of observing $m_i$ when $m_j$ exists in a tradition, $p (m_i|m_j)$. If $p (m_i|m_j)$ is larger than $p (m_j|m_i)$, $m_i$ precedes $m_j$ in a tradition. 
We specifically target these asymmetric relationships and assign a directional arrow from $m_i$ to $m_j$ \cite{jo2020extracting}.  

In this way, motifs that commonly act as precursors emerge as the roots of the motif tree, with subsequent motifs materializing as leaves in Fig. 5. 
These leaves are geographically placed based on their dispersion, positing that localized motifs will behave differently from global motifs.
Fig. 5 displays only ubiquitous motifs to ensure the statistical power for asymmetric conditional probability for directions, as well as simplifying visualization, but it is critical to note that these conditional probabilities are derived from the full, unfiltered empirical data set $M^e_t$ (inset of Fig. 5B). 

Our findings identify three motifs at the apex of Fig. 5 that act as keystones, creating branching paths of subsequent motifs: these are `Tasks of the in-laws,' `Figure on lunar disc,' and `Trickster-fox, jackal or coyote.' 
The motifs on the top layers are those highly populated motifs (red marks in the inset of Fig. 5B), and motifs on the bottom of the tree are those located in less populated areas which are nested within those rich traditions (toward the right in the inset of Fig. 5B). The implication is their high informational value in the mythological tradition. For instance, `Figure on lunar disc' is not intrinsically rich in motif composition but becomes so when paired with others. On the contrary, the core motifs in Fig. S\ref{fig:backbone_entire} are not ubiquitously found; they are relatively rare but assume a lead role when bundled, acting as prime movers in the rich tapestry of global mythologies.
All these results indicate the deeply nested hierarchical structure in our narrative traditions.

\section*{Discussion}

In our study, the coarse-graining methodology we employ uncovers the geographical and thematic nesting of global mythologies across multiple scales. 
This multiscale method yields several key insights, unpacking how the theoretical approach depends on the choice of analytic scale, either implicitly or explicitly. 
At the finest-possible scale, each mythological tradition becomes a distinct latent myth, demanding its own tailored theoretical framework for its cultural and historical evolution. 
As we broaden our analytic scale by coarse-graining, there is a trade-off as we sacrifice detail in favor of greater generality, gradually converging toward an overarching theory. 

Our study delineates the salient scales at which this occurs. The initial transition is at the biogeographic scale, a point at which granularity is sacrificed for a more general understanding. 
The second transition becomes apparent at the continental scale, where Fig. 4A indicates a significant decline in the auto-correlation between mythological traditions at a characteristic distance of $2,725$km, just below the average size of a continent. This distance also correlates with a weak retention rate, as noted when $|\alpha| = 2$ in Fig. 1C (SI Section~\ref{si:sim_decay_ind}) ~\cite{kavvas1981stochastic,nekola1999distance}.

What is so remarkable about these results lies in the robust spatial structures we uncover, even when employing a model-free approach. 
This approach treats motifs as random variables, disregarding factors like geography, history, or story lineage, not to mention evolutionary mechanisms.
It seems like the statistical structure left behind by cultural mechanisms is so strong enough that this structure exhibits a clear anthropological signature, bounded in space by language families, biogeography, and continents. We show the temporal and spatial structure others have reported about specific myths \cite{dhuy2013cosmic, dhuy2013phylogenetic, berezkin2005cosmic, berezkin2005continental, berezkin2007dwarfs, berezkin2007earth, tehrani2013phylogeny}, are general features of this entire database. What sets our work apart is its generalizability: We demonstrate that not only do motifs cluster geographically \cite{thuillard2018large}, but also fit into a more expansive spatial hierarchy that encompasses the entire database (See SI Sec.~\ref{si:clustering_results}).  

The discrete hierarchical clustering we find identifies a handful of keystone motifs found in all clusters at all scales---`Trickster, fox, jackal, or coyote', `Tasks of the in-laws', and `Figures on lunar disk'--- highlighting the global nature of this hierarchical structure. Moreover, the depth of these keystones suggests an evolutionary structure. If taken literally, this evolutionary history would imply that either the hunter-gatherer societies of Middle-to-Late Stone Age Africa that eventually seeded the human biogeographic expansion across the planet carried with them tales of tricksters, in-laws, and figures on the lunar disk as they spread, which have been retained ever since. Or, alternatively, human societies around the planet have a universal tendency to create stories about features common to their experience, including canines, in-laws, and the moon, to the extent that such stories seem to converge on these roots. The analyses we conduct here cannot discriminate between these two hypotheses.

Interestingly, however, subjecting our data to the Neighbor-Net -- a tool frequently employed in evolutionary biology to reconstruct phylogenetic networks where simple tree-like structures cannot be assumed ~\cite{bryant2004neighbor} -- our analysis produced distinct networks shown in Fig. 3 and Fig. S\ref{si:fig:neighbornet} in SI Section~\ref{si:neighbornet}. 
Using two distance methods, both networks are largely consistent with an out-of-Africa model, flowing from Africa to Eurasia, the Americas, and eventually Oceania. These network flows also align with existing evolutionary frameworks for global mythological traditions and find further support from genetic and archaeological evidence~\cite{armitage2011southern,henn2012great,nielsen2017tracing,witzel2012origins}. Our significance testing of 1,000 randomized networks confirms the robustness of this spatial structure. If correct, these results would support a model of deep ancestry rather than convergence.

The mythological record we see around the world today in traditional societies is a palimpsest of motifs aggregated over time and space over the course of human history. Some have asked how deep particular stories within that palimpsest go. Others have asked how such stories evolve over time and space. Here we asked, how is this palimpsest structured? Myths, tales, and folklore have been shared among human societies for as long as humans have existed, with the oldest stretching back into the Paleolithic \cite{dhuy2013cosmic, berezkin2005cosmic}. This is reflected not only in phylogenetically ancient myths but in Paleolithic traditions of rock art, ornamentation, and statuary, undoubtedly the archaeological residue of ancient belief systems. The deep structure we report here is the result of these cultural evolutionary processes playing out over tens of thousands of years.

Mythology has always played a central role in anthropology because it is such an evocative human universal, encoding the complex links between people, each other, and their environments into stories that continue to play a central role in the human world \cite{michalopoulos2021folklore, jackson2023supernatural}. It is of little surprise then that comparative mythology offers such a rich source of information about cultural evolutionary processes ~\cite{dhuy2019folk, thuillard2018large, thuillard2018computational, dhuy2013phylogenetic, tehrani2013phylogeny, bortolini2017inferring, berezkin2016spread, berezkin2008people, vuong2018cultural, tangherlini2013folklore, tangherlini2016, abello2012computational, berezkin2015spread, witzel2012origins}.
Comparative mythology has long proposed that motifs show regional and even global patterns. Here, we have shown the depth of this structure. In a similar vein to historical linguists who have identified ``ultraconserved'' elements in words, languages, and beliefs~\cite{pagel2013ultraconserved, Youn16022016, swadesh1955towards, calude2011we, grollemund2015bantu, greenhill2010shape, Greenhill17102017, bellah2011religion, eliade1959sacred, norenzayan2013big, botero2014ecology}, local mythological traditions also appear to harbor enduring features that reflect the rich tapestry of human cultural diversity.

\section*{Data and Methods} \label{sec: method}

\textbf{Coarse-graining Mythological Traditions}

The CorEx algorithm assigns mythological traditions into groups, $\alpha$, in a manner that maximizes the total explanatory power of latent myths. 
This explanatory power is quantified using mutual information. Specifically, the sum of multivariate mutual information between the traditions within each group and the group's corresponding latent myth (SI Section~\ref{si:corex}) \cite{NIPS2014_5580}. 

Given a specific scale $|\alpha|$, we assign $t$ to one of $\{\alpha_1, \alpha_2, ...\}$ to maximize the function, 
$\max_{\{\alpha, M^{l}_{\alpha}\}} \sum_{i=1}^{|\alpha|}[\sum_{t\in\alpha_{i}}I(M_{t}^{e}:M^{l}_{\alpha_{i}})-I(M_{\alpha_{i}}^{e}: M^{l}_{\alpha_{i}})]$, 
where $M^{e}_{t}$ is empirically observed motif in tradition $t$ and $M^{l}_{\alpha_{i}}$ is a latent myth of traditions assigned to $\alpha_i$. 
$I(X:Y)$ represents mutual information between two binary vectors $X$ and $Y$, expressed as $I(X:Y) = \sum_{x\in X}\sum_{y\in Y}p(x,y)\log{\frac{p(x,y)}{p(x)p(y)}}$ where $p(x)$ is the probability for $x$ in $X$ and $p(x,y)$ is the probability for the combination $(x,y)$ in the binary vectors $X$ and $Y$. In case of our data where $x$ and $y$ take presence or absence in $M^e_t$ and $M^l_\alpha$, $p(x)$ indicates the fraction of motifs in $M^e_t$, $p(y)$ the fraction of motifs in $M^l_\alpha$, and $p(x,y)$ the fraction of motifs that appear in both. Therefore, the information $p(x,y)$ is normalized by random expectation $p(x)p(y)$. 

As we increase the value of $|\alpha|$, the explanatory power also rises. This is because each group's latent myth needs to explain the small group size. For example, in the extreme case, $|\alpha|$ equals $|t|$, the maximum possible group size, the explanatory power reaches its peak. This is because $I(X:X)$, the unit of explanatory power, is at its maximum when a variable is compared with itself. 

\noindent\textbf{Quantify Nested Structure: Retention rate, Ubiquity, and Dispersion}\\
In the analysis of mythological traditions, we employ a multiscale, coarse-grained approach. 
The number of groups varies with scale, expressed as $|\alpha| \neq |\alpha'|$ where the prime indicates a group at an alternate scale. Given this, a one-to-one comparison between groups at different scales is unfeasible. Instead, we adopt a one-to-many mapping function. 
For instance, several $\alpha'_i$ groups will correspond to a single $\alpha_k$ if $|\alpha'| > |\alpha|$, and this is where the nesting structure arises. 

The basis for this mapping is shared membership, evaluated along two dimensions, traditions and motifs, captured in $t \in \alpha$ and $m \in M^l_\alpha$, respectively. 
The nesting structure in the tradition dimension is straightforward: traditions of the same group are not only geographically clustered, but the smaller clusters are also subsets of larger ones, as evidenced in Fig. 1A. The motif dimension is more complex, requiring additional calculations to quantify the mapping function's effectiveness through what we term the $r_{\alpha_i, \alpha'_k}$. 

We initiate this by calculating the proportion of motifs shared between two groups $\alpha_i$ and $\alpha'_k$ at scales $|\alpha|$ and $|\alpha'|$, respectively.
Suppose $\alpha'_k$ is nested to $\alpha_i$. Then, the most of latent myths would still be observed in the $M^{l}_{\alpha_i}$, represented as $r_{\alpha_i, \alpha'_k} = |M^{l}_{\alpha_i} \cap M^{l}_{\alpha'_k}| / |M^{l}_{\alpha'_k}|$. 
Accordingly, we then examine all conceivable pairs of groups at different scales along the branches between groups of different scales in Fig. 1C.
In case the coarse-graining is not nesting structure across different scales, the retention rate would be low. 
This is because the membership of motifs to latent myths ($m \in M^l_\alpha$) is recalculated at each scaling level.  Consequently, when traditions are aggregated at broader scales, a motif $m_i$ found at one level may not necessarily persist at the next. Therefore, the prominence of high retention rates, as shown in Fig. 1C, is particularly noteworthy. 

In the worst-case scenario, if a motif doesn't remain persistent across scaling levels, there would be no retention rate, potentially undermining our ability to discern a hierarchically nested structure. As a closing observation for future exploration, it's worth noting that the nested structure illustrated in Fig. 1A is not monolithic nesting, akin to a Russian doll. Instead, it displays a modular nested structure, shown as a branching tree in Fig. 1, and this is the reason why the upper-triangular fill of  $M^e_t$ is fuzzy in Fig. 5B inset \cite{mones2012hierarchy}.

Ubiquity ($U (m_i)$) is defined as the number of latent myths in which motif $m_i$ appears at the scale of biogeography, that is, $|\{\alpha_k | m_i \in M^l_{\alpha_k}\}|$ (SI Section~\ref{si:ubiquity}). Fig. 4B compares $U(m_i$ with its average radius of traditions $\{t | m_i \in M^e_t\}$ (SI Section~\ref{si:standard_distance}). 
While the ubiquity captures motif abundance across latent myths at the biogeography scale, a motif's dispersion ($D (m_i)$) accounts for the depth of the tree by calculating the total branch length that connects $\{ \alpha_i | m_i \in M^l_{\alpha_i}\}$ in which a motif is found.  
To avoid the factors of ubiquity in the dispersion measure, we normalize the total length by the minimum total branch length of other motifs of the same ubiquity (SI Section~\ref{si:dispersion}).

\noindent\textbf{Conditional Probabilities for Motif Hierarchy Structure}\\
The conditional probability that infers the directionality operates on the empirically observed tradition-motif matrix $M^e_t$. 
We calculate conditional probabilities of every pair of motifs in the matrix to infer directions between two motifs, as illustrated in Fig. 5A \cite{jo2020extracting}. Fig. 5B presents a backbone structure of the aggregated motif pairs, according to \cite{Serrano6483} (SI Section~\ref{si:conditional}).

\clearpage
\bibliography{scibib}
\bibliographystyle{Science}

\section*{Acknowledgments}
The authors thank Yuri Berezkin for sharing his dataset and insightful comments. The authors also thank Julien d'Huy, Manvir Singh, Robert Walker, Briggs Buchanan, and Brian Uzzi for insightful comments on earlier versions of this manuscript. H.Y. would like to acknowledge the support of the National Science Foundation Grant Award Number 2133863 and the NRF Global Humanities and Social Sciences Convergence Research Program (2024S1A5C3A02042671). 

\end{document}


\title{Supplementary Information: Deep nested structure of mythological traditions worldwide }
\author{Hyunuk Kim}
\affiliation{Department of Management and Entrepreneurship, Martha and Spencer Love School of Business, Elon University, Elon, NC 27244, USA}
\affiliation{Northwestern Institute on Complex Systems, Evanston, IL 60208, USA}
\author{Marcus J. Hamilton}
\email{marcus.hamilton@utsa.edu}
\affiliation{Department of Anthropology, University of Texas at San Antonio, San Antonio, TX 78249, USA}
\affiliation{School of Data Science, University of Texas at San Antonio, San Antonio, TX 78207, USA}
\author{Woo-Sung Jung}
\affiliation{Department of Industrial and Management Engineering, Pohang University of Science and Technology, Pohang 37673, Korea}
\affiliation{Department of Physics, Pohang University of Science and Technology, Pohang 37673, Korea}
\author{Hyejin Youn}
\email{hyejin.youn@kellogg.northwestern.edu}
\affiliation{Northwestern Institute on Complex Systems, Evanston, IL 60208, USA}
\affiliation{Department of Management \& Organizations, Kellogg School of Management, Northwestern University, Evanston, IL 60208, USA}
\affiliation{Santa Fe Institute, Santa Fe, NM 87501, USA}

\date{\today}
\maketitle

\tableofcontents

\clearpage

\section{Latent structure of motif co-occurrences in narratives}

\subsection{Motif database}
\label{si:data}

Myths and folktales are universal in human societies and essential to understanding how human cultural diversity has evolved over the planet. We analyzed the global structure of myths and folktales with a database that identifies the ethnolinguistic traditions in which particular narrative motifs are observed~\cite{berezkin2017peopling}. Constructed by Yuri E. Berezkin, the database includes information about motifs appearing across traditions. This allows us to exam larger-scale patterns of motif co-occurrences. 

From the data, we constructed the presence-absence matrix $W$ of 979 traditions and 2,718 motifs. Three motifs -- `Beware of cut off nails' (C32C), `Competitions and difficult tasks' (K27), and `The turtle is a tricky failure' (M29K1) -- are not included in the matrix by following Berezkin's previous studies. $W_{ij}$ is 1 if a tradition $i$ has a motif $j$ and 0 otherwise. 

\begin{figure}[h]
\centerline{\includegraphics[width=0.55\textwidth]{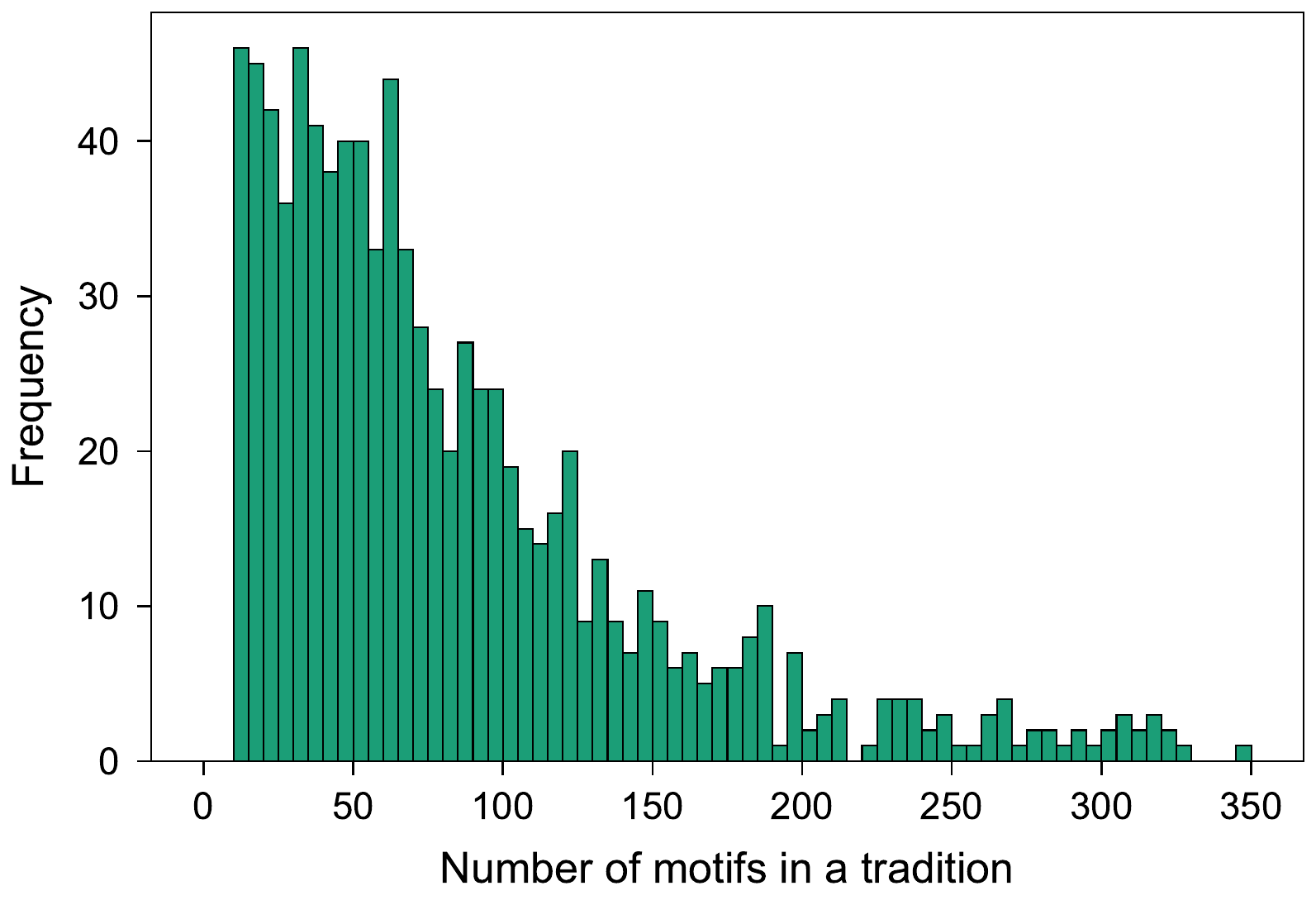}}
\caption{Histogram for the number of motifs in a tradition.}
\label{si:fig:hist_num_motifs}
\end{figure}

\begin{figure}
\centerline{\includegraphics[width=0.55\textwidth]{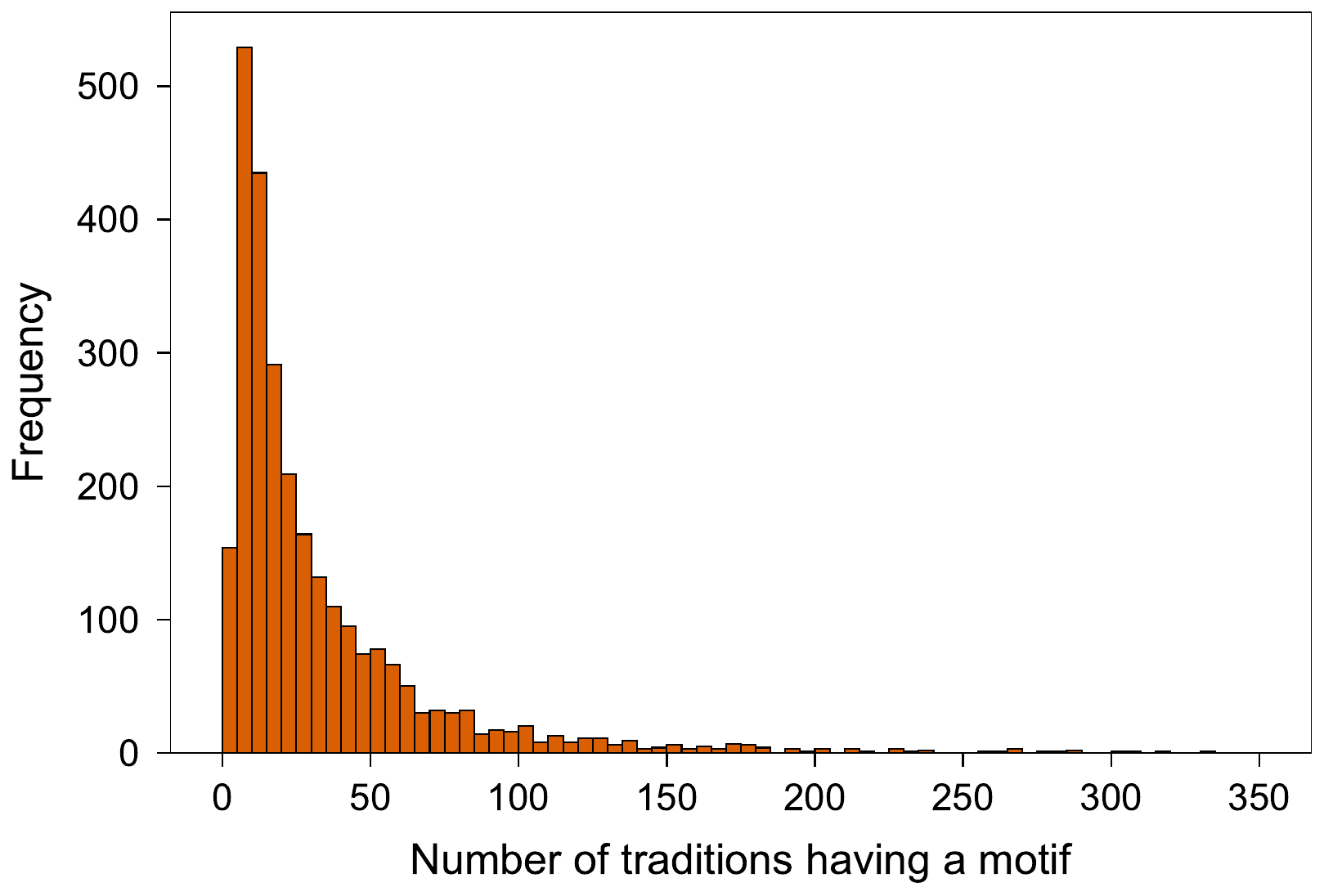}}
\caption{Histogram for the number of traditions having a motif.}
\label{si:fig:hist_motif_apperances}
\end{figure}

In order to reduce noise, we limited our analyses to the 921 traditions that have at least ten motifs. In the refined database there is an average of 96 motifs per tradition but this distribution is highly skewed (FIG \ref{si:fig:hist_num_motifs}). The number of traditions having any particular motif is also widely distributed with a mean value of 33 (FIG \ref{si:fig:hist_motif_apperances}). The five most frequent motifs are `Tasks of the in-laws' (K27N), `Figure on lunar disc' (A32), `Trickster-fox, jackal or coyote' (M29B), `Male sun and female moon' (A3), and `The Sun and the Moon are males' (A5). They are found in 368, 361, 354, 332, and 316 traditions respectively. 

\subsection{Multi-scale clusters of motifs}
\label{si:corex}

Each tradition has a distinct motif configuration that is recorded in the binary format, which makes comparing traditions of different sizes problematic. Consequently, the previous study \cite{berezkin2017peopling}, utilizing principal component analysis, explains only a small fraction of the data, focusing mainly on the spatial distribution of specific motifs. This subsample explanation, although rich in context, loses complex patterns of motif co-occurrences. To understand the structure better, we aggregated traditions into clusters through an information-theoretic learning algorithm called ``Correlation Explanation'' (CorEx)~\cite{NIPS2014_5580}. For binary variables, CorEx finds clusters better than other methods such as \textit{k}-means clustering \cite{NIPS2014_5580}. 

CorEx reduces a high dimensional data into a fixed number of clusters to best explain the existing correlations within the data. Correlation of this algorithm means the amount of shared information, which is called total correlation in information theory. In our study, CorEx works on the presence-absence matrix of 921 traditions and 2,718 motifs. $TC(X)$, the total correlation of a set $X$, is calculated as Equation \ref{si:eq:tc}, where $H(X)$ is the entropy of $X$ defined as $H(X)\equiv-\sum_{x}{p(x)\log{p(x)}}$. $x$ is 0 or 1, and $p(x)$ is a probability of $x$. The total correlation after conditioning a latent factor $Y$, $TC(X|Y)$, can be represented by adding the conditional term on both sides of the equation of $TC(X)$ (Equation \ref{si:eq:tcy}).

\begin{equation}
\label{si:eq:tc}
    TC(X)=\sum_{i=1}^{n}H(X_{i})-H(X)
\end{equation}
\begin{equation}
\label{si:eq:tcy}
    TC(X|Y)=\sum_{i=1}^{n}H(X_{i}|Y)-H(X|Y)
\end{equation}

$TC(X;Y)$, which is $TC(X)-TC(X|Y)$, indicates the amount of information that a latent factor $Y$ explains X. CorEx maximizes $\sum_{j=1}^{K}TC(X_{G_j};Y_j)=\sum_{j=1}^{K}[TC(X_{G_j})-TC(X_{G_j}|Y_j)]$ by updating $K$ latent factors ($Y_1, Y_2, ..., Y_K$) and their associated cluster assignments. $X_{G_j}$ is a group of traditions classified into a cluster $j$ that corresponds to a latent factor $Y_j$, a binary vector what we call the latent myth of the cluster $j$. A latent myth implies which motifs anchor co-occurrences within a cluster. For each trial, the algorithm updates $K$ latent factors ($Y_1, Y_2, ..., Y_K$) and cluster assignments until it meets convergence conditions. We stopped a CorEx trial if it reaches 200 iterations or it doesn't increase total correlation by $10^{-4}$. This process was repeated 1,000 times as the algorithm returns different cluster assignments depending on initial conditions. We used a publicly available Python library for CorEx implementation~\cite{gallagher2017anchored} (see details at https://github.com/gregversteeg/corex\_topic).

\begin{figure}
\begin{center}
\subfloat[][Total correlation by the number of clusters.]{
\includegraphics[width=0.46\textwidth]{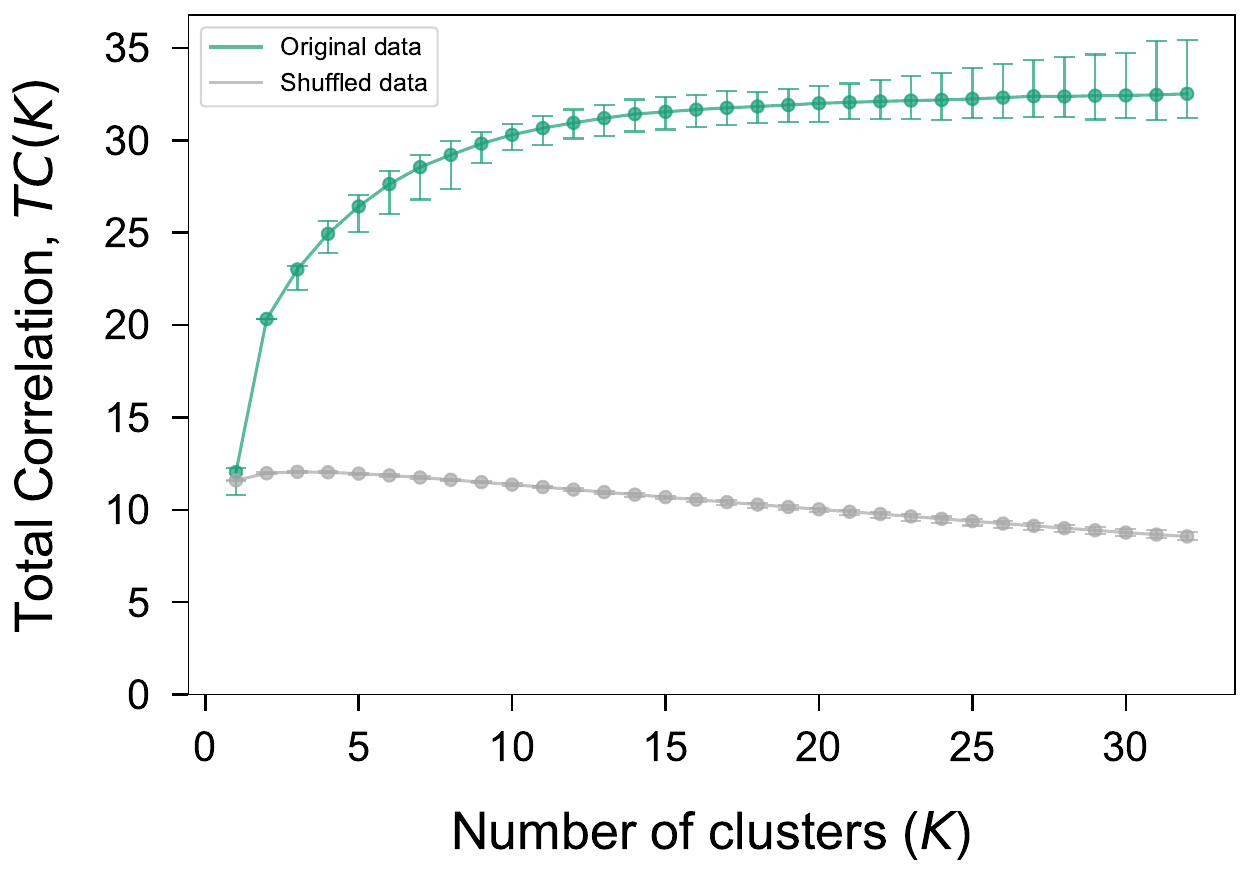}}
\subfloat[][The first derivative of total correlation.]{
\includegraphics[width=0.49\textwidth]{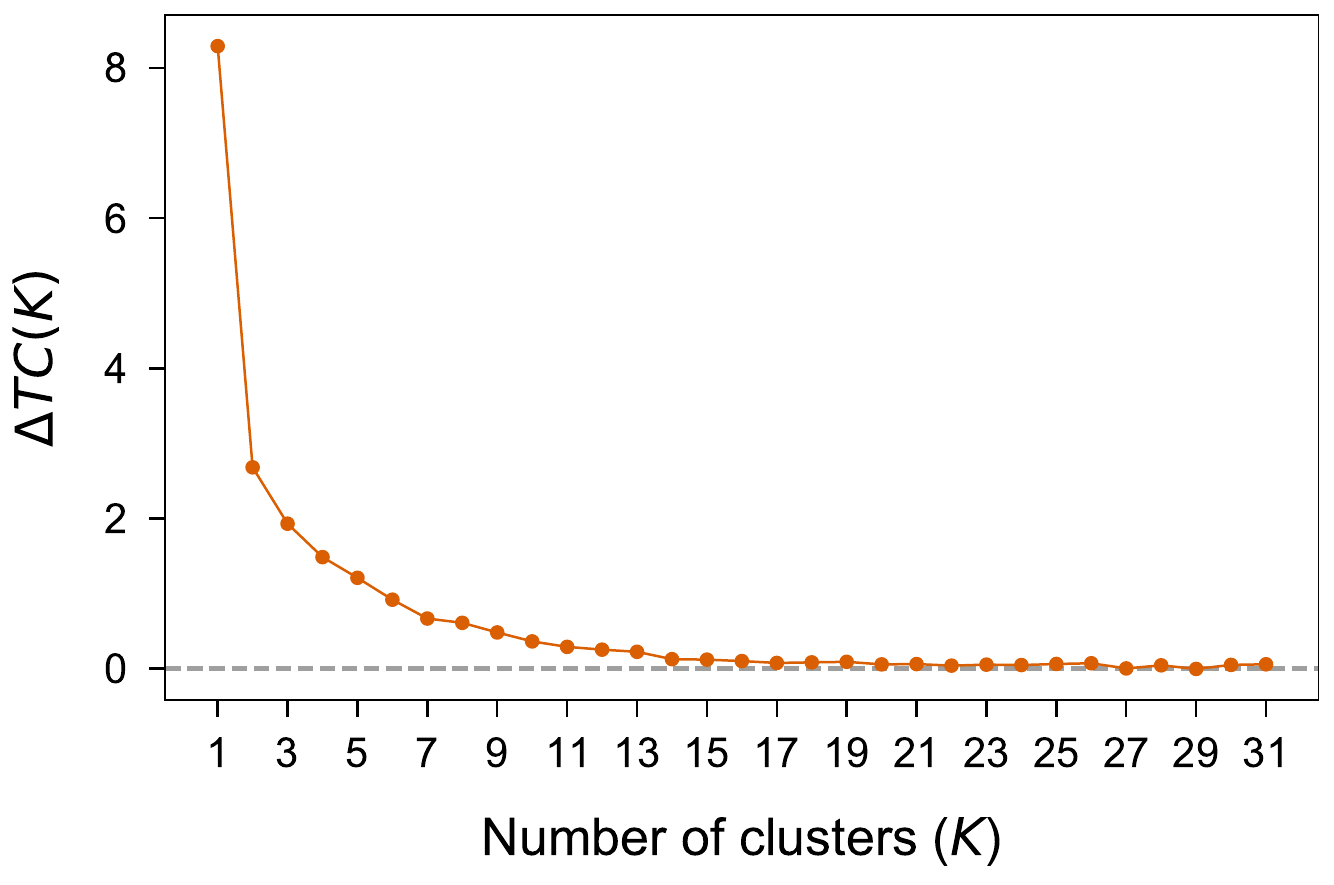}}
\end{center}
\caption{(A) The amount of total correlation as a function of the number of clusters ranging from 1 to 32. The green points and the associated bars are means and 95\% errors of 1,000 clustering results for the original data, while the gray points and bars are the results of 1,000 shuffled data where correlation is removed. (B) The increment of total correlation by the number of clusters.}
\label{si:fig:corex:clusters}
\end{figure}

\subsubsection{Geographical separations}
\label{si:clustering_results}

\begin{figure}[!h]
\centerline{\includegraphics[width=0.7\textwidth]{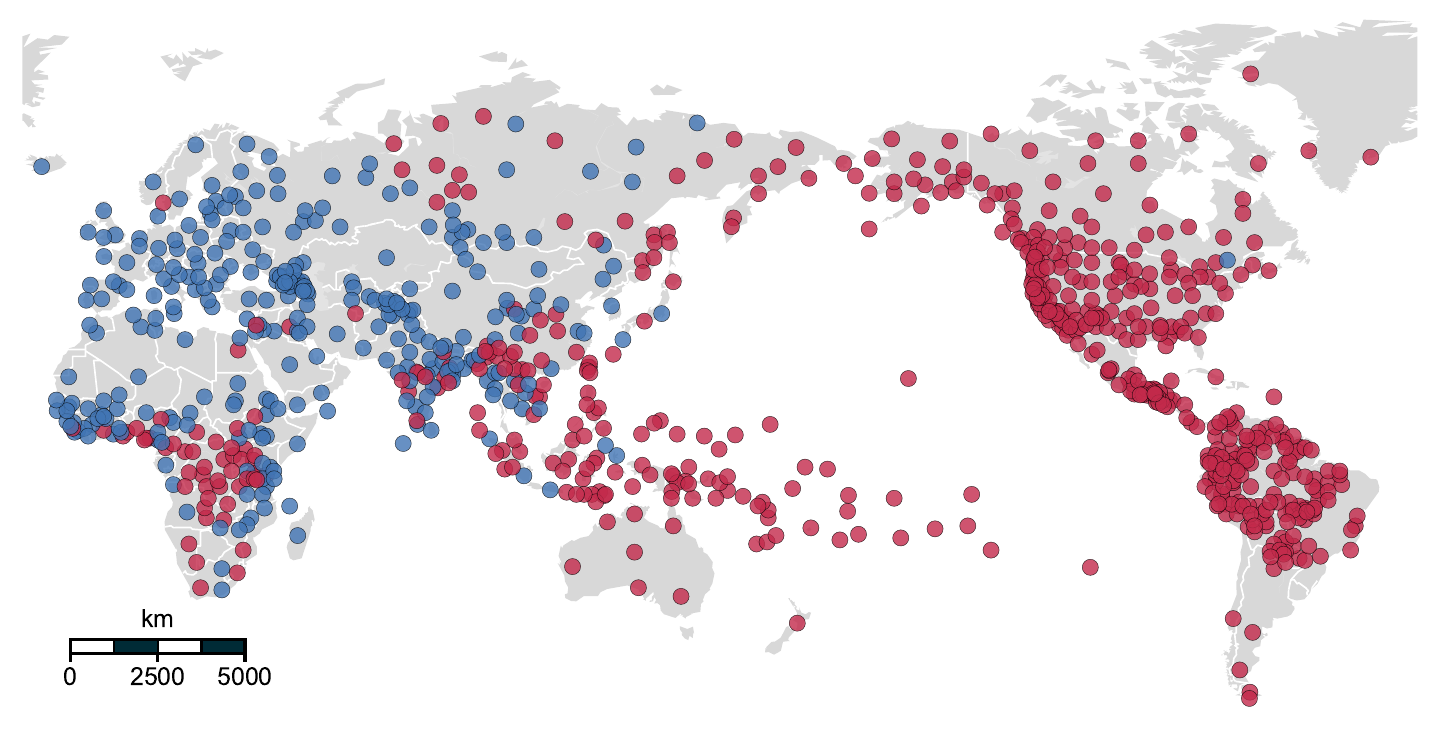}}
\caption{The best CorEx clustering result out of 1,000 attempts for two clusters. A blue tradition in northeast North America is Attikamek.}
\label{si:fig:corex2}
\end{figure}

\begin{figure}
\centerline{\includegraphics[width=0.7\textwidth]{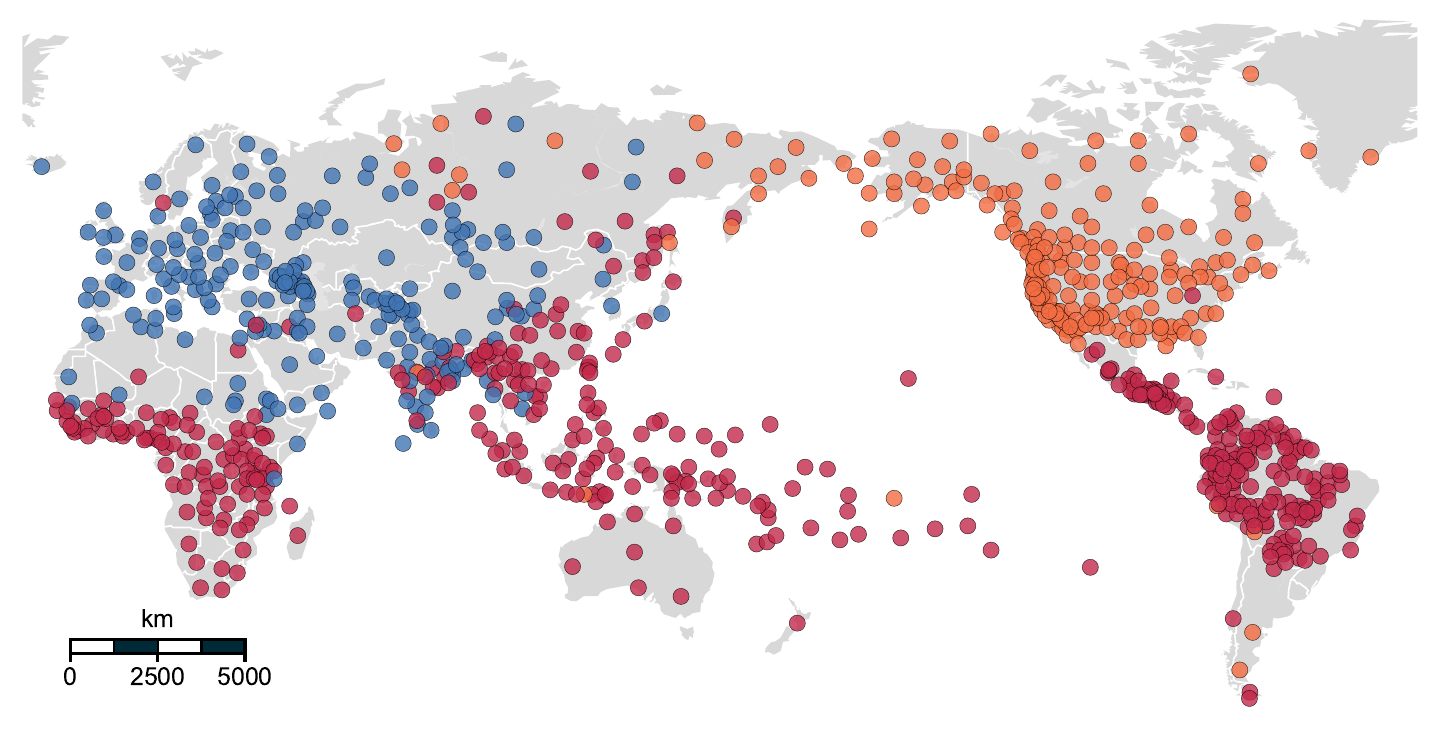}}
\caption{The best CorEx clustering result out of 1,000 attempts for three clusters.}
\label{si:fig:corex3}
\end{figure}

\begin{figure}
\centerline{\includegraphics[width=0.7\textwidth]{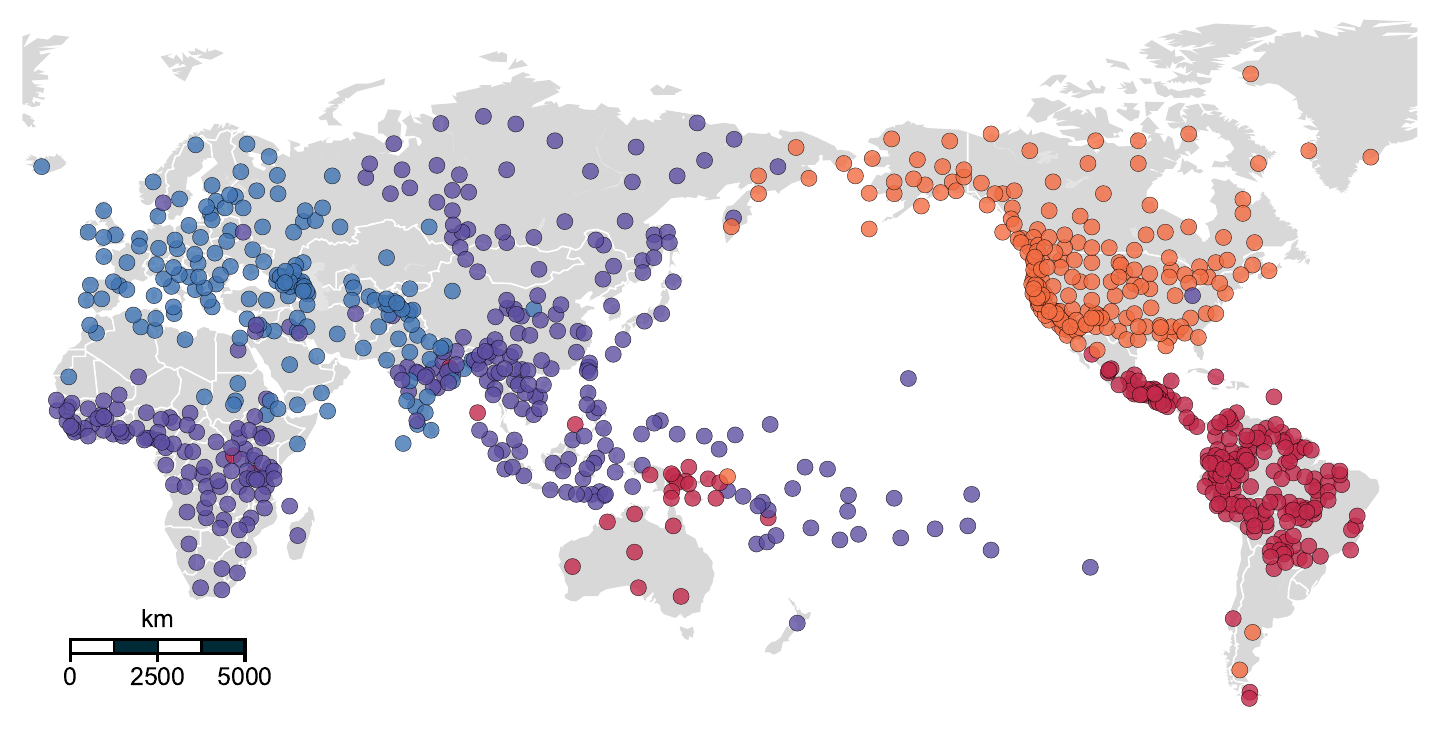}}
\caption{The best CorEx clustering result out of 1,000 attempts for four clusters.}
\label{si:fig:corex4}
\end{figure}

\begin{figure}
\centerline{\includegraphics[width=0.7\textwidth]{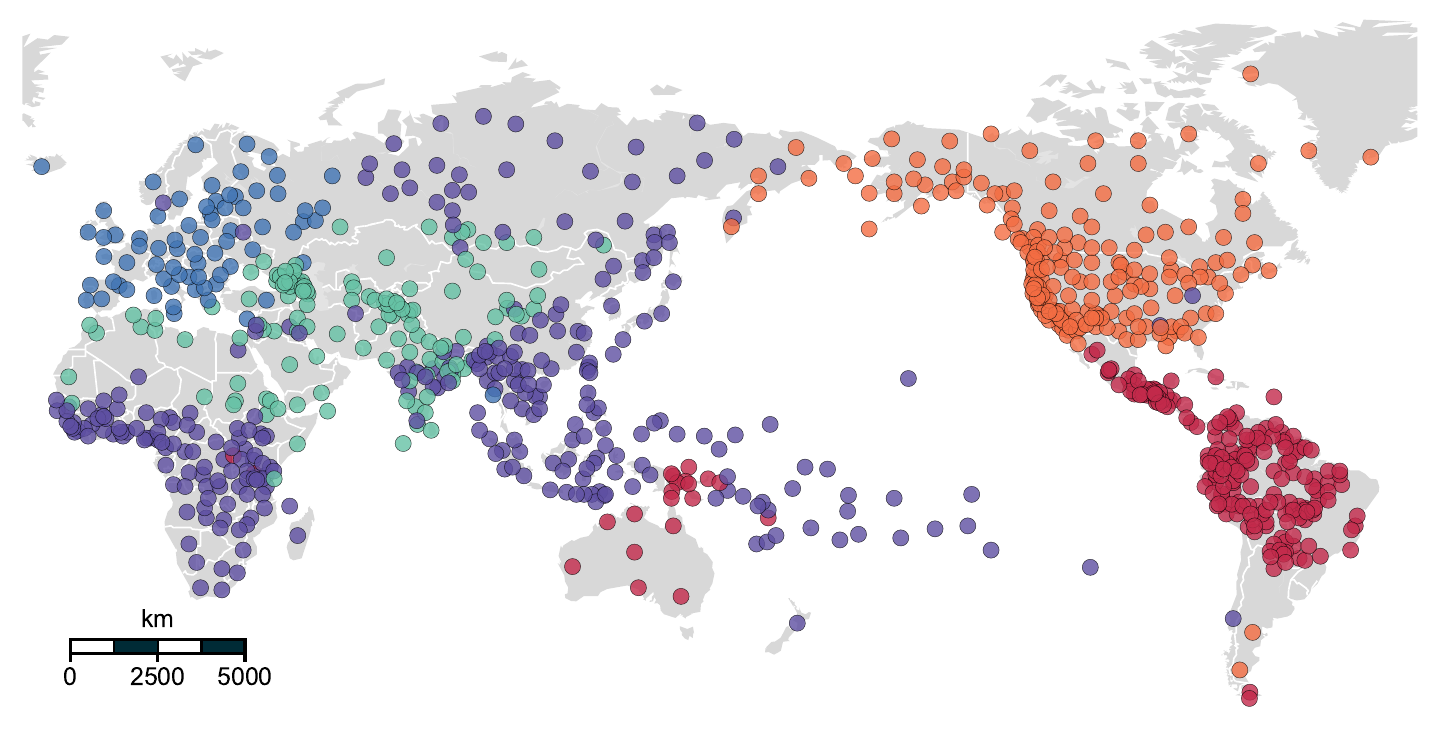}}
\caption{The best CorEx clustering result out of 1,000 attempts for five clusters.}
\label{si:fig:corex5}
\end{figure}

\begin{figure}
\centerline{\includegraphics[width=0.7\textwidth]{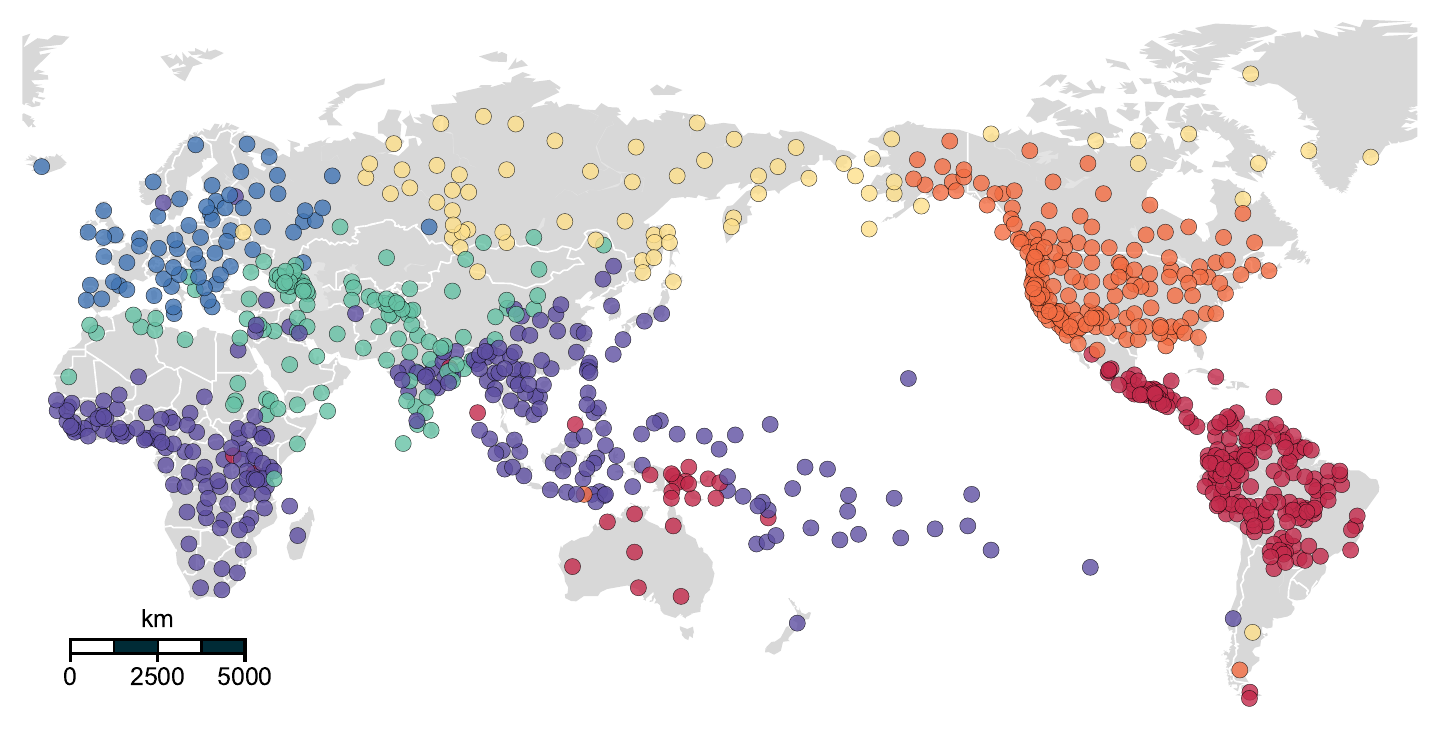}}
\caption{The best CorEx clustering result out of 1,000 attempts for six clusters.}
\label{si:fig:corex6}
\end{figure}

\begin{figure}
\centerline{\includegraphics[width=0.7\textwidth]{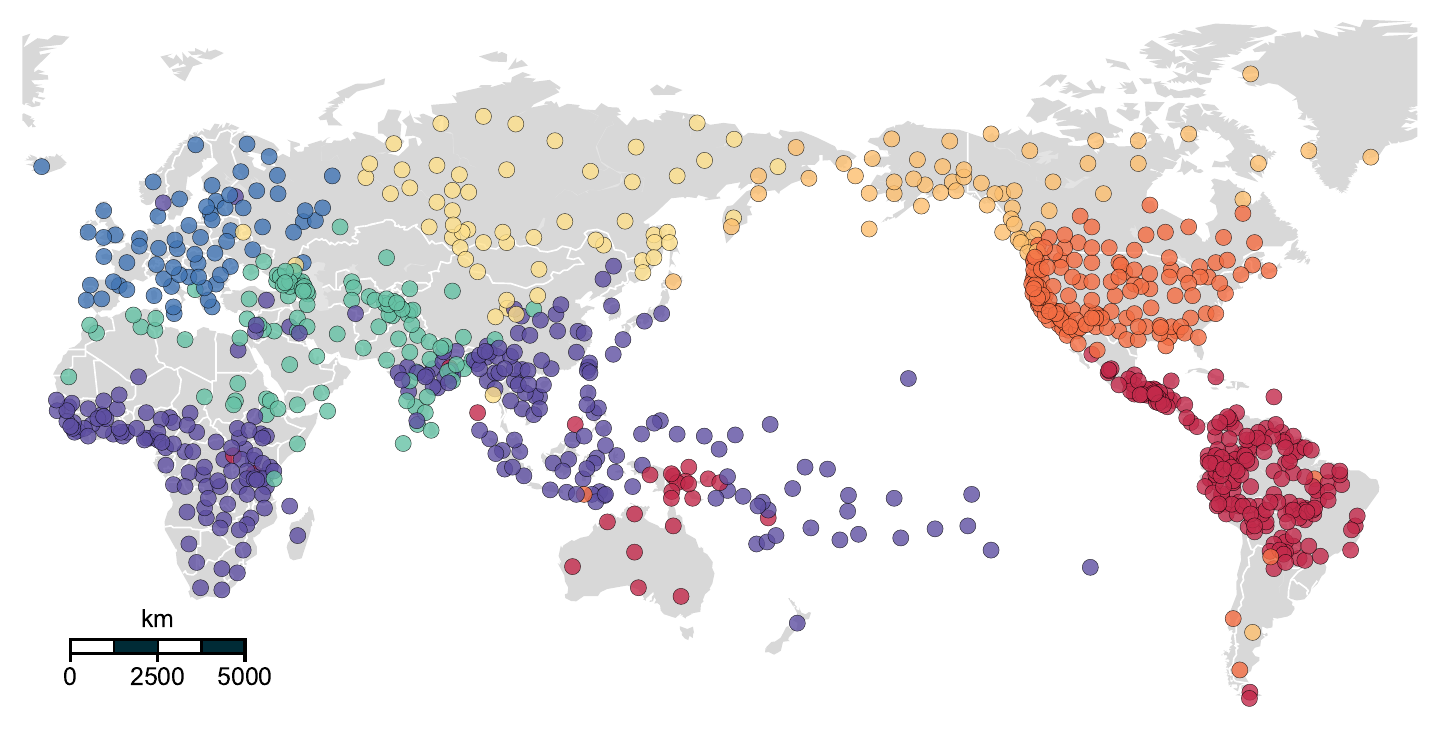}}
\caption{The best CorEx clustering result out of 1,000 attempts for seven clusters.}
\label{si:fig:corex7}
\end{figure}

\begin{figure}
\centerline{\includegraphics[width=0.7\textwidth]{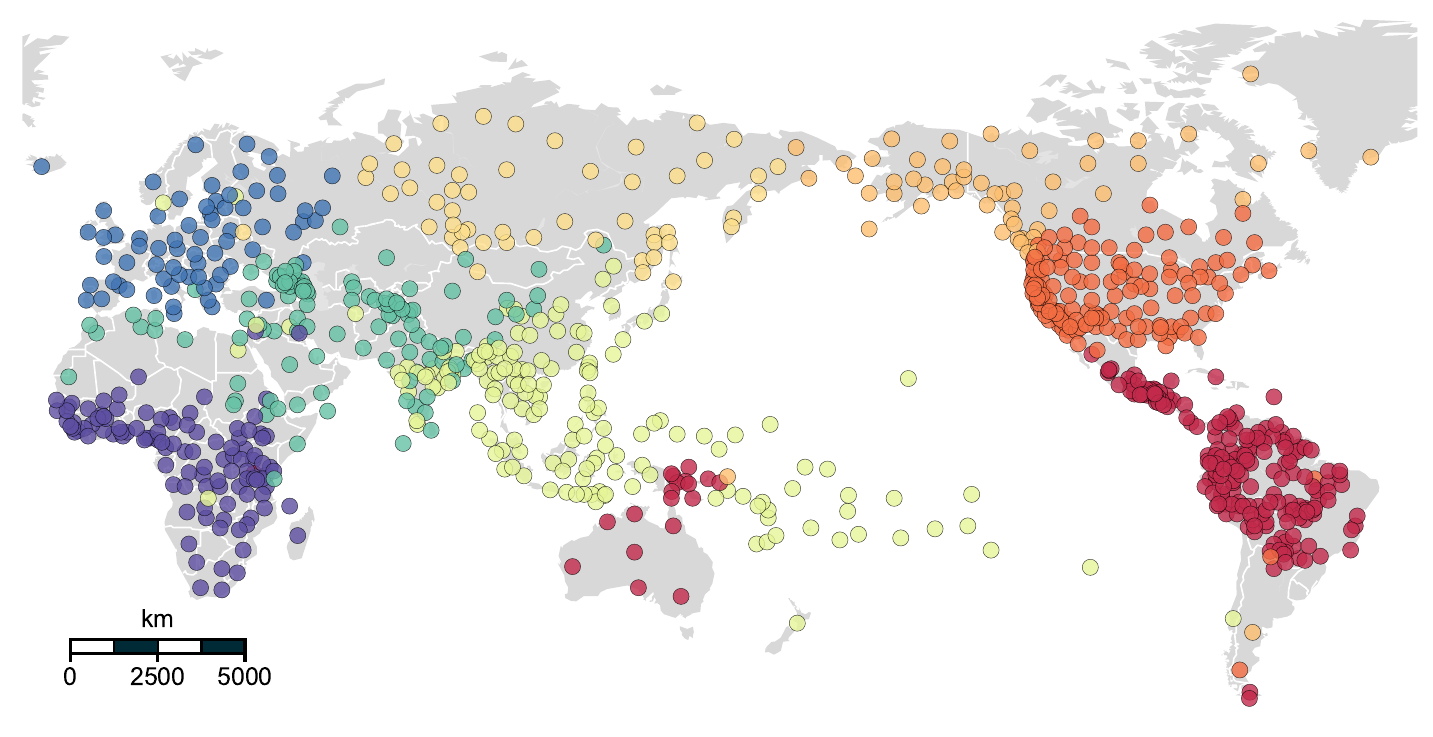}}
\caption{The best CorEx clustering result out of 1,000 attempts for eight clusters.}
\label{si:fig:corex8}
\end{figure}

\begin{figure}
\centerline{\includegraphics[width=0.7\textwidth]{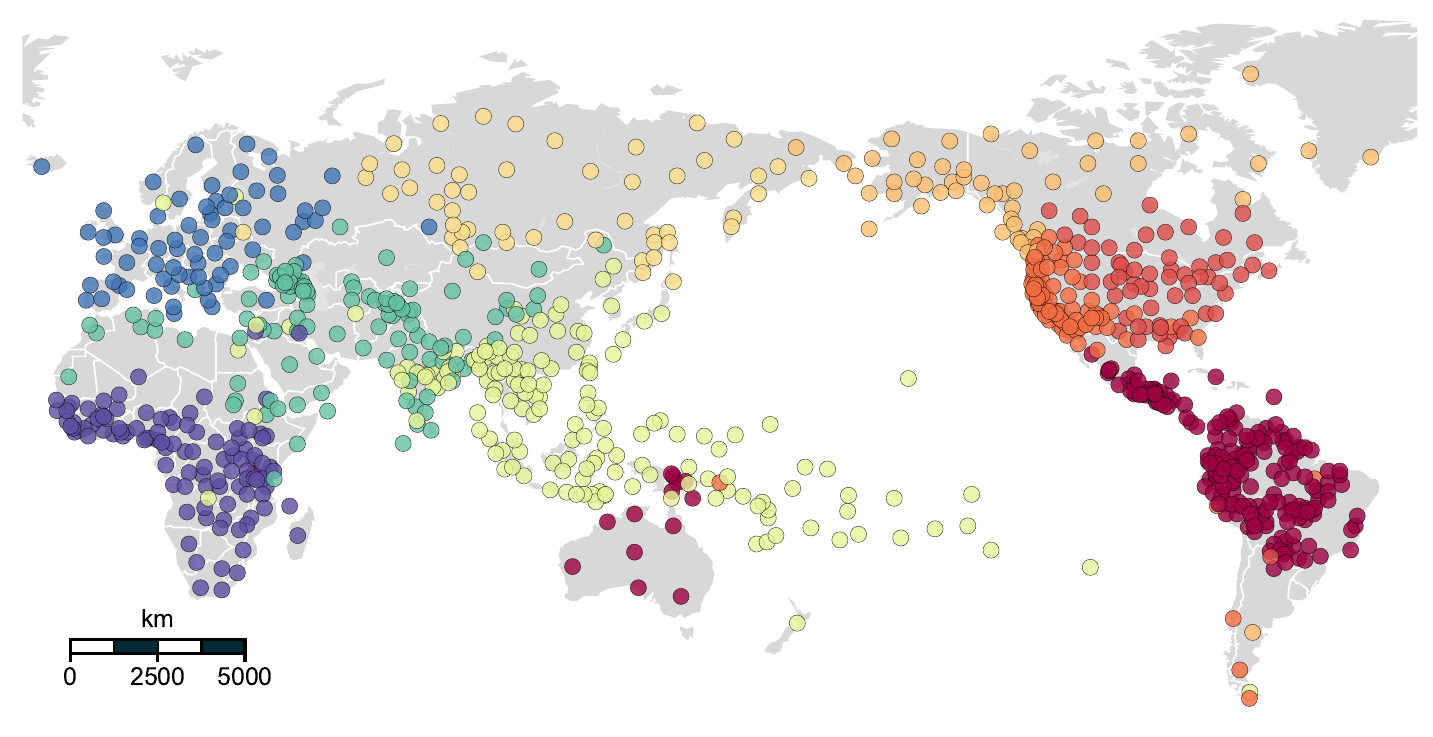}}
\caption{The best CorEx clustering result out of 1,000 attempts for nine clusters.}
\label{si:fig:corex9}
\end{figure}

\begin{figure}
\centerline{\includegraphics[width=0.7\textwidth]{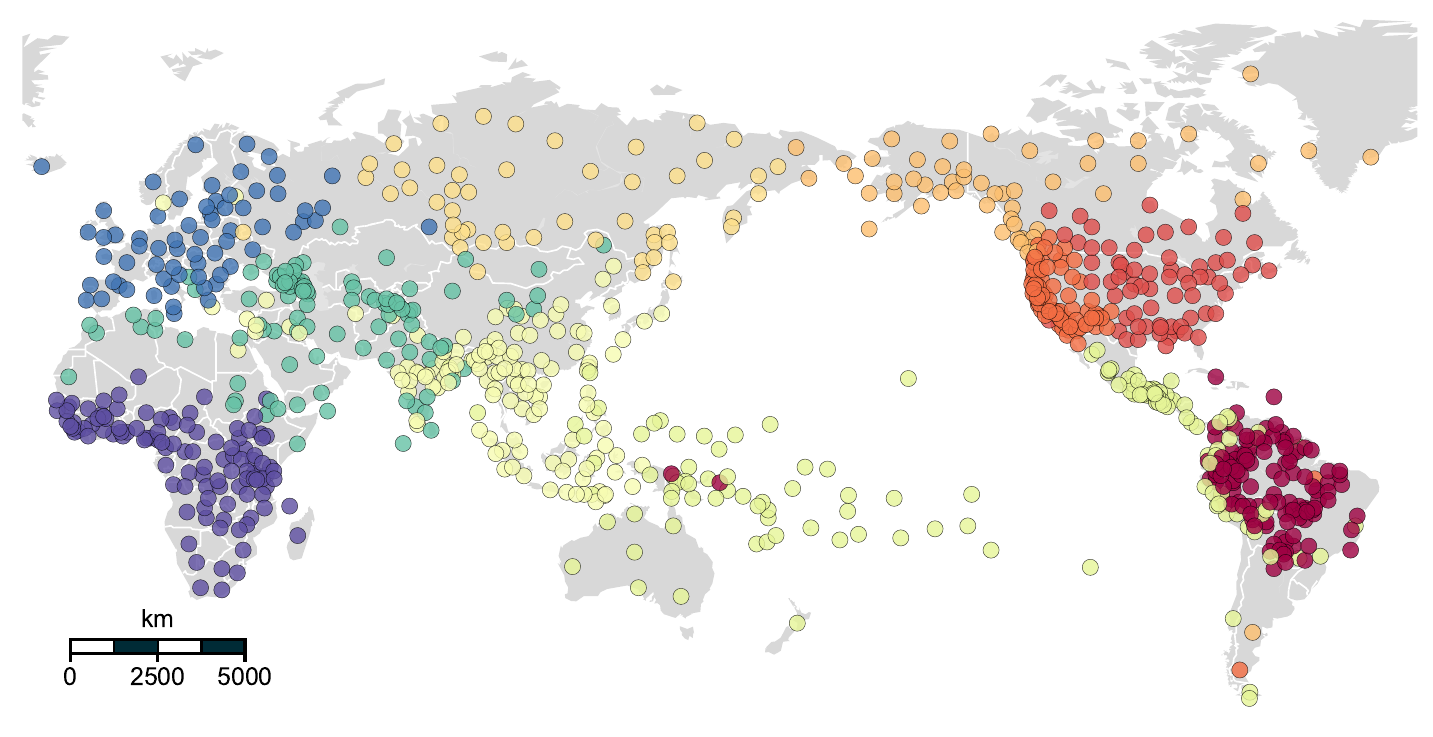}}
\caption{The best CorEx clustering result out of 1,000 attempts for ten clusters.}
\label{si:fig:corex10}
\end{figure}

\begin{figure}
\centerline{\includegraphics[width=0.7\textwidth]{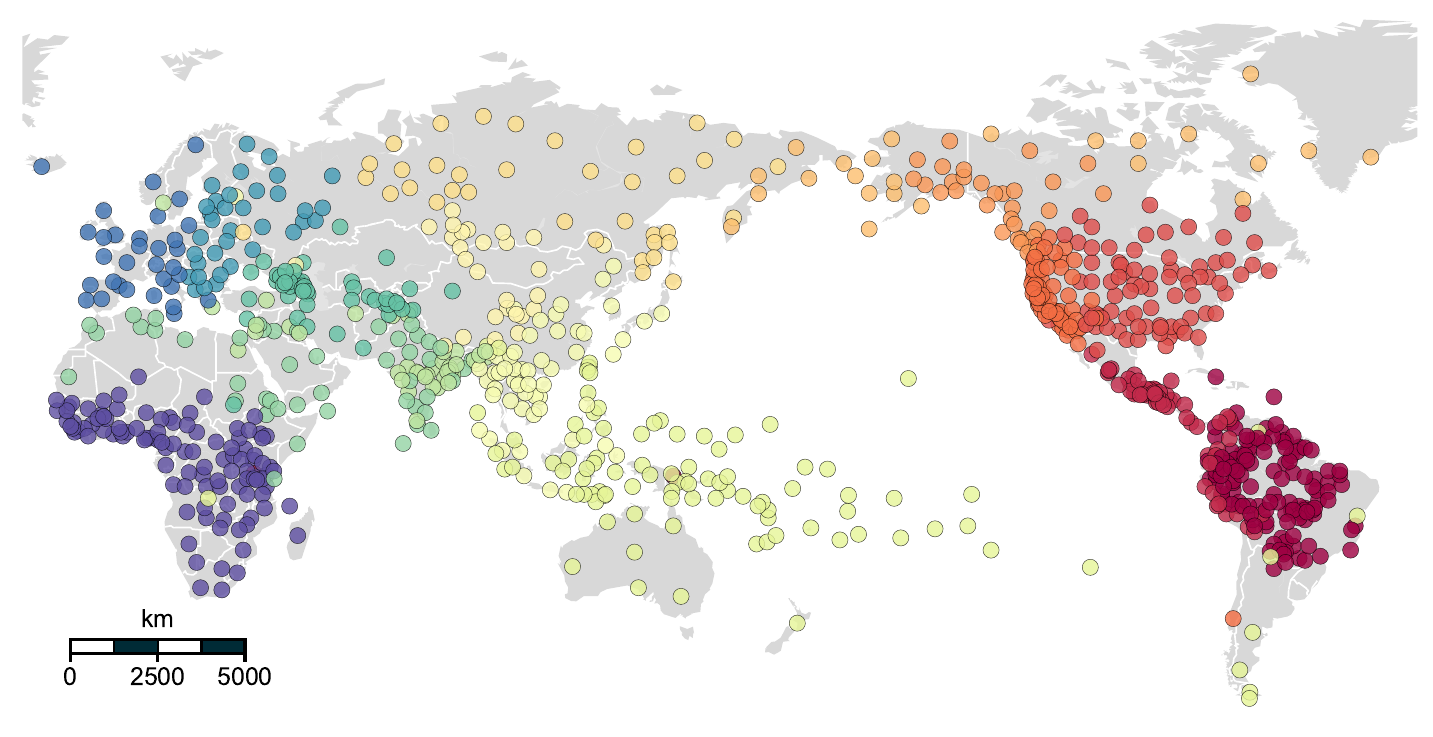}}
\caption{The best CorEx clustering result out of 1,000 attempts for sixteen clusters.}
\label{si:fig:corex16}
\end{figure}

The amount of total correlation increases by the number of clusters (FIG \ref{si:fig:corex:clusters}A) and saturates around fifteen clusters (FIG \ref{si:fig:corex:clusters}B). The points and the associated bars in FIG \ref{si:fig:corex:clusters}A are means and 95\% errors of 1,000 independent clustering results for each number of clusters. Interestingly, the clusters that the algorithm detects are nested in space (FIG \ref{si:fig:corex2} -- \ref{si:fig:corex16}). For example, the New World cluster at K=2 (FIG \ref{si:fig:corex2}) is split into three clusters, the Afro-Eurasian, North American, and Latin American clusters at K=4 (FIG \ref{si:fig:corex4}). For a larger number of clusters, the Latin American cluster is decomposed into the Central and South American clusters (FIG \ref{si:fig:corex8}). We listed the clustering results for different number of clusters from 2 to 10, and 16 in this document (FIG \ref{si:fig:corex2} -- \ref{si:fig:corex16}). 

The clustering results indicate that geographically close traditions tend to have similar motifs. The identified clusters are embedded in space even though no prior geographical information are given in the clustering procedure. These patterns are found in all cases regardless the number of clusters up to 32. We hereafter mainly focus on the result of the sixteen clusters. The sixteen clusters are 1) Sub-Saharan Africa, 2) Western Europe, 3) Eastern Europe, 4) Southwest Asia, 5) Afro-Asia, 6) Indian Subcontinent, 7) Oceania, 8) Southeast Asia, 9) Eastern Asia, 10) Northern Asia, 11) Arctic, 12) Northwest Coast, 13) Eastern North America, 14) Western North America, 15) Central America, and 16) South America.

\clearpage

\clearpage

\clearpage

\clearpage

\paragraph{Bootstrapping data to remove correlations}
\label{si:bootstrap}

\begin{figure}[h]
\begin{center}
\subfloat[][K=2]{
\includegraphics[width=0.47\textwidth]{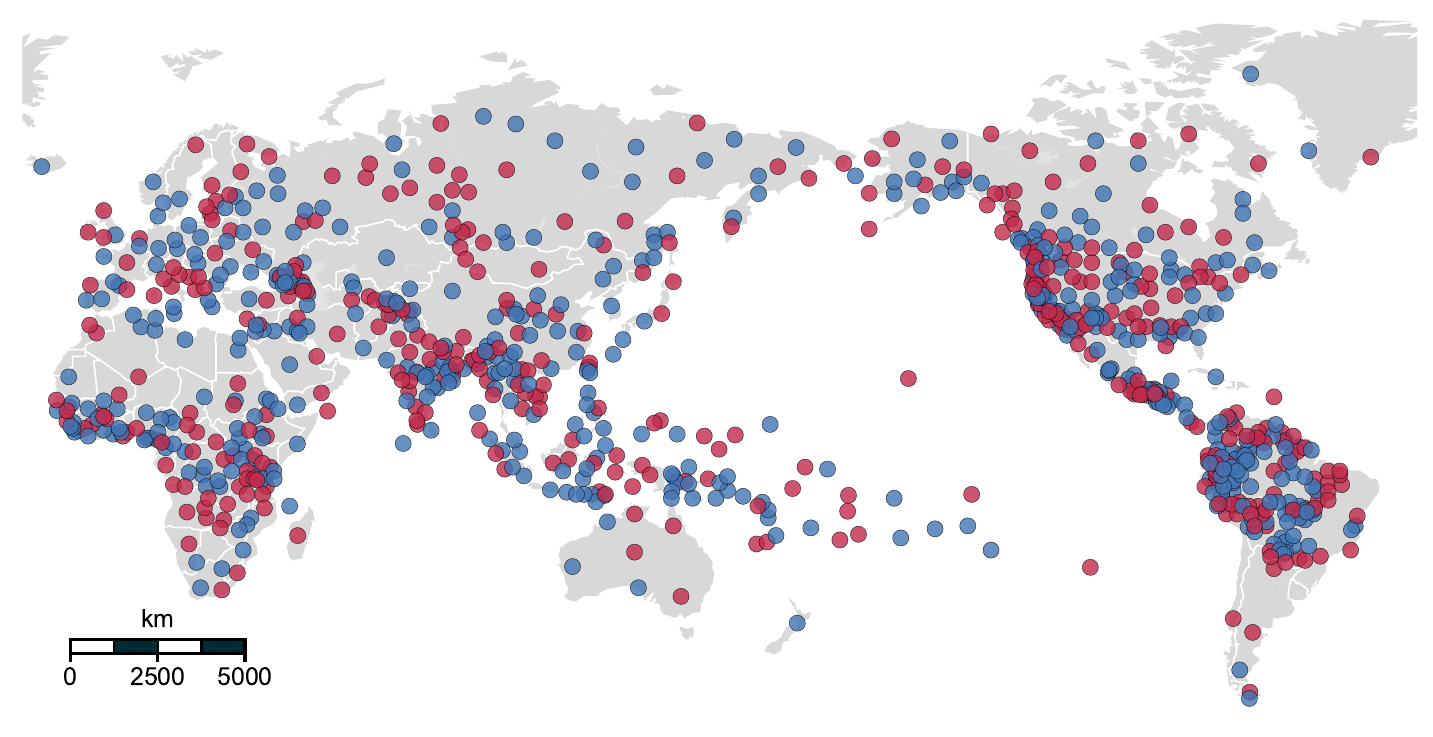}}
\subfloat[][K=4]{
\includegraphics[width=0.47\textwidth]{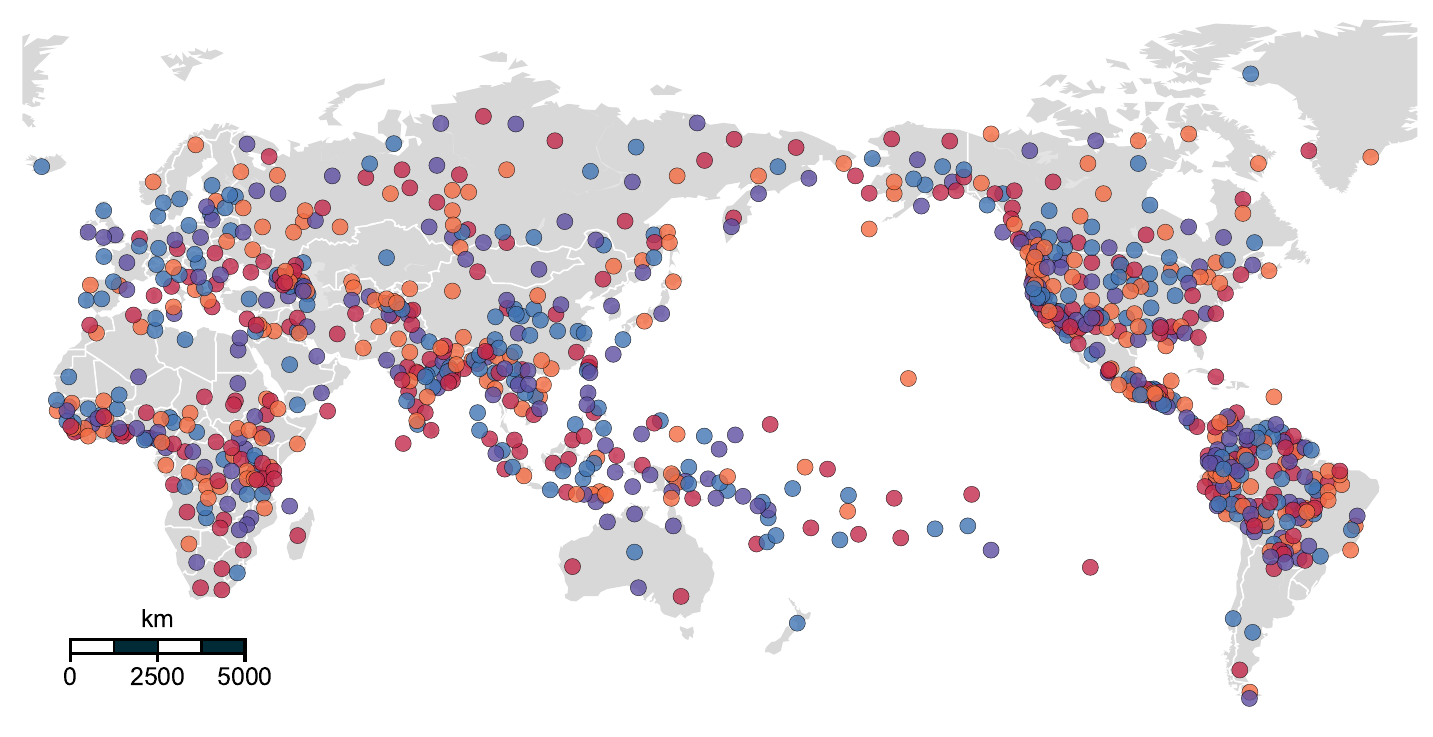}}

\subfloat[][K=8]{
\includegraphics[width=0.47\textwidth]{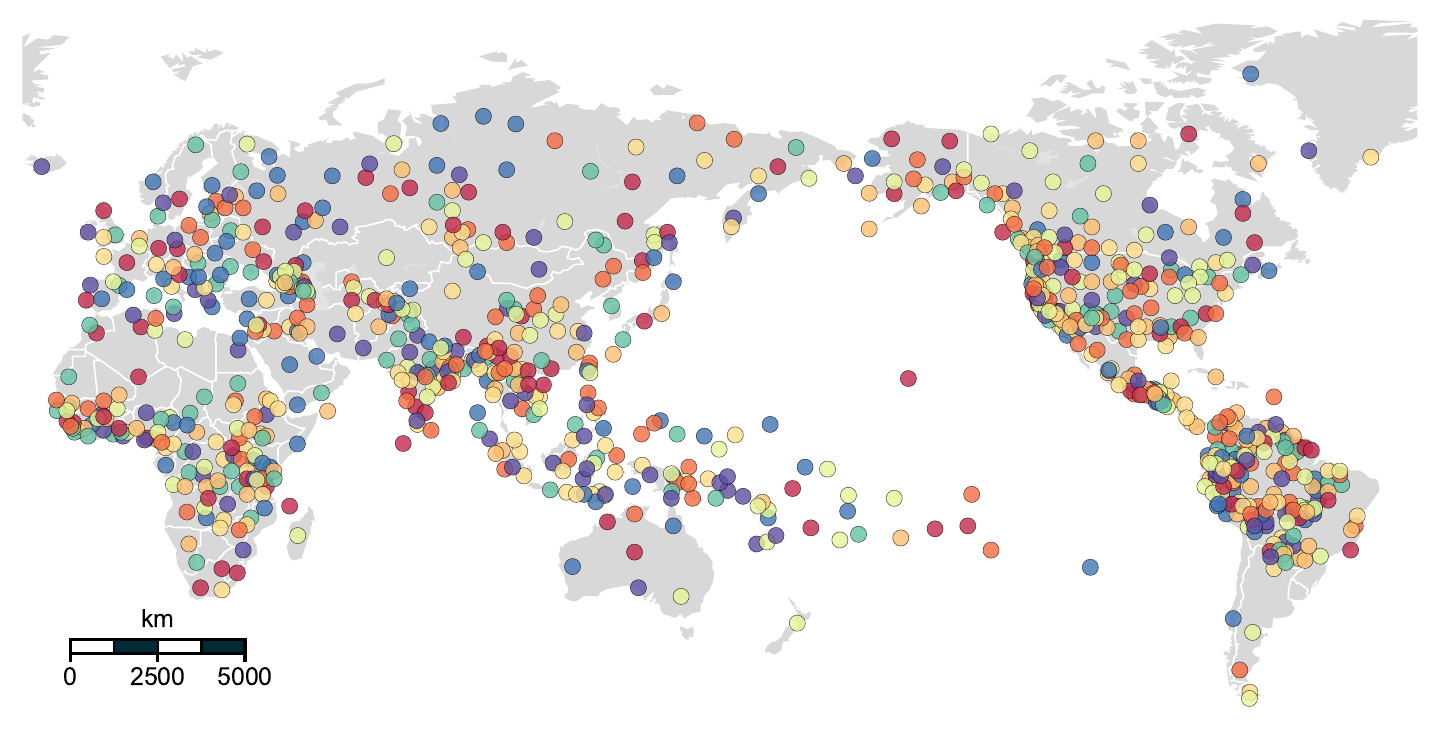}}
\subfloat[][K=16]{
\includegraphics[width=0.47\textwidth]{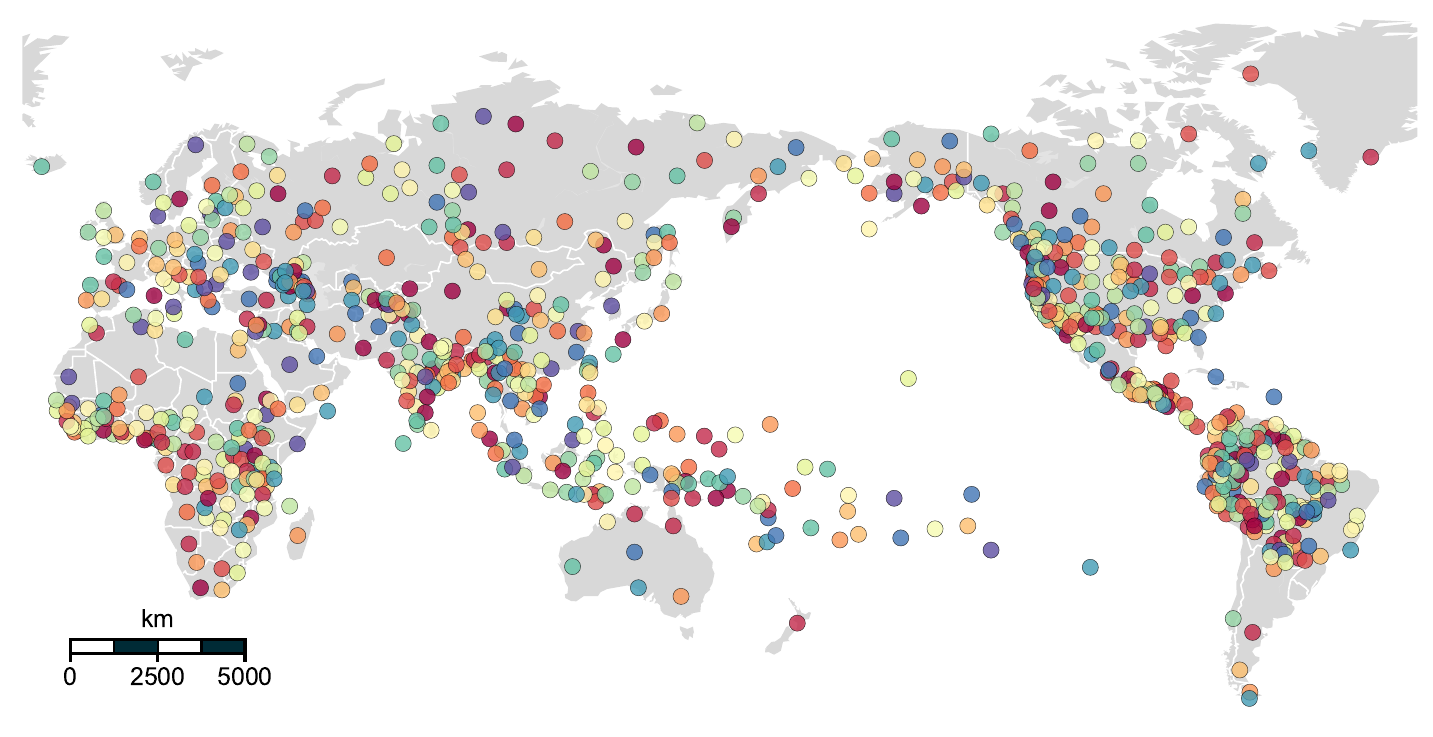}}
\end{center}
\caption{CorEx clusters identified in bootstrapped matrices.}
\label{si:fig:bootstrap}
\end{figure}

Are motif co-occurrences generated by chance? The answer is no, as the total correlations in the original motif data are significantly higher than total correlations in 1,000 bootstrapped data for the scales from 2 to 32 (FIG \ref{si:fig:corex:clusters}A). In order to generate bootstrapped data, we used the Curveball algorithm~\cite{strona2014fast}, which swaps two motif presences of two randomly selected traditions at each iteration for preserving row and column totals (i.e., fixing the number of traditions having a motif and the number of motifs that a tradition has). We sufficiently shuffled our motif presence-absence matrix by running 100,000 iterations. Clustering results drawn in the world map are also strikingly different from the CorEx results on the original data (FIG \ref{si:fig:bootstrap}). There are no geographic clusters in the bootstrapped data, and it suggests that the observed motif co-occurrences cannot be generated by random diffusion of motifs. 

\clearpage

\paragraph{Comparison with \textit{k}-means clustering}
\label{si:kmeans}

\begin{figure}[!h]
\centerline{\includegraphics[width=0.7\textwidth]{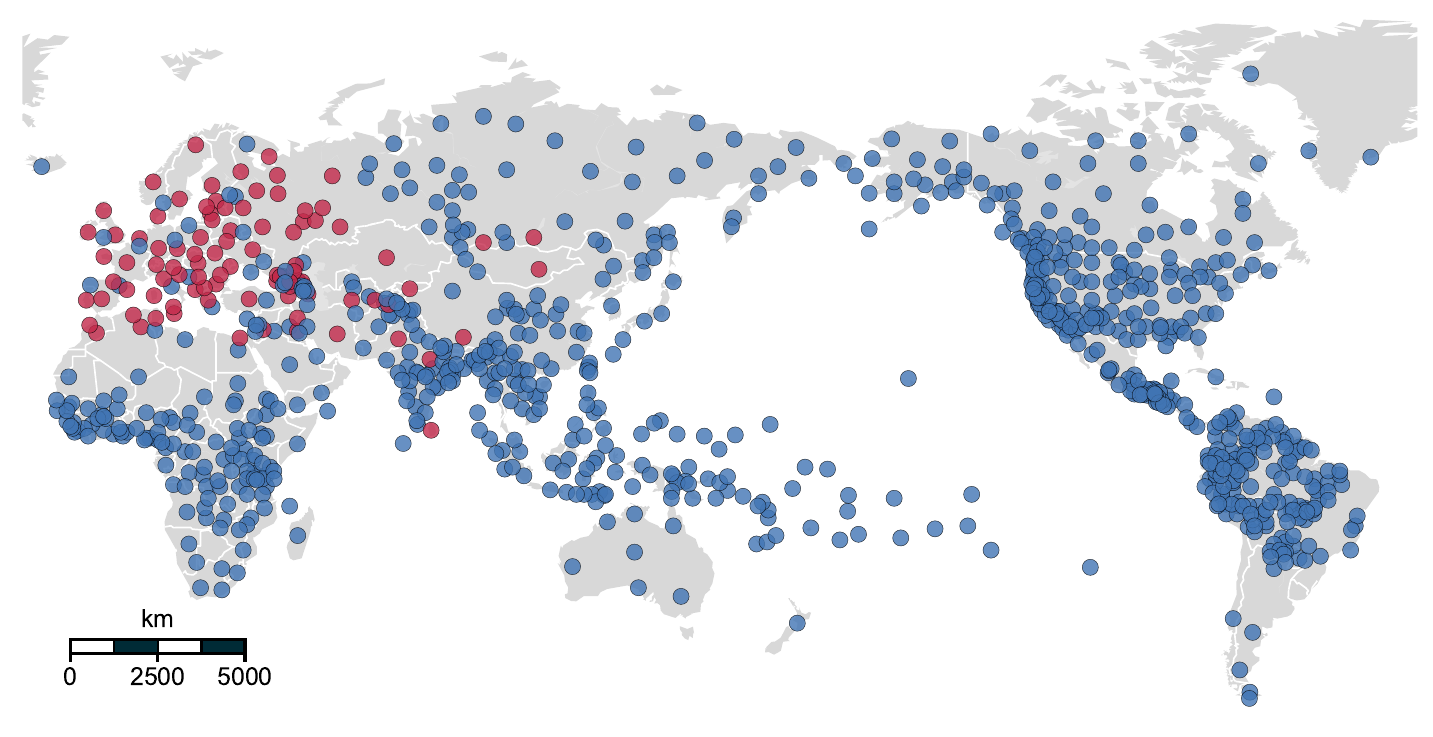}}
\caption{A \textit{k}-means clustering result for two clusters.}
\label{si:fig:kmeans2}
\end{figure}

\begin{figure}
\centerline{\includegraphics[width=0.7\textwidth]{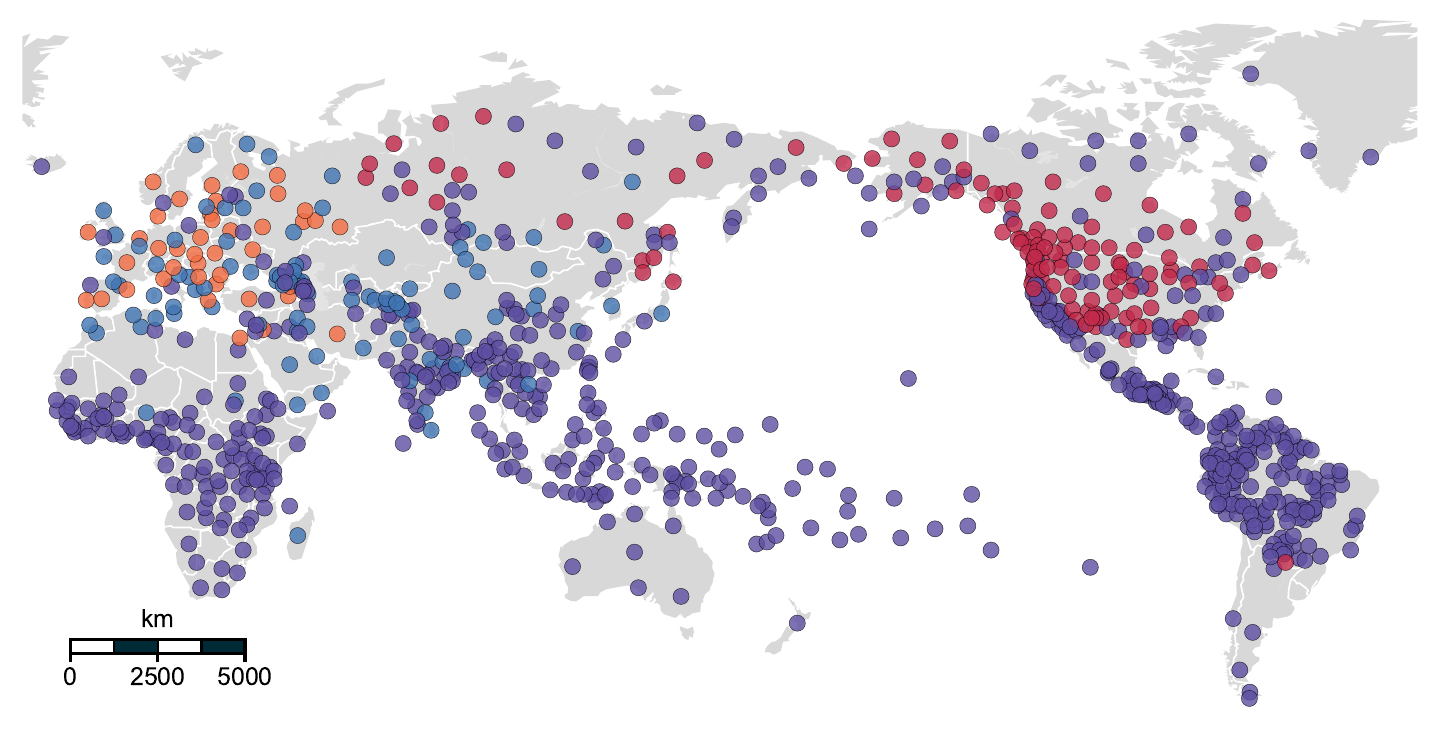}}
\caption{A \textit{k}-means clustering result for four clusters.}
\label{si:fig:kmeans4}
\end{figure}

\begin{figure}
\centerline{\includegraphics[width=0.7\textwidth]{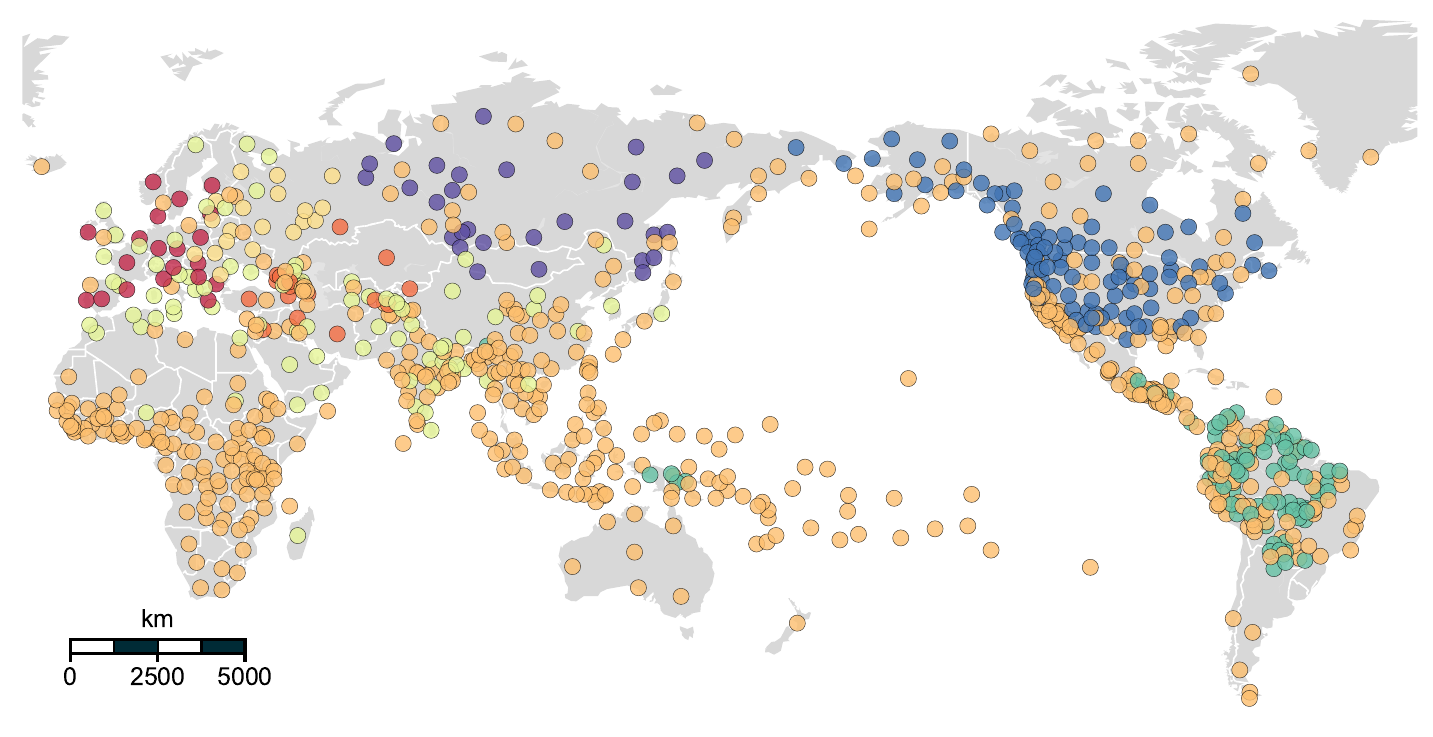}}
\caption{A \textit{k}-means clustering result for eight clusters.}
\label{si:fig:kmeans8}
\end{figure}

\begin{figure}
\centerline{\includegraphics[width=0.7\textwidth]{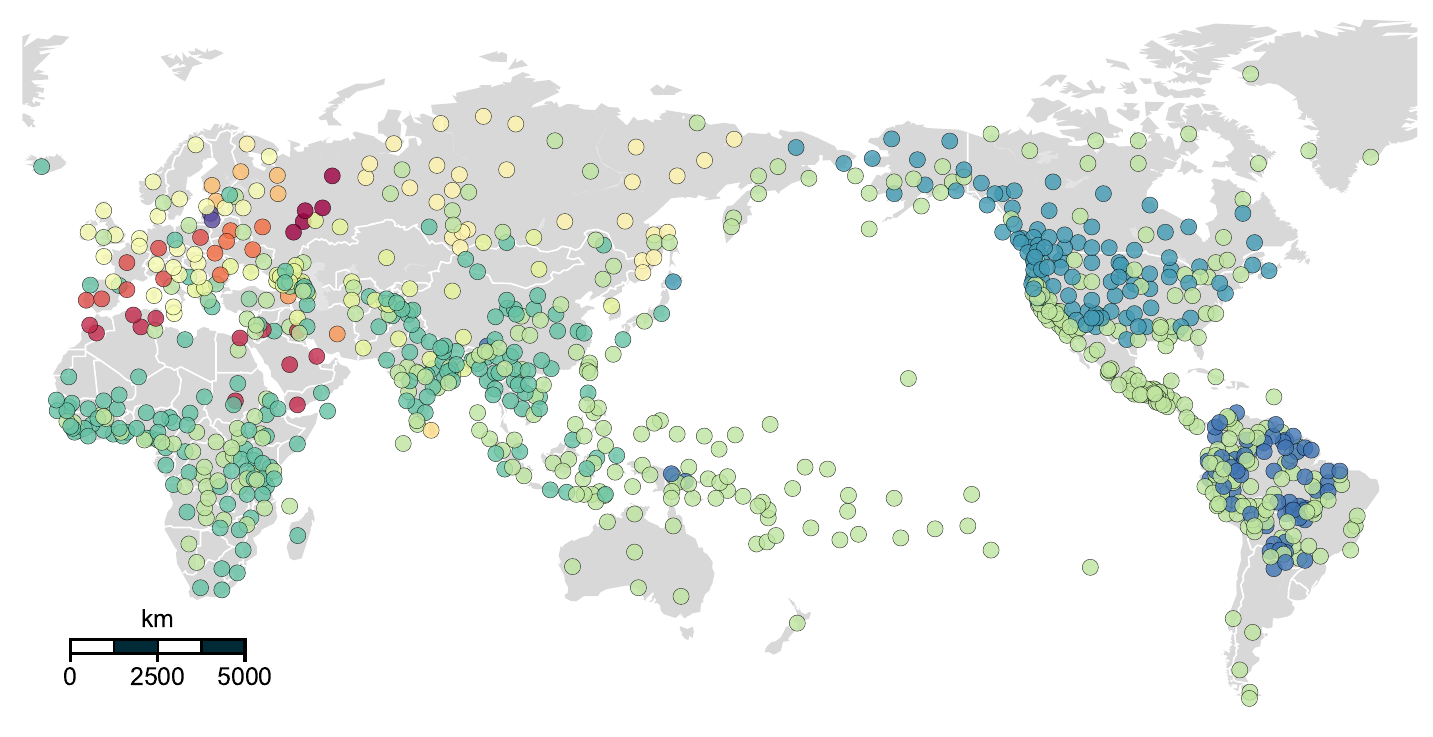}}
\caption{A \textit{k}-means clustering result for sixteen clusters.}
\label{si:fig:kmeans16}
\end{figure}

While CorEx identifies the clusters nested in space, it is questionable whether CorEx is the only way to obtain the clusters. Other popular clustering methods, such as \textit{k}-means clustering, could yield similar results as CorEx did. Here, we ran \textit{k}-means clustering for $k=2,4,8,16$ to confirm that CorEx is suitable to reveal latent structures within the motif database recorded in the binary format (FIG \ref{si:fig:kmeans2} -- \ref{si:fig:kmeans16}). The \textit{k}-means clustering results are quite different from the CorEx results. Specifically, at K=2, \textit{k}-means clustering separates traditions in the Indo-European region from others. This suggests that the traditions colored in red at FIG \ref{si:fig:kmeans2} have similar motif configurations compared to other traditions of different regions. At K=4 (FIG \ref{si:fig:kmeans4}), we have two clusters in Europe but they are not clearly separated in space. The nested structure that CorEx generated - the East and West European clusters - are not observed. At K=8 (FIG \ref{si:fig:kmeans8}), we have North and South American clusters. However, many traditions across the world are in the single cluster colored in orange. It would be due to the principle of \textit{k}-means clustering algorithm that groups traditions sharing many motifs until it has \textit{k} clusters. This tendency is observed even at K=16 (FIG \ref{si:fig:kmeans16}). There are many small clusters in Europe, whereas a large number of traditions are grouped into the green cluster. Thus, \textit{k}-means clustering does not identify the nested structure at all scales so that could limit our understanding of myths and folktales. CorEx seems to work better for the motif database that intrinsically contains false positives and false negatives as the database was constructed by a person who reviewed existing literature. CorEx extracts latent structures encoded in the motifs which significantly co-occur at various scales. 

\clearpage

\clearpage
\subsubsection{Overlaps with biogeography}
\label{si:bioregion}

\begin{figure}
\centerline{\includegraphics[width=0.75\textwidth]{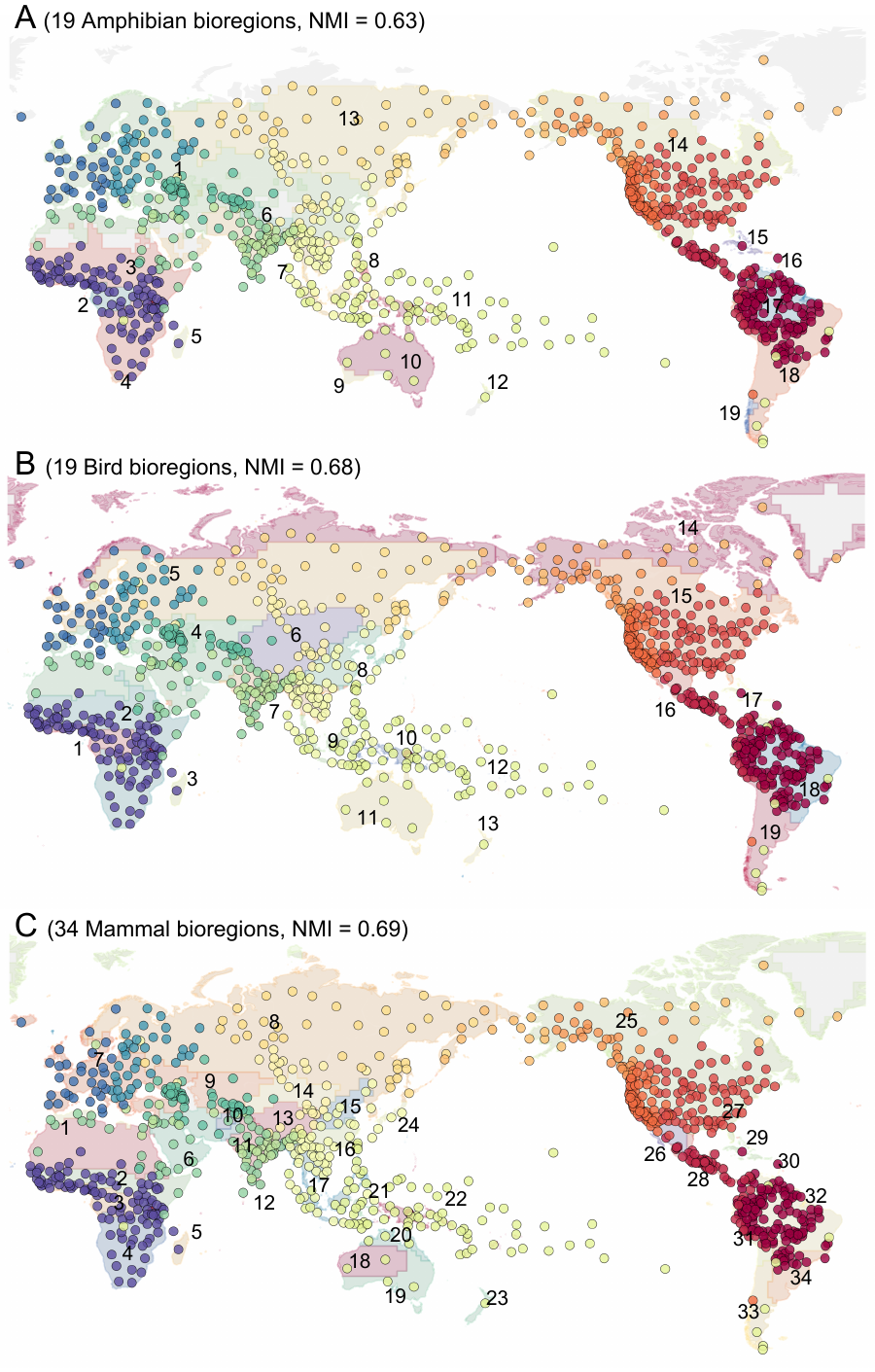}}
\caption{Traditions in the motif database are plotted over bioregions~\cite{Holt74} of (A) amphibians, (B) birds, and (C) mammals. Average normalized mutual information values are 0.63, 0.68, and 0.69 for each map, indicating significant overlaps between the clusters and bioregions.}
\label{si:fig:bioregion}
\end{figure}

The identified clusters seem to be separated by natural barriers such as mountains and seas. This observation led us to think about the connections between the clusters and bioregions, areas of similar ecological properties. The updated Wallace zoogeographic maps \cite{Holt74} of amphibians, birds, and mammals are taken as references to quantify the extent of overlaps between bioregions and the clusters. The zoogeographic maps for three classes have 19, 19, and 34 bioregions, respectively. For 921 ethnolinguistic traditions, we compared their clusters and bioregions as FIG \ref{si:fig:bioregion}. 

To quantify the overlaps, we calculated normalized mutual information (NMI) between them. NMI is often used to evaluate the consistencey of two cluster assignments \cite{schutze2008introduction}. For the set of clusters $A$ and the set of bioregions $B$, NMI is given as
\begin{equation}
    NMI(A,B) = \frac{2 \times I(A,B)}{H(A)+H(B)},
\end{equation}

where $I$ is mutual information and $H$ is entropy function. $H(A)=-\sum_{i}{p(A_i)\log{p(A_i)}}$ and $H(B)=-\sum_{j}{p(B_j)\log{p(B_j)}}$ are calculated with probabilities $p(A_i)$ and $p(B_j)$ that a tradition is in a cluster $i$ and a bioregion $j$ respectively. Mutual information $I$ is expressed as Equation \ref{si:eq:mi}. $p(A_i \cap B_j)$ are the probability that a tradition is classified into both cluster $i$ and bioregion $j$. Traditions of no bioregion information were excluded in NMI calculation.

\begin{equation}
\label{si:eq:mi}
    I(A,B) = \sum_{i}\sum_{j}p(A_i \cap B_j)\log{\frac{p(A_i \cap B_j)}{p(A_i) p(B_j)}}
\end{equation}

Averaged NMI values for amphibian, bird, and mammal bioregions are 0.63, 0.68, and 0.69 (FIG \ref{si:fig:bioregion}) that are significantly higher compared to that for 1,000 randomly shuffled pairs. Maximum NMI for randomly shuffled pairs in three animal classes are less than 0.12, implying that the clusters are coupled with bioregions, and biogeography would affect the structure of global myths and folktales.

\subsubsection{Ubiquity}
\label{si:ubiquity}

Motifs are found across the clusters that are associated with bioregions. We quantified the extent to which a motif is ubiquitous based on the learned CorEx model of the sixteen clusters. ``Ubiquity'' of a motif $m$ ($U_m$) is defined as the number of latent myths that contain the motif (TABLE \ref{si:tab:ubiquity_stat}). Note that $U_m$ can be $0$ even though some traditions have the motif $m$ if the motif is not important in a cluster's narrative structure with respect to motif co-occurrences. Motifs of similar frequency (i.e., the number of traditions having a motif) could have different ubiquity values depending on their spatial distributions. 

\begin{table}
    \caption{Distribution of ubiquity}
    \label{si:tab:ubiquity_stat}
    \begin{center}
    \small
    \begin{tabular}{|l|l|l|l|l|l|l|l|l|l|l|l|l|l|l|l|l|l|}
        \hline Ubiquity & 0&1&2&3&4&5&6&7&8&9&10&11&12&13&14&15&16 \\ \hline
        Number of motifs & 344 & 897 & 500 & 307 & 212 & 132 & 99 & 67 & 40 & 39 & 17 & 15 & 11 & 8 & 11 & 15 & 4 \\ \hline
    \end{tabular}
    \end{center}
\end{table}

\begin{figure}[!ht]
\centerline{\includegraphics[width=0.7\textwidth]{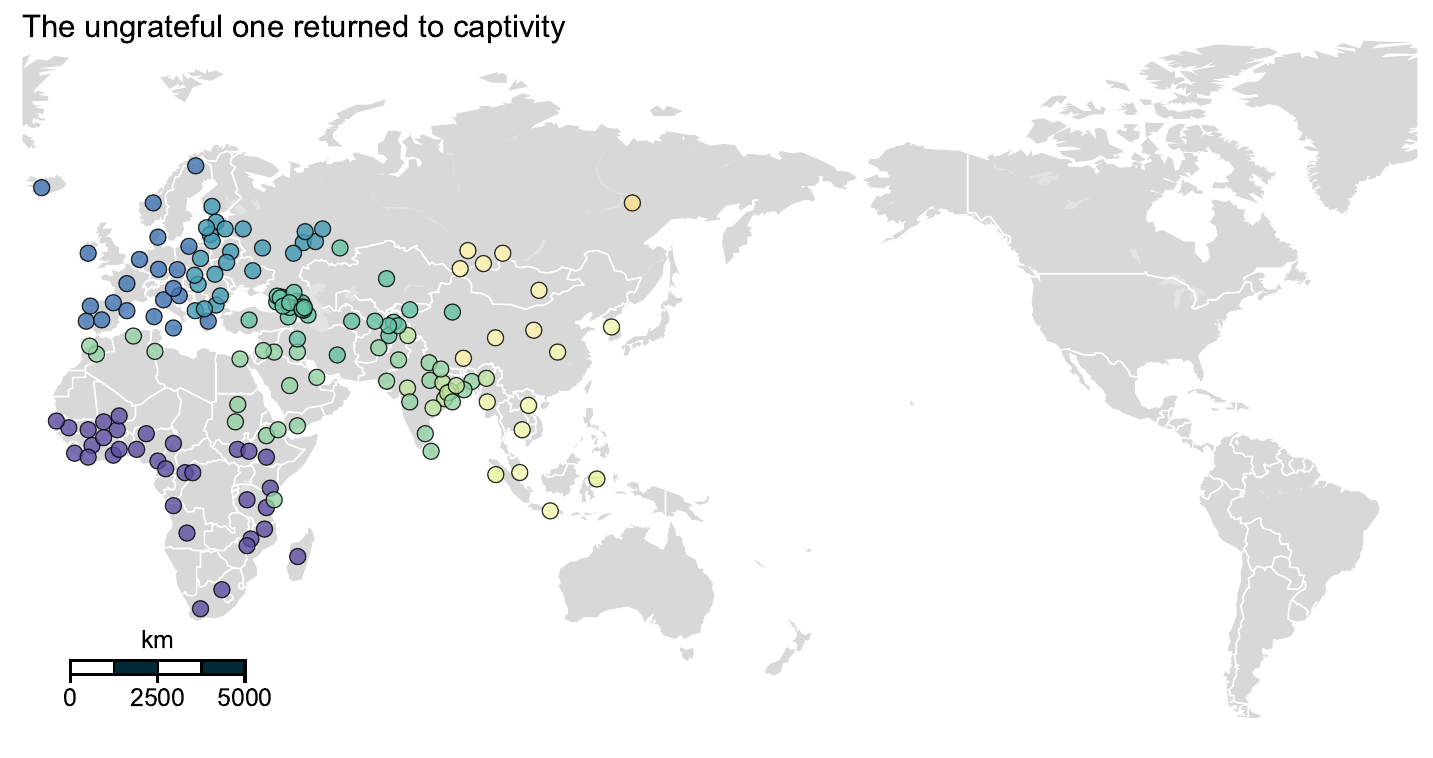}}
\caption{The spatial distribution of `The ungrateful one returned to captivity' which ubiquity is 8 and frequency is 163.}
\label{si:fig:ubiq_example1}
\end{figure}

\begin{figure}
\centerline{\includegraphics[width=0.7\textwidth]{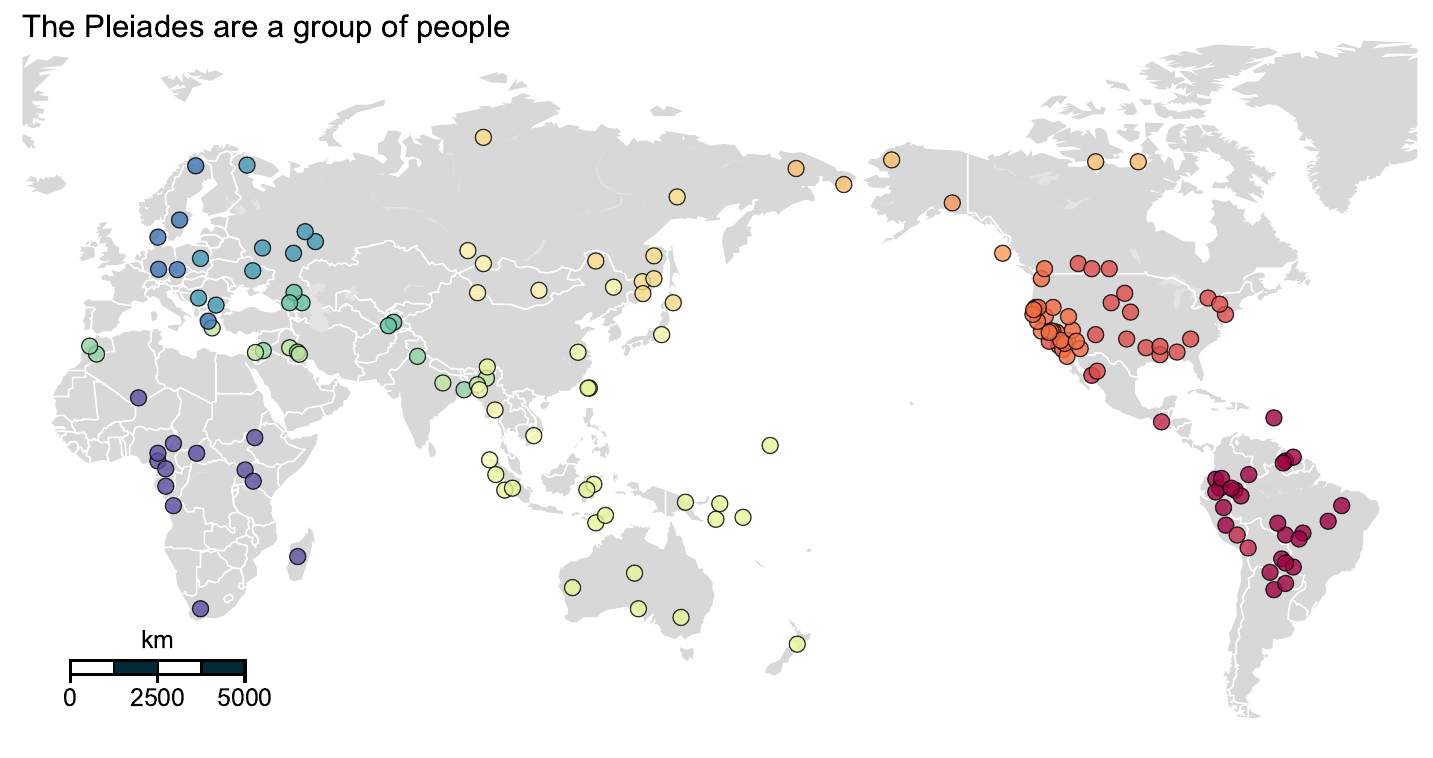}}
\caption{The spatial distribution of `The pleiades are a group of people' which ubiquity is 14 and frequency is 164.}
\label{si:fig:ubiq_example2}
\end{figure}

\begin{figure}
\centerline{\includegraphics[width=0.45\textwidth]{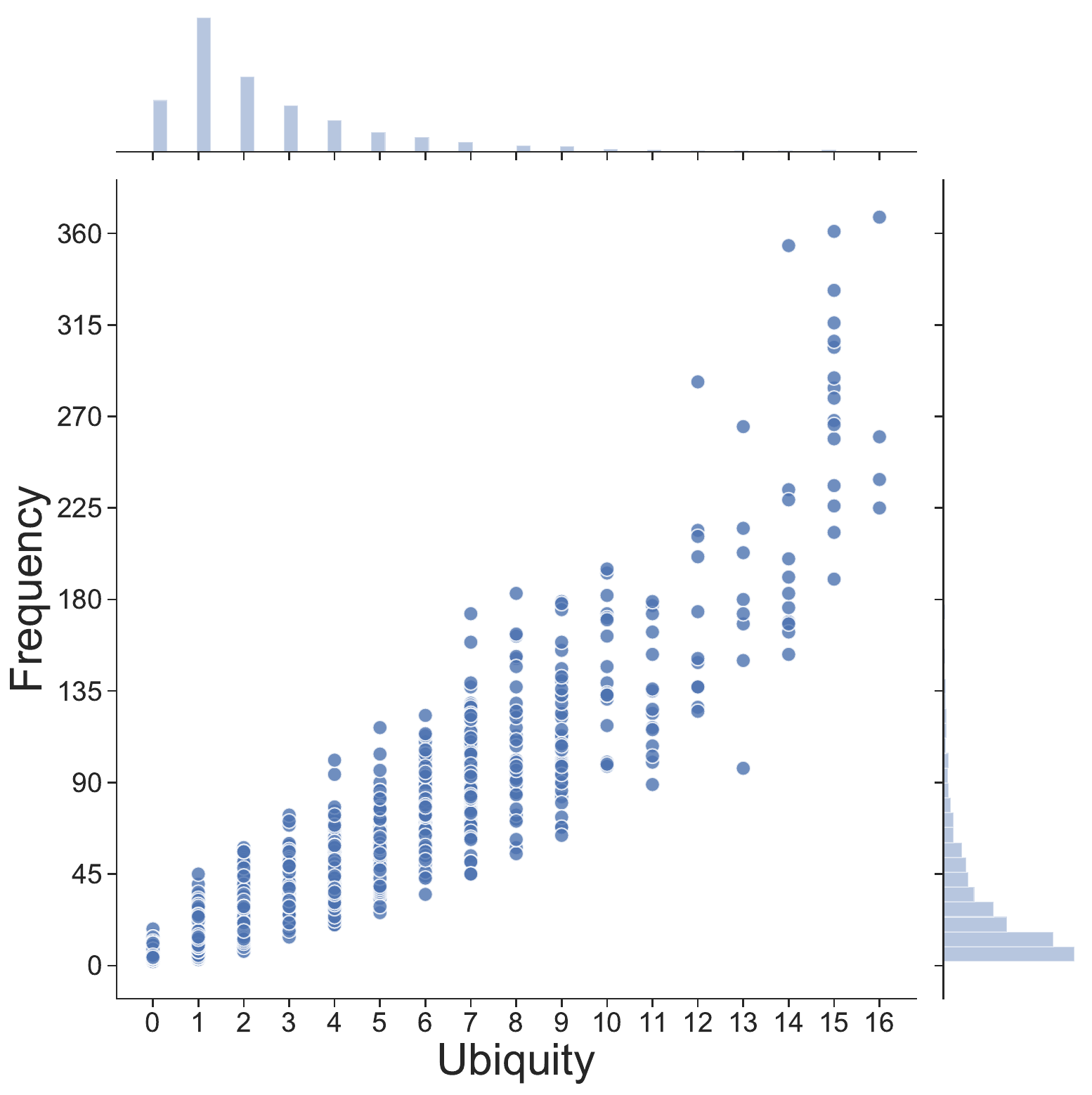}}
\caption{Ubiquity and frequency are highly correlated by definition but show different spatial features of motifs. Motifs can be localized or widespread even they are found at the same number of traditions across the world.}
\label{si:fig:ubiq}
\end{figure}

For instance, `The ungrateful one returned to captivity' (M156) and `The pleiades are a group of people' (I100B), which are found in 163 and 164 traditions respectively, show that our ubiquity captures how a motif is localized or spread with less sampling bias. `The ungrateful one returned to captivity'' is concentrated in the Old World ($U=8$, FIG \ref{si:fig:ubiq_example1}) while `The pleiades are a group of people'' is broadly distributed as it appears in both the Old and New World ($U=14$, FIG \ref{si:fig:ubiq_example2}). The ubiquity index and frequency are highly correlated by definition (FIG \ref{si:fig:ubiq}), but provide different perspectives on myths and folktales.

\subsubsection{Dispersion}
\label{si:dispersion}

\begin{figure}
    \centering
    \includegraphics[width=\textwidth]{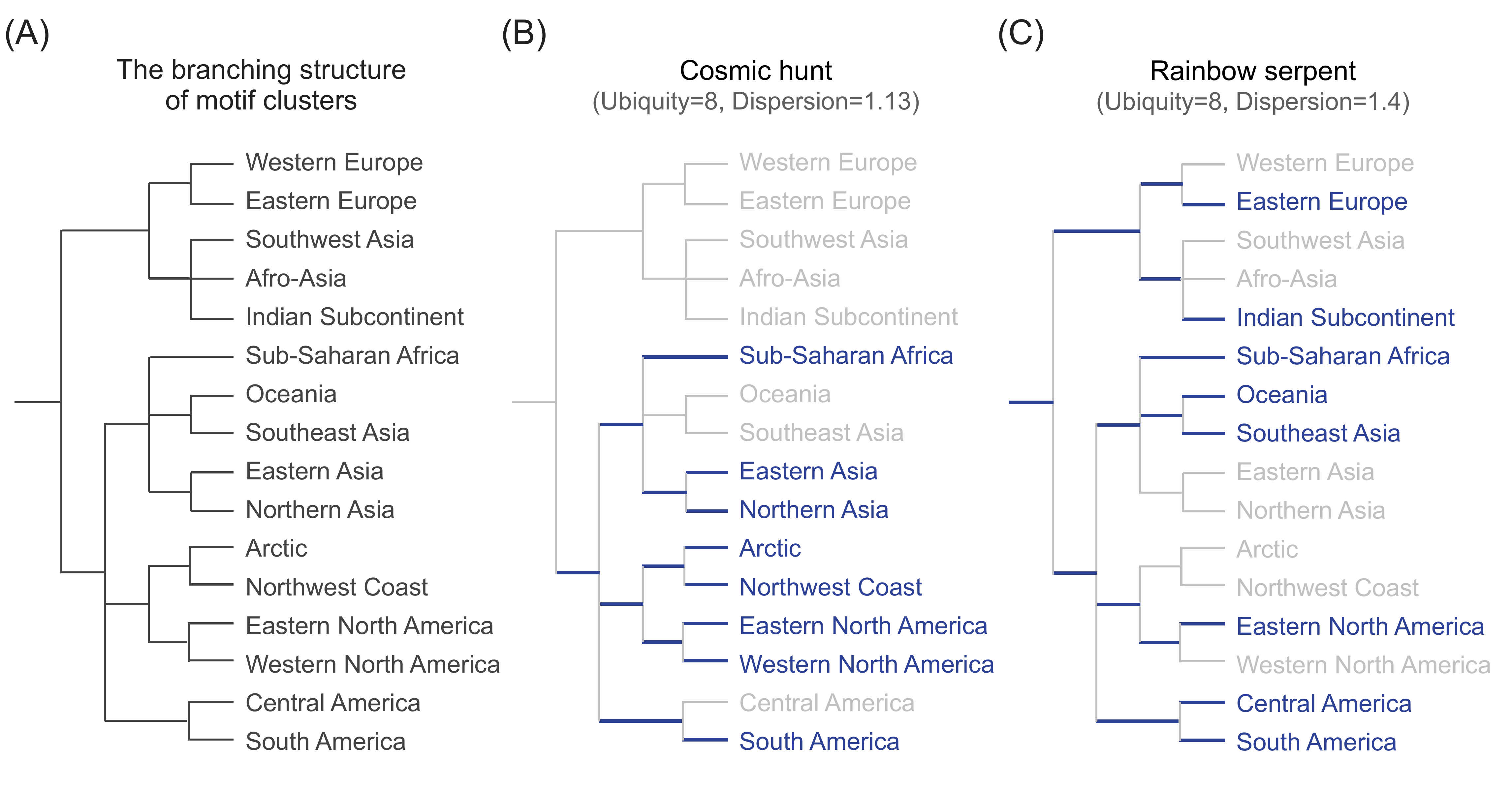}
    \caption{(A) The branching structure of motif clusters. (B) The sub-dendrogram consisting of latent myths that `Cosmic hunt' appears are colored in blue. (C)
    The sub-dendrogram consisting of latent myths that `Rainbow serpent' appears are colored in blue.}
    \label{fig:tree_dispersion}
\end{figure}

While ubiquity is a simple, effective measure for summarizing how a particular motif is spread across the clusters, it still does not differentiate the spatial distributions of motifs of the same ubiquity. Suppose we have two motifs A and B of $U=2$, where A is found in the Western and Eastern European latent myths and B is found in the Sub-Saharan African and South American latent myths. Their spatial distributions are clearly different, but their ubiquity values are same. To address this issue, we define an additional metric ``Dispersion'', which is calculated based on the branching tree structure of the 16 motif clusters that is constructed by linking the best clusters at K=2,4,8,16 (FIG \ref{fig:tree_dispersion}a). Dispersion is the ratio of the total branch length of the sub-dendrogram comprising latent myths containing a particular motif to the minimum theoretical branch length of sub-dendrograms having the same number of latent myths. In case of the `Cosmic hunt' (FIG \ref{fig:tree_dispersion}b), there are eight latent myths in which the motif is found, and these latent myths are connected with the branches of length 17 colored in blue, while the minimum theoretical branch length for the same ubiquity is 15. Thus, its dispersion becomes 17/15=1.13. Another example motif `Rainbow serpent' is also found in eight latent myths, but its dispersion is 1.4 because this motif is more widespread than `Cosmic hunt'. 

\begin{figure}[!h]
    \centering
    \includegraphics[width=0.8\textwidth]{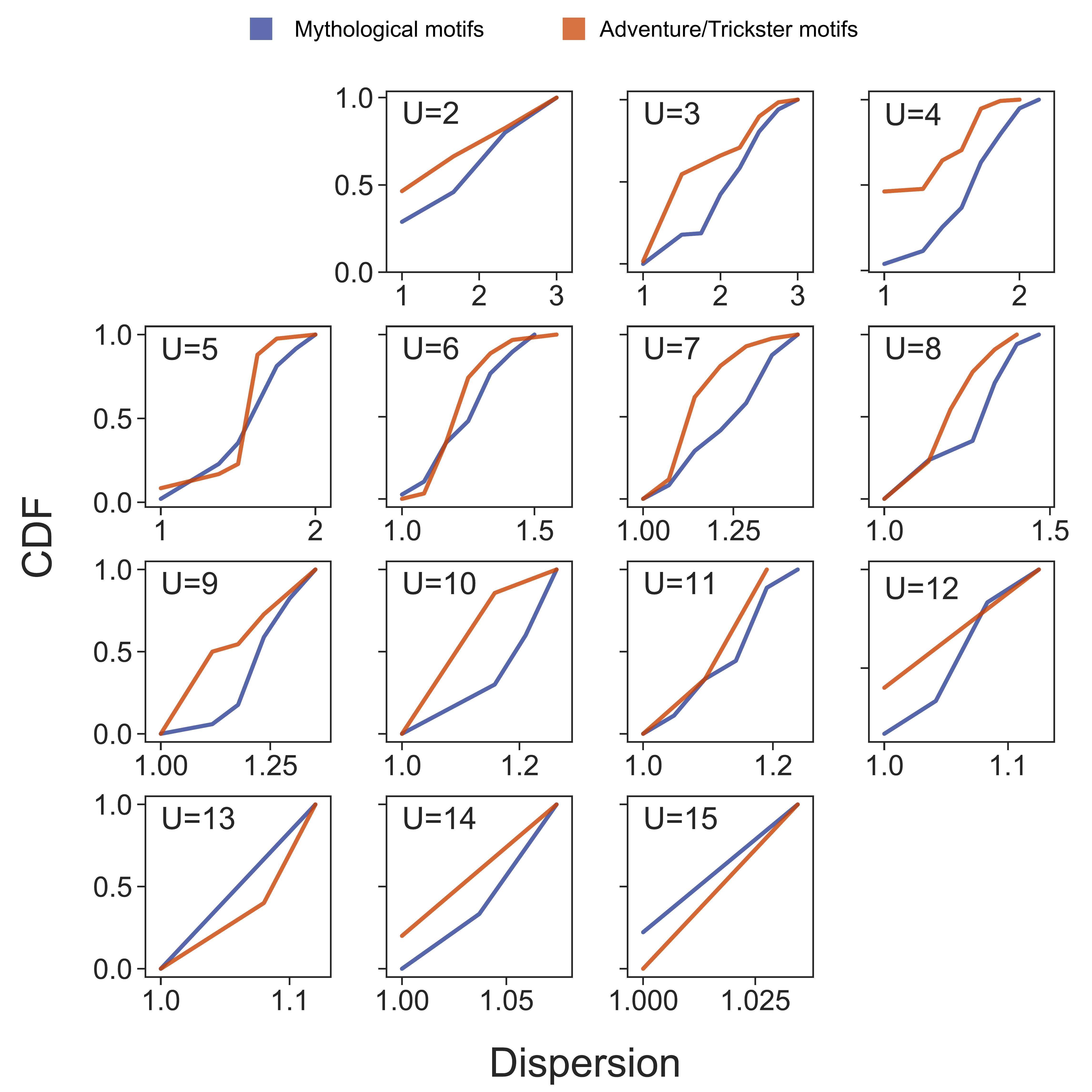}
    \caption{The cumulative distributions of dispersion by motif type and ubiquity}
    \label{fig:dispersion_cdf}
\end{figure}

From this simple comparison, we can notice that dispersion provides a meaningful spatial information in addition to ubiquity. To better investigate how motifs are dispersed across the clusters, we calculated dispersion for each motif and plot the cumulative distributions of dispersion by ubiquity and motif type -- Mythological motifs about cosmology and etiology; and Adventure/Trickster motifs (FIG \ref{fig:dispersion_cdf}). Interestingly, mythological motifs seem to be more dispersed than adventure/trickster motifs. This pattern is clear for ubiquity less than 5 and between 7 and 10. This difference in dispersion implies that the histories of mythological and adventure/trickster motifs are related to their spatial distributions. Our ubiquity and depth measure would help to filter out motifs that have a rare dispersion pattern. These motifs have interesting implications for anthropologists and folklorists who wish to interpret motif dispersal with qualitative and comparative studies. 90 motifs of ubiquity 2 and 10 motifs of ubiquity 3 that have the maximum dispersion for each ubiquity are given in the Table~\ref{si:tab:max_dispersion_motifs}.  

\begin{table}[!h]
    \caption{Motifs of ubiquity 2 or 3 that are maximally dispersed.}
    \label{si:tab:max_dispersion_motifs}
    \begin{center}
    \scriptsize
    \begin{tabular}{|
p{1.5cm}|p{10cm}|p{1.5cm}|}
        \hline
        Motif & Description & Ubiquity \\ \hline
        A35C & Scars and wounds on the Moon's face & 2 \\ 
        A4A & The Sun dazzles eyes & 2 \\
        B110 & 	If by back, the ravines, if by head the red flowers & 2 \\
        B115 & Evergreen trees & 2 \\
        B3 & A primeval swamp & 2 \\
        B3C & An attempt to drown God & 2 \\
        B3D & Earth from worm’s excrements & 2 \\
        B46A1 & Stars of Big Dipper are robbers & 2 \\
        B49 & Muddled request & 2 \\
        B52C & Earth bigger than sky & 2 \\
        C23 & Tree eclipses sky-light & 2 \\
        C6E & The diver is a crustacean & 2 \\
        C8B & Siblings change their looks and marry & 2 \\
        E36 & Hard covering of the body & 2 \\
        F11 & Biting penis & 2 \\
        F23 & Origin of menses: sexual act & 2 \\
        F24 & Origin of menses: a bite & 2 \\
        F40A & Husband of the first women & 2 \\
        F5 & Brides for first men & 2 \\
        F97 & The prohibited fruit: origin of sex & 2 \\
        F9A1 & The pike's mouth & 2 \\
        H36G & Muddled message: how many meals a day & 2 \\
        H36G1 & Bull distorts the message & 2 \\
        H46 & The dog's part & 2 \\
        H48 & Daughters of evil spirit & 2 \\
        H54 & The eyelids of Viy & 2 \\
        I126 & Constellation of the funerals & 2 \\
        I128 & Ursa major is a dipper & 2 \\
        I13D & Hibernating with snakes & 2 \\
        I41A & Rainbow from anthill or termite nest & 2 \\
        I45C & Not to count stars & 2 \\
        I85A & Animals walk around Polaris & 2 \\
        I87AC & Bone in the eye & 2 \\
        I98B & The Pleiades are a duck's nest & 2 \\
        J25A & Son of the grave & 2 \\
        J32E & The stolen foals & 2 \\
        J63 & Son saves, daughter betrays & 2 \\
        K100E & Aggressive stories & 2 \\
        K102A3 & The tooth of death & 2 \\
        K116B & The lecherous holy man and the maiden in the box & 2 \\ \hline
    \end{tabular}
    \end{center}
\end{table}

\begin{table}[!h]
    \begin{center}
    \scriptsize
    \begin{tabular}{|
p{1.5cm}|p{10cm}|p{1.5cm}|}
        \hline
        Motif & Description & Ubiquity \\ \hline
        K125 & House utensil betrays its master & 2 \\
        K126 & Wolf pays for the eaten up horse & 2 \\
        K12A & A strained bow & 2 \\
        K148 & The stolen colts & 2 \\
        K159 & Peas poured under the feet & 2 \\
        K15B & Substituted barrel of water & 2 \\
        K27L1 & To be frozen in the ice & 2 \\
        K27W & Monster brought by the hero kills the task-giver & 2 \\
        K27Z5 & An agreement to marry the would be born children & 2 \\
        K27ZY & Hero between two ogresses & 2 \\
        K27ZZ & The outcast queens and the ogress queen & 2 \\
        K27ZZ1 & Only the youngest queen's child survives & 2 \\
        K32G1 & Forty horses or forty knives? & 2 \\
        K33B & Friends abandon a pretty girl & 2 \\
        K35B & The most delicious dish & 2 \\
        K40 & One laughs, another weeps & 2 \\
        K48 & Singing bird of the hero & 2 \\
        K62A & Quarrel of mouse and bird & 2 \\
        K62A1 & A man cures the wounded eagle & 2 \\
        K77BB & The goat's weapons & 2 \\
        K89D & Person hides turning into a needle & 2 \\
        K90 & The black and the red ones & 2 \\
        K91 & An invisible battle & 2 \\
        L10A & Demon comes to hunter's camp-fire & 2 \\
        L116 & Singing girl in a bag & 2 \\
        L6A & Asks to be carried & 2 \\
        L72C & Obstacle flight: the thrown mirror & 2 \\
        L81A2 & Demon comes to drink blood of a girl & 2 \\
        L87 & Taste of blood & 2 \\
        M114C & To protect from rain by his own body & 2 \\
        M119 & Demonstrated many times & 2 \\
        M125 & Eating his own eyes & 2 \\
        M126 & The speaking skull & 2 \\
        M153A & The washed pig & 2 \\  M153B & Wolf rides a horse & 2 \\
        M187 & Snail is a participant of the race & 2 \\
        M198A1 & The eldest: it is round, the middle: it is hard, the youngest: it is a nut! & 2 \\
        M198B4 & The pretended diviner: names of the thieves & 2 \\
        M199A & Extracting brain from the earth & 2 \\
        M25 & Banquet in the sky & 2 \\ \hline
    \end{tabular}
    \end{center}
\end{table}

\begin{table}[!h]
    \begin{center}
    \scriptsize
    \begin{tabular}{|
p{1.5cm}|p{10cm}|p{1.5cm}|}
        \hline
        Motif & Description & Ubiquity \\ \hline
        M29W1 & The leopard is a failure & 2 \\
        M29X & The hyena is a failure  & 2 \\
        M33 & Anus closed & 2 \\
        M39A6A & Clever daughter-in-law of the imprisoned khan & 2 \\
        M60A1 & Herdsman explains how to ferry & 2 \\
        M60A2 & False servant licks soles & 2 \\
        M75B2 & Bird tries to avert predetermined marriage & 2 \\
        M83A & Crying for his dead children & 2 \\
        N29 & Before the water starts to boil & 2 \\
        N3 & Hungry fingers & 2 \\ \hline
        A45 & The insulted Moon & 3 \\
        B42F & Ursa major is an ungulate & 3 \\
        B53 & Creatures or objects from cut off genitals & 3 \\
        E4 & Creation from cuticle & 3 \\
        F80 & Primeval people have no genitals & 3 \\
        G23B & Alive being turns into nations & 3 \\
        I14A & Eject digested food through the mouth & 3 \\
        K24A & Supernatural male hides clothes of human girl & 3 \\
        K56A8 & Swallowed by boa & 3 \\
        L34 & Burning hair & 3 \\ \hline
    \end{tabular}
    \end{center}
\end{table}

\subsubsection{Multiscale backbone of motifs}
\label{si:backbone}

\begin{figure}[!h]
    \centering
    \includegraphics[width=\textwidth]{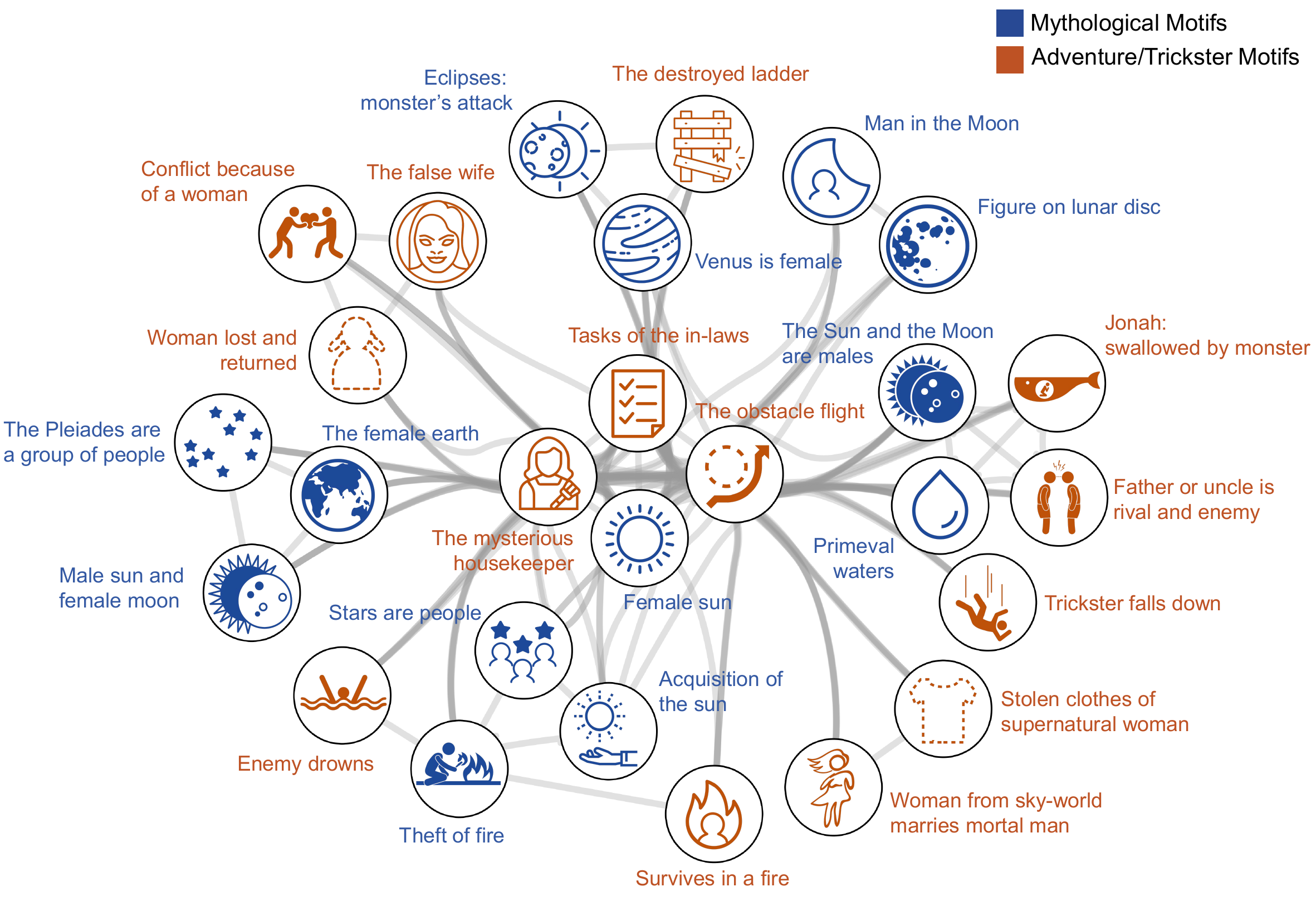}
    \caption{The multiscale backbone constructed from the 16 latent myths.}
    \label{fig:backbone_entire}
\end{figure}

To reveal important motif co-occurrences, we built a multiscale backbone of the sixteeen latent myths as follows. First, motifs were connected if they co-occur in a latent myth. Link weight hence denotes the number of latent myths containing both two motifs. However, this weighted network was too dense to interpret the structure. To resolve this issue, by following the paper introducing the multiscale backbone method \cite{Serrano6483}, we extracted links of which normalized weight $w_{ij}/s_{i}$ is statistically significant compared to the null model that uniformly distributes the sum of link weights of a node over its neighbors, where $w_{ij}$ is link weight between node $i$ and $j$ and $s_i$ is node strength that is the sum of link weights associated with node $i$. The probability $\alpha_{ij}$ that we observe the normalized weight in random counterparts is calculated as $(1-w_{ij}/s_{i})^{k_i-1}$ where $k_i$ is node degree of node $i$. The link between $i$ and $j$ is included in the backbone if $\alpha_{ij}$ is smaller than $\alpha$, a threshold to determine statistical significance of a link. We set $\alpha=0.006$ for the backbone of latent myths. This backbone has 27 motifs and 126 links (FIG \ref{fig:backbone_entire}). Four motifs -- Female sun, The mysterious housekeeper, The obstacle flight, and Tasks of the in-laws -- are found in all latent myths, and other motifs in the backbone are included at least 13 latent myths. These characteristics imply that the backbone represents core motif co-occurrences well. 

\subsubsection{Conditional dependencies between motifs}
\label{si:conditional}
The multiscale backbone (FIG~\ref{fig:backbone_entire}) implies that some motifs serve as foundations of narratives so that are more likely to be connected with others. We calculated conditional probabilities between motifs to infer their dependencies. Specifically, a method identifying significant conditional dependencies~\cite{jo2020extracting} was adopted in our analysis. First, we constructed a co-occurrence network of motifs by converting a tradition's motif incidence into a complete network and then accumulating them to one large weighted network. Link weight in this network is equal to the number of traditions having a pair of motifs. This network consists of 2,718 nodes and 1,523,024 links. Next, the $z$-scores of all links were computed from a hypergeometric distribution with a mean $\mu$ and a variance $\sigma^2$ as follows,
\begin{equation}
    \mu = \frac{n(m_i)n(m_j)}{N}
\end{equation}
\begin{equation}
    \sigma^2 = \frac{n(m_i)n(m_j)}{N}\frac{N-n(m_i)}{N}\frac{N-n(m_j)}{N-1}
\end{equation}
where $N$ is the total number of traditions, $m_i$ is a motif $i$, and $n(m_i)$ is the number of traditions having $i$. We selected only significant links of which $z$-score, $(n(m_i, m_j)-\mu)/\sigma$, is higher than $z_{th}=3.3$, which corresponds to the 99.9th percentile. This process yields a network of 2,718 nodes and 527,641 links. 

\begin{figure}[!h]
    \centering
    \includegraphics[width=\textwidth]{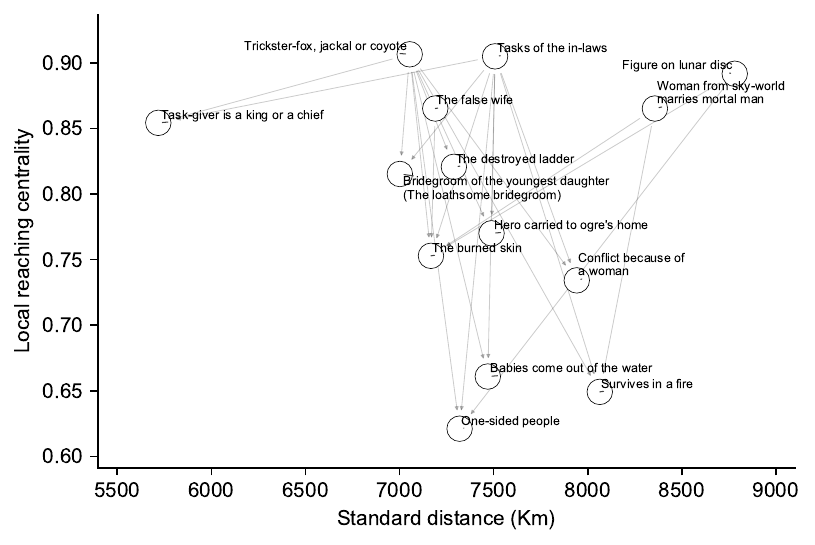}
    \caption{A hierarchical structure of high-ubiquity motifs ($U>12$)}
    \label{fig:cond_dependencies}
\end{figure}

Based on the pruned network, we infer the dependencies between motifs by comparing conditional probabilities. We assume that a directionality exists from $m_i$ to $m_j$ (i.e., $m_i$ precedes $m_j$, $m_i$ is at a higher lever of a hierarchy) if $p(m_i|m_j)-p(m_j|m_i)$ is higher than $\alpha_{th} \times k_{max}/\min(k_i, k_j)$ where $k_i$ is the degree (i.e., the number of links) of $i$ in the pruned network, $k_{max}$ is the maximum degree, and $\alpha_{th}$ is a threshold to filter significant dependencies from the pruned network. $\alpha_{th}$ is set to 0.5 in our paper. These steps created a directed network of 1,284 nodes and 142,569 links. Local reaching centralities were calculated to capture nodes' importance in a hierarchy~\cite{mones2012hierarchy}. Local reaching centrality of $i$ is the proportion of nodes reachable from $i$ through directed links. 

Although we identified significant dependencies, it is still challenging to explore the structure. For this reason, we intentionally focused on motifs of ubiquity higher than 12 and reduced the network to have 14 nodes and 22 links as shown in FIG~\ref{fig:cond_dependencies}. These nodes are positioned by local reaching centrality and standard distance, which is the root mean square of geographical distances between traditions and their centroid for each motif. Details of standard distance can be found at SI Section~\ref{si:standard_distance}.

\subsubsection{Environmental and ethnographic motifs}
\label{si:qje}

\begin{figure}[!h]
    \centering
    \includegraphics[width=0.65\textwidth]{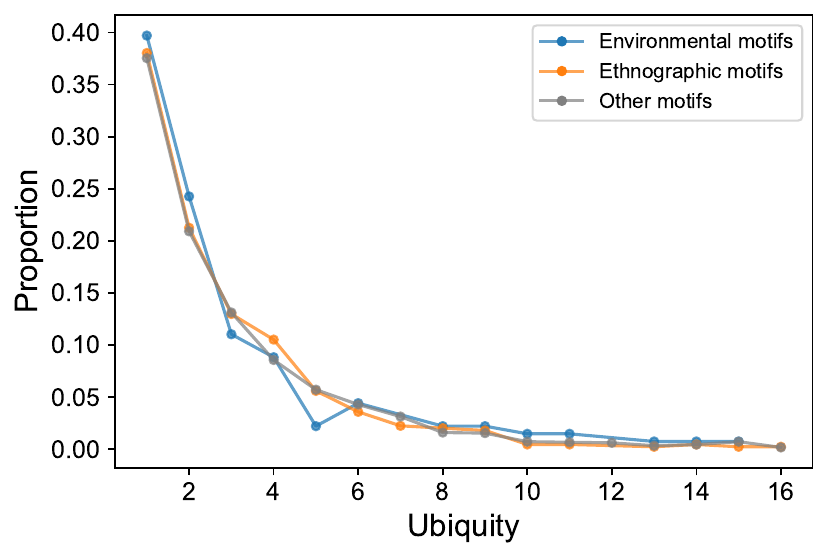}
    \caption{Ubiquity distributions for environmental, ethnographic, and other motifs}
    \label{fig:qje_ubiquity}
\end{figure}

\begin{figure}[!h]
    \centering
    \includegraphics[width=0.65\textwidth]{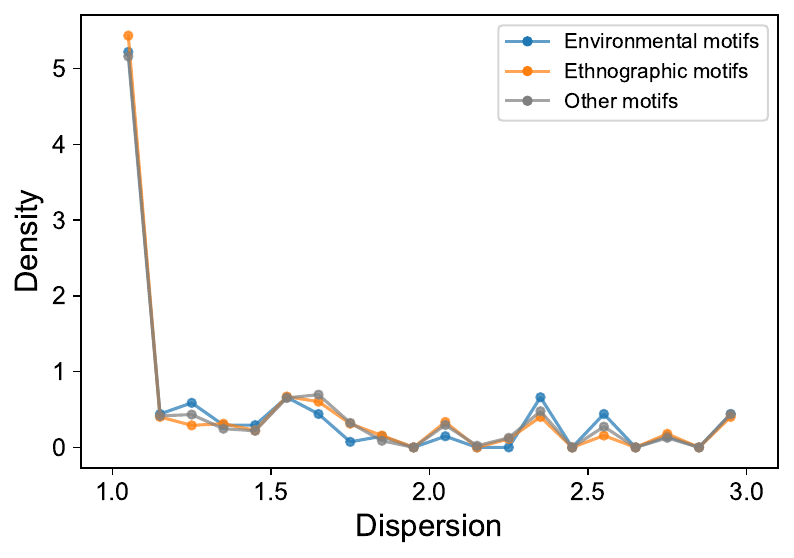}
    \caption{Dispersion distributions for environmental, ethnographic, and other motifs}
    \label{fig:qje_dispersion}
\end{figure}

There are environmentally and culturally relevant motifs identified and validated in~\cite{michalopoulos2021folklore}. Environmental motifs include those with Earthquake, Storm, Frozen, Crops, and River, and ethnographic motifs include those with Mode of Subsistence, Family Structure, and Political Centralization~\cite{DVN/IXOHKB_2021}. 
These motifs account for about 22\% of the motif database, and four of them -- `The destroyed ladder', `Woman from sky-world marries mortal man', `Task of the in-laws', and `Father or uncle is rival and enemy' -- are found FIG~\ref{fig:backbone_entire}, a multiscale backbone network. FIG~\ref{fig:qje_ubiquity} and FIG~\ref{fig:qje_dispersion} show that these motifs behave just like any other motifs in ubiquity and dispersion values. The observations suggest that the environmental and ethnographic motifs are no more important than any other motifs in the deep structure we found. In other words, environmental and ethnographic attributes explain a small fraction of the global mythologies. The complex network of motif co-occurrences can only be understood by our bottom-up approach.

\clearpage
\section{Correlations at various geographic scales}

\subsection{Mantel test at the continent scale}
\label{si:mantel_test}

Mantel test that checks statistical significance of a correlation between two matrices can be used to confirm whether motif co-occurrence patterns are similar across continents. We followed the paper by Youn et al. \cite{Youn16022016} which found the universal semantic structure, and compared the Pearson correlation of commute distances~\cite{Chandra1996} of two continents' motif co-occurrence matrices. Commute distance is the expected number of steps when a random walker traverses over a network with a probability in proportion to link weights. 

Mantel test was done as follows. First, motif co-occurrence networks were constructed for five continents: Africa, Eurasia, Oceania, North America, and Latin America. Link weight is equal to the number of traditions having two motifs in each continent. In order to make a random walker could reach any motifs in the network, a small weight $1/n(n-1)$ was added to all possible links even they were not connected yet, where $n$ is the number of motifs. Next, the modified co-occurrence network was transformed into a new network of which link weight is commute distance, a product of resistance distance ($\Omega$) and the sum of inverse link weights. Resistance distance between motif $i$ and $j$, $\Omega_{i,j}$, is equal to $\Gamma_{i,i}+\Gamma_{j,j}-2\Gamma_{i,j}$, where $\Gamma$ is the Moore-Penrose inverse of the Laplacian matrix of the given modified network. Commute distances larger than the number of motifs are excluded when calculating correlation because they would be artifacts originating from the process that adds small weight to make networks fully connected. 

For an empirical correlation of two continents, we tested its statistical significance in two ways: 1) randomly shuffling commute distances of a continent 1,000 times, and 2) generating 100 artificial groups of the same size with target continents. $p_1$ and $p_2$ are the fraction of random samples having a larger correlation than the empirical value. If $p_1$ is lower than 0.01, two continents are significantly correlated compared to random permutations. If $p_2$ is higher than 0.99, empirical motif co-occurrence patterns are less correlated than synthetic continents. The Pearson correlations of commute distances are significantly high compared to permuted samples ($p_1<0.01$, TABLE \ref{si:tab:mantel_test_geography}) for geographically adjacent continents (e.g., Africa and Eurasia). These significant correlations disappear when we shuffled their geographic information ($p_2>0.99$). These bootstrap experiments support that adjacent continents share some motifs but have distinct co-occurrence patterns.

\begin{table}
    \caption{Commute distances calculated from motif co-occurrence matrices of adjacent continents are correlated relative to random permutations (p-value$<$0.01), while these correlations disappear for randomly generated groups of the same size. This result suggests that adjacent continents share some motifs, but have different motif co-occurrence patterns.}
    \begin{center}
    \label{si:tab:mantel_test_geography}
    \begin{tabular}{|l|l|l|l|}
        \hline
        Geography & $r$ & $p_1$ & $p_2$ \\ \hline
        Africa vs. Eurasia & 0.21 &0.000 &1.00 \\ \hline
        Africa vs. Oceania & 0.10 &0.050 &1.00 \\ \hline
        Africa vs. North America & 0.01 &0.390 &1.00 \\ \hline
        Africa vs. Latin America & 0.01 &0.443 &1.00 \\ \hline
        Eurasia vs. Oceania & 0.04 &0.163 &1.00 \\ \hline
        Eurasia vs. North America & 0.08 &0.002 &1.00 \\ \hline
        Eurasia vs. Latin America & 0.01 &0.427 &1.00 \\ \hline
        Oceania vs. North America & 0.06 &0.133 &1.00 \\ \hline
        Oceania vs. Latin America & 0.09 &0.075 &1.00 \\ \hline
        North America vs. Latin America & -0.01 &0.548 &1.00 \\ \hline
    \end{tabular}
    \end{center}
\end{table}

\subsection{Phylogenetic network of the clusters of motifs}
\label{si:neighbornet}

\begin{figure}
    \begin{center}
    \subfloat[][Neighbor-Net with the cosine distance]{\includegraphics[width=0.44\textwidth]{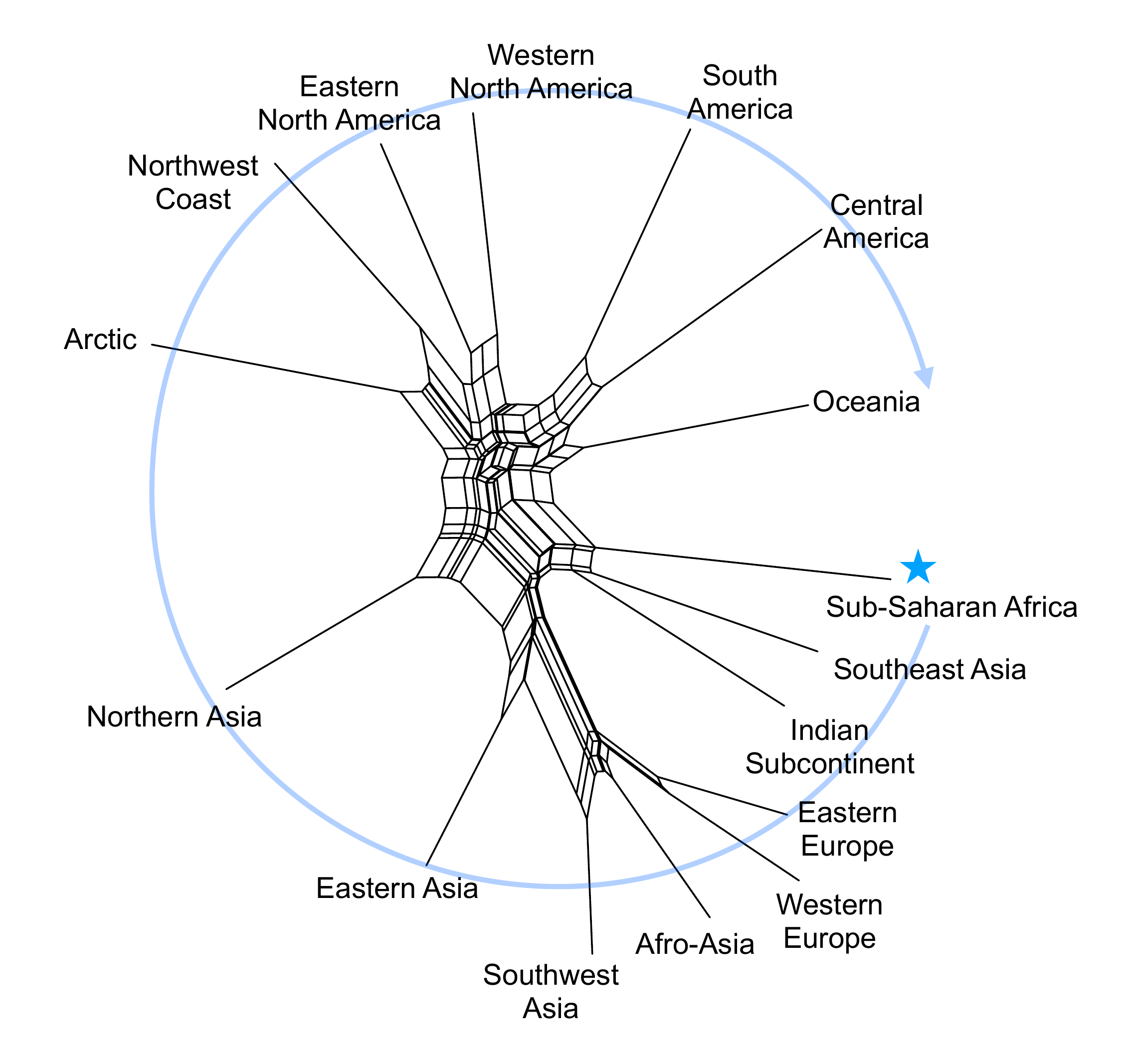}}
    \subfloat[][Neighbor-Net with the Jaccard distance]{
    \includegraphics[width=0.44\textwidth]{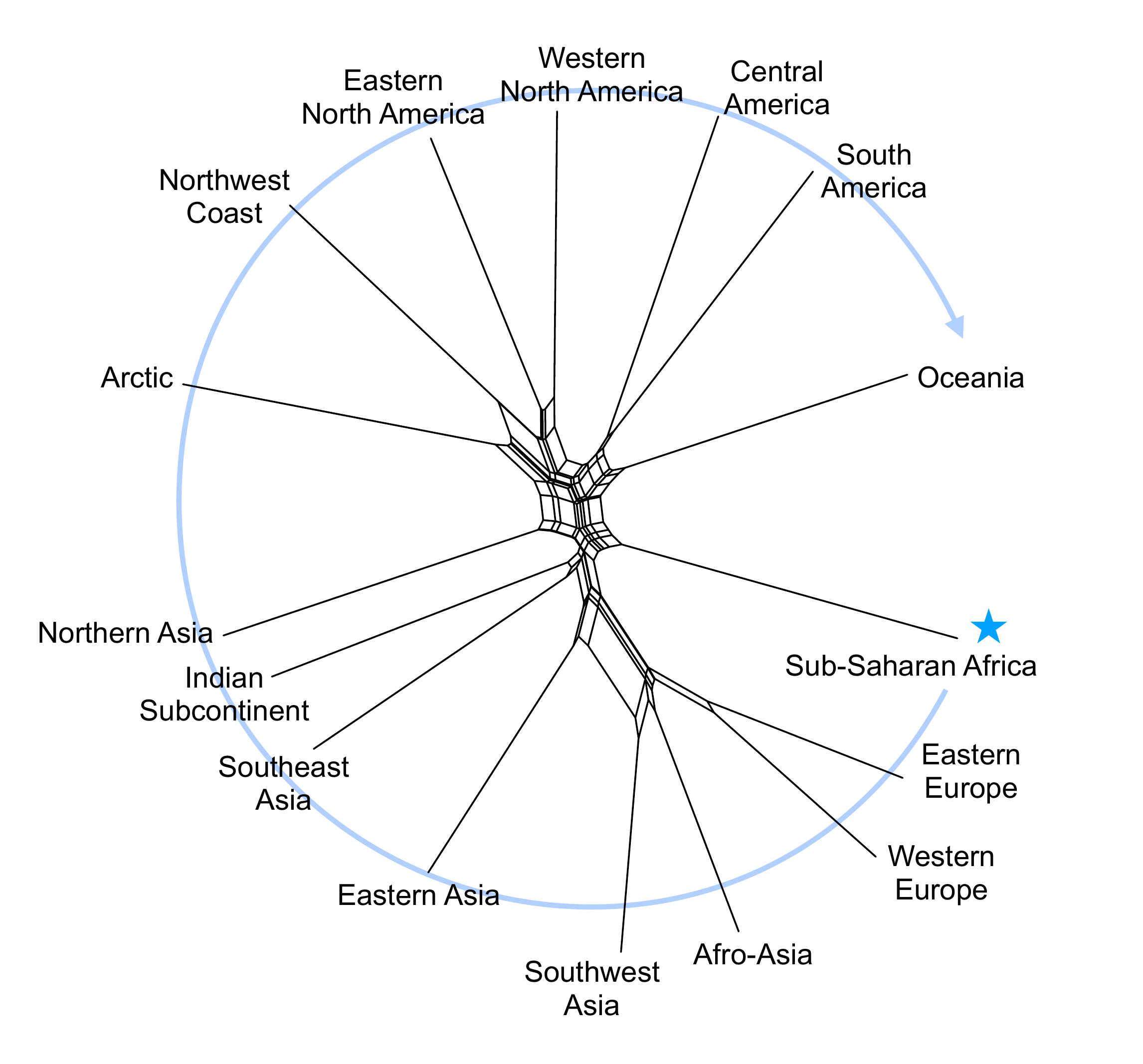}}
    \end{center}
    \caption{Phylogenetic networks are constructed through the Neighbor-Net algorithm~\cite{neighbornet} with (A) the cosine and (B) the Jaccard distance between the sixteen latent myths. Starting from Sub-Saharan Africa, the cluster alignment resembles the prehistoric human dispersal out of Africa.}
    \label{si:fig:neighbornet}
\end{figure}

Latent myths indicate which motifs anchor co-occurrences in the clusters of motifs. All clusters are comparable as they have enough motifs (more than 200) in their latent myths. To investigate the extent to which clusters are similar, we built a phylogenetic network based on cosine distance between latent myths with the Neighbor-Net algorithm \cite{neighbornet}, an agglomerative method that allows more than one parent for each leaf (FIG \ref{si:fig:neighbornet}A). Edge length in the Neighbor-Net algorithm represents how much two leaves are different. Interestingly, while we did not impose any geographical information in clustering, the detected clusters are grouped into the Old and New World, and the sequence of clusters is in good agreement with the prehistoric dispersal of humans out of Africa. This alignment is significant compared to 1,000 random permutations. The structure of the Neighbor-Net is similar even though we use the Jaccard distance (FIG \ref{si:fig:neighbornet}B). These motif co-occurrence patterns imply that myths and folktales would vary along early human dispersal and are much alike for geographically adjacent traditions. 

A phylogenetic network of the flows between these clusters are consistent with the temporal and spatial sequence of dispersal events as humans expanded out of Africa, resulting in the structures we find at the continent scale. The phylogenetic network of the latent myths replicates the spatial structure in Fig 2A (Fig 2D and SI \ref{si:neighbornet}); a test of 1,000 randomized networks confirms the significance of this spatial structure (p-value$<$0.001) where geographically adjacent clusters have the most similar latent myths. In Fig 2D, moving clockwise from Sub-Saharan Africa, the flows first lead to western Eurasia (i.e., branches including Europe and north Africa/central Asia, followed by eastern Eurasia (South/East Asia and northern Asia), then the northern New World (northern, western, and eastern North America), the southern New World (Central and South America), and finally Austronesia. 

\subsection{Analysis of molecular variance}
\label{si:amova}

The variances that the mythological clusters and biogeography explain were investigated through ``Analysis of Molecular Variance'' (AMOVA) \cite{excoffier1992analysis} that decomposes variances into hierarchical levels, and estimates their statistical significance from random permutations. We set three levels for AMOVA decomposition, 1) among the clusters of motif (or bioregions), 2) among language families within the clusters (or bioregions), and 3) within language families, because language family affiliation has been known as an important factor of cultural diffusion \cite{bortolini2017inferring}. Language family information were collected from the Berezkin's database. 
   
We first focus on language families that have at least five traditions within a cluster. The selected 29 language families are Afrasian, Algic, Altaic, Arawak, Australian, Austroasiatic, Austronesian, Chibchan, Dravidian, Escoaleut, Hokan, Indoeuropean, Macro Ge, Mayan, Na-Dene, Niger-Kongo, Nilo-Saharan, North Caucasian, Oto-Mangean, Paleoasiatic, Pano-Tacana, Penuti, Salishan, Sino-Tibetan, Sioux-Katawba, Tucano, Tupian, Uralic, and Uto-Aztecan. 715 traditions in one of these 29 language families are distributed well across the clusters. The variances between two traditions are measured by the Jaccard distance. R package poppr 2.8.2 \cite{kamvar2014poppr} is used for AMOVA analysis.

\subsubsection{Clusters of motifs / Language families}

While most variances are found within language families as known in previous studies \cite{Ross20123065, bortolini2017inferring}, variances among the clusters and among language families within clusters are also significant compared to 1,000 random permutations (TABLE \ref{si:tab:amova_mythcluster}). It is noted that more variances lie among clusters than among language families within clusters. The variances of ``among clusters'' component ($\sim6.49\%$) are higher than those of ``among language families within clusters'' component ($\sim3.36\%$).

\begin{table}[!h]
    \caption{Summary of the AMOVA results with the clusters of motifs.}
    \label{si:tab:amova_mythcluster}
    \begin{center}
    \small
    \begin{tabular}{|l|l|l|l|l|}
        \hline
        AMOVA & Variances & \% total & p-value & Compared to random \\ \hline
        \multicolumn{5}{ |c| }{Clusters of motifs / Language families (715 traditions / 29 language families)} \\ \hline
        Among clusters & 0.031 & 6.49 & $<$0.001 & Greater\\ \hline
        Among language families within clusters & 0.016 & 3.36 & $<$0.001 & Greater\\ \hline
        Within language families & 0.435 & 90.15 & $<$0.001 & Less\\ \hline
    \end{tabular}
    \end{center}
\end{table}

\begin{table}
    \caption{Summary of the AMOVA results with bioregions.}
    \label{si:tab:amova_bioregion}
    \begin{center}
    \small
    \begin{tabular}{|l|l|l|l|l|}
        \hline
        AMOVA & Variance & \% total & p-value & Compared to random \\ \hline
        \multicolumn{5}{ |c| }{Amphibian bioregions (643 traditions / 11 bioregions / 27 language families)} \\ \hline
        Among bioregions & 0.021 & 4.35 & $<$0.001 & Greater\\ \hline
        Among language families within bioregions & 0.020 & 4.20 & $<$0.001 & Greater\\ \hline
        Within language families & 0.442 & 91.44 & $<$0.001 & Less\\ \hline
        \multicolumn{5}{ |c| }{Bird bioregions (651 traditions / 15 bioregions / 27 language families)} \\ \hline
        Among bioregions & 0.025 & 5.08 & $<$0.001 & Greater\\ \hline
        Among language families within bioregions & 0.018 & 3.76 & $<$0.001 & Greater\\ \hline
        Within language families & 0.440 & 91.16 & $<$0.001 & Less\\ \hline
        \multicolumn{5}{ |c| }{Mammal bioregions (616 traditions / 23 bioregions / 27 language families)} \\ \hline
        Among bioregions & 0.023 & 4.86 & $<$0.001 & Greater\\ \hline
        Among language families within bioregions & 0.022 & 4.51 & $<$0.001 & Greater\\ \hline
        Within language families & 0.437 & 90.63 & $<$0.001 & Less\\ \hline
    \end{tabular}
    \end{center}
\end{table}

\subsubsection{Bioregions / Language families}

The same procedure was applied to the three bioregions of the updated Wallace zoogeographic maps \cite{Holt74} -- Amphibian, Bird, and Mammal bioregions. Language families having at least five traditions within a cluster were chosen. These bioregions explain variances less than the clusters, and these variances still remain about the same level as language families (TABLE \ref{si:tab:amova_bioregion}). AMOVA results emphasize the role of biogeography on the global mythologies.

\subsection{Motif similarity between traditions}
\label{si:sim_decay}

\subsubsection{Similarity decaying by geographical distance}
\label{si:sim_decay_ind}

\begin{figure}
\subfloat[][Jaccard similarity decaying]{\includegraphics[width=0.44\textwidth]{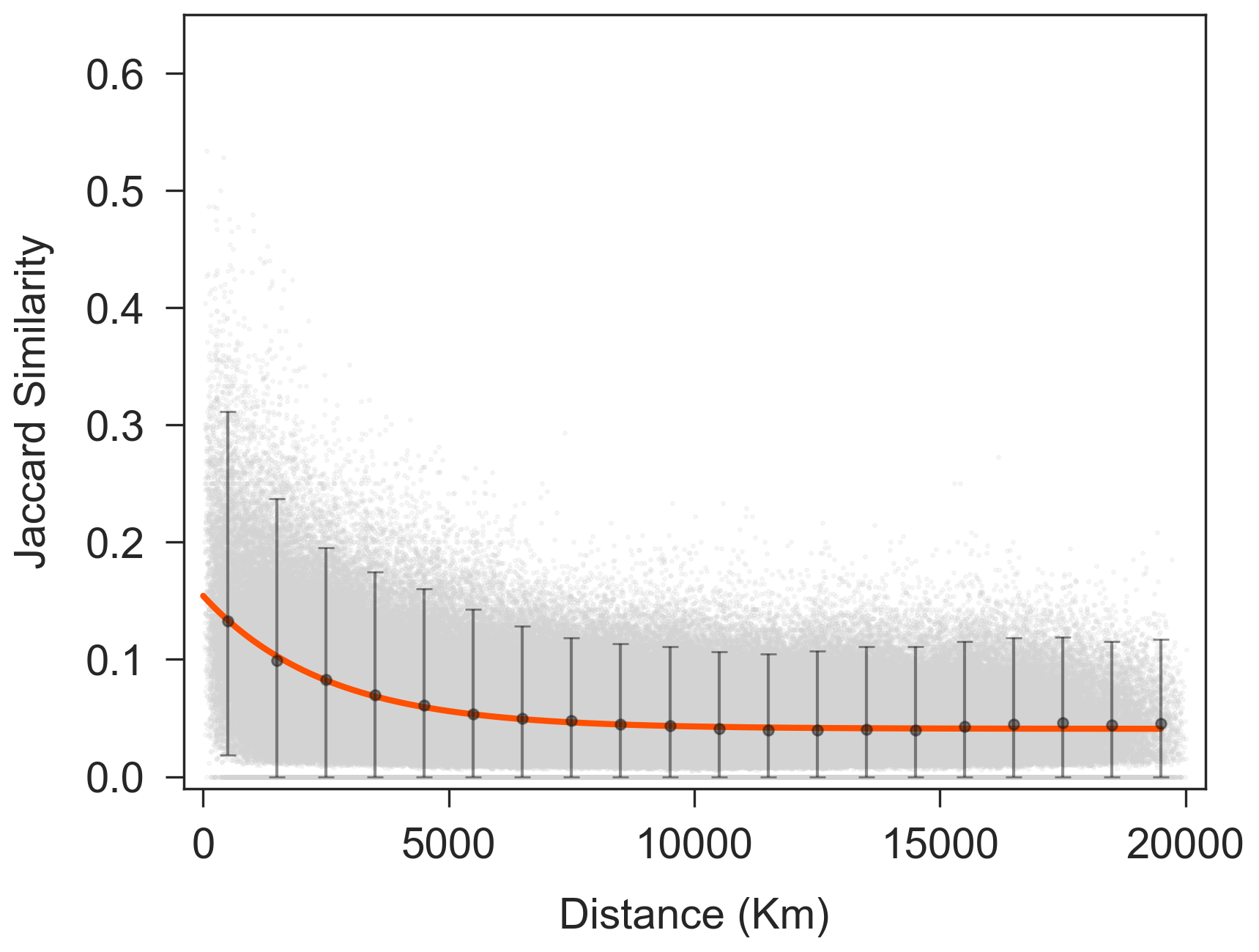}}
\subfloat[][Spatial autocorrelation]{\includegraphics[width=0.45\textwidth]{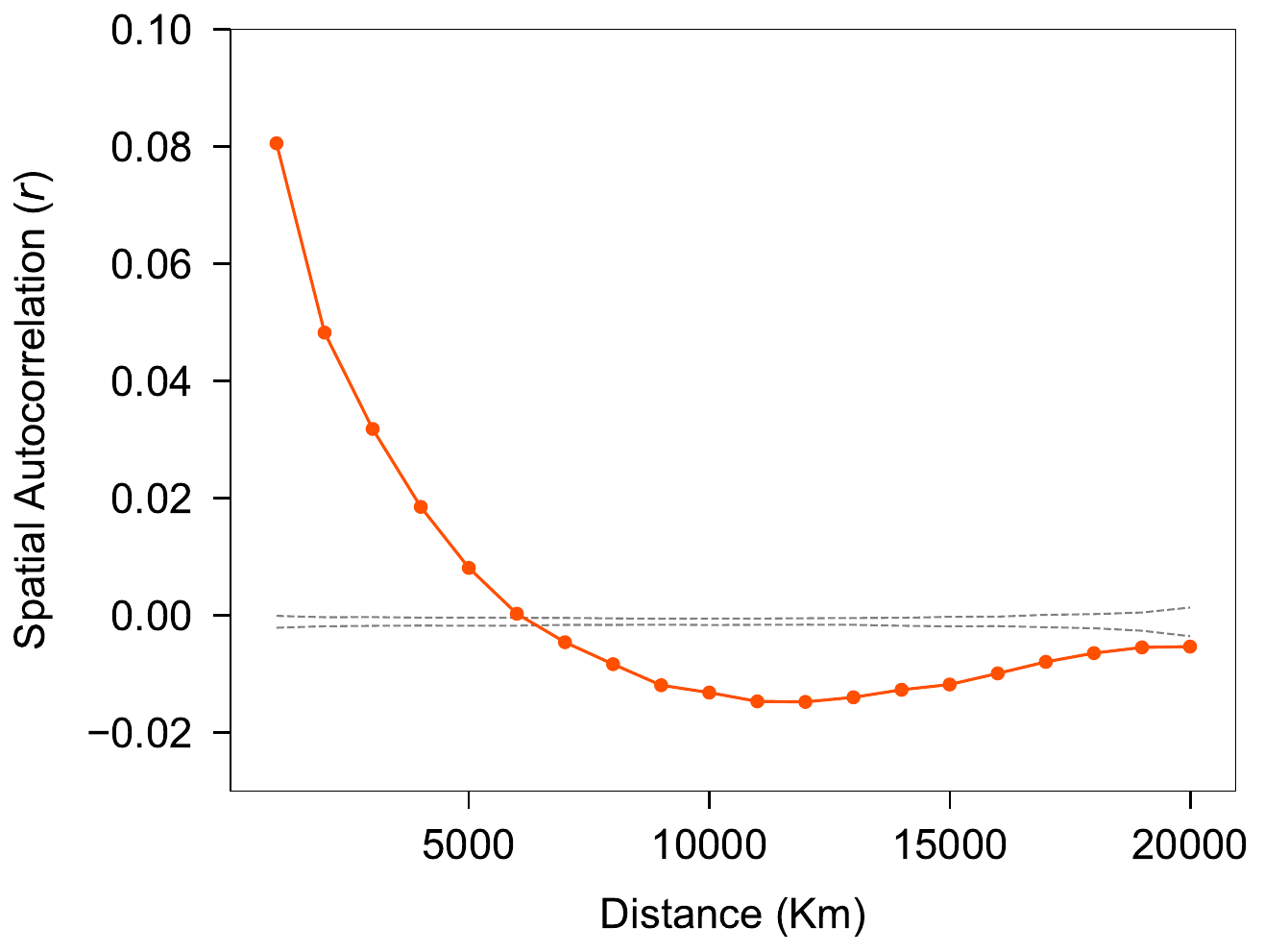}}
\caption{(A) Jaccard similarity decaying by geographical distance. Similarity decays exponentially ($e^{-x/x_{0}}$) with the characteristic distance $x_0$ that is about 2,725km. Orange line is the fitted function of the exponential similarity decaying. The black dots are the mean values of 1,000km bins and the associated bars are 95\% errors. (B) Spatial autocorrelation of motif distributions. In close distance less than 5,000km, spatial autocorrelation is significantly high compared to 1,000 random samples but rapidly decays after 5,000km, indicating motif co-occurrences are distinct at the continent and cluster scales. Gray dashed lines are 99\% confidence intervals of random samples.}
\label{si:fig:exp_decay_global}
\end{figure}

The multi-scale clusters of motifs imply that geographically adjacent traditions share some motifs. We plotted the Jaccard similarity of all tradition pairs as a function of the Haversine distance (FIG \ref{si:fig:exp_decay_global}A) and fitted similarity values to an exponential function $y=ae^{-x/x_{0}}+b$. The Jaccard similarity between two traditions $A$ and $B$ is calculated as $|A\cap B|/|A\cup B|$ where $|A\cap B|$ is the number of motifs that two traditions share, and $|A\cup B|$ is the number of motifs that are found in at least one tradition. The Jaccard similarity decays exponentially up to 5,000km, about the size of continents (orange line). Characteristic distance, $x_{0}$, is about 2,725km. The black points are the mean similarity values of 1,000km bins, and the associated bars are 95\% errors. Non-zero background similarity ($b=0.026$) is attributed to common motifs that exist across the clusters. Moreover, the decreasing trend of spatial autocorrelation is significant compared to 1,000 permuted samples which exclude the effect of geographical proximity on autocorrelation (FIG \ref{si:fig:exp_decay_global}B). Spatial autocorrelation is calculated through R package PopGenReport \cite{adamack2014popgenreport} that implements the paper of Smouse and Peakall \cite{smouse1999spatial}. 

\begin{figure}
\centerline{\includegraphics[width=0.85\textwidth]{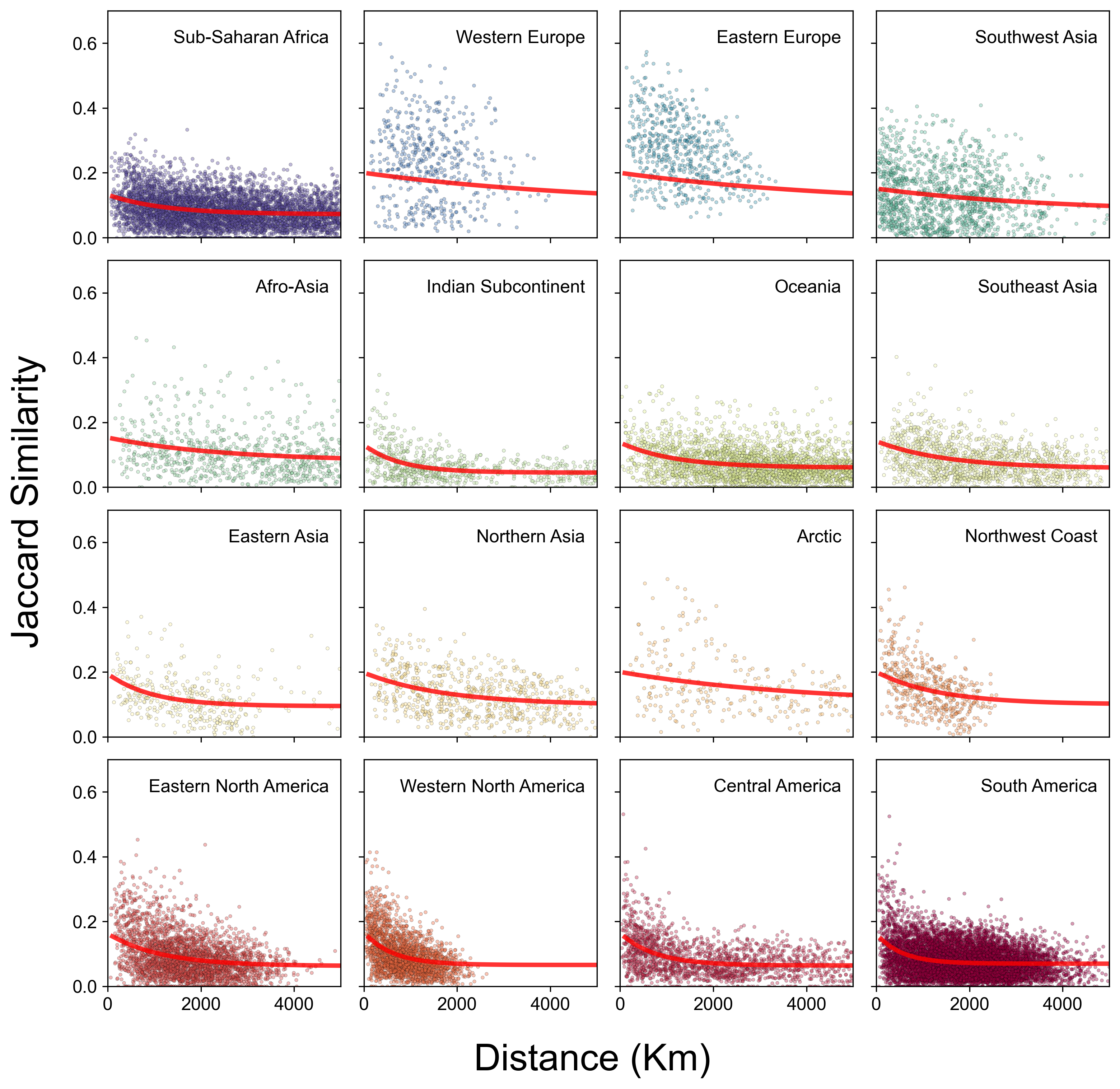}}
\caption{Similarity decaying within clusters. Geographically adjacent traditions have similar motifs while the extent of decaying varies by clusters. The red lines are fitted exponential functions.}
\label{si:fig:exp_decay_cluster}
\end{figure}

We also plotted motif similarity by geographical distance for each cluster to check whether this pattern holds within the clusters. While decaying rate varies by cluster, we found clear decreasing trends in all cases (FIG~\ref{si:fig:exp_decay_cluster}). This pattern is in agreement with ``isolation-by-distance'' \cite{wright1943isolation, Ross20123065}.

\subsubsection{Standard distance of motif dispersal}
\label{si:standard_distance}

A motif of $U=1$ anchors motif co-occurrences only in one cluster, so it tends to be localized within close distances. This pattern would also be found in $U=2$ if similarity decays by distance. We checked whether this assumption holds by comparing the ubiquity and standard distance which is a statistic that represents dispersion of points around their geometric center (Equation.~\ref{si:eq:sd}). 

\begin{equation}
\label{si:eq:sd}
    \text{Standard Distance} = \sqrt{\frac{\sum_{i=1}^{n}{(\text{Distance from the geometric center to a tradition $i$}})^2}{n}}
\end{equation}

\begin{figure}
\centerline{\includegraphics[width=0.8\textwidth]{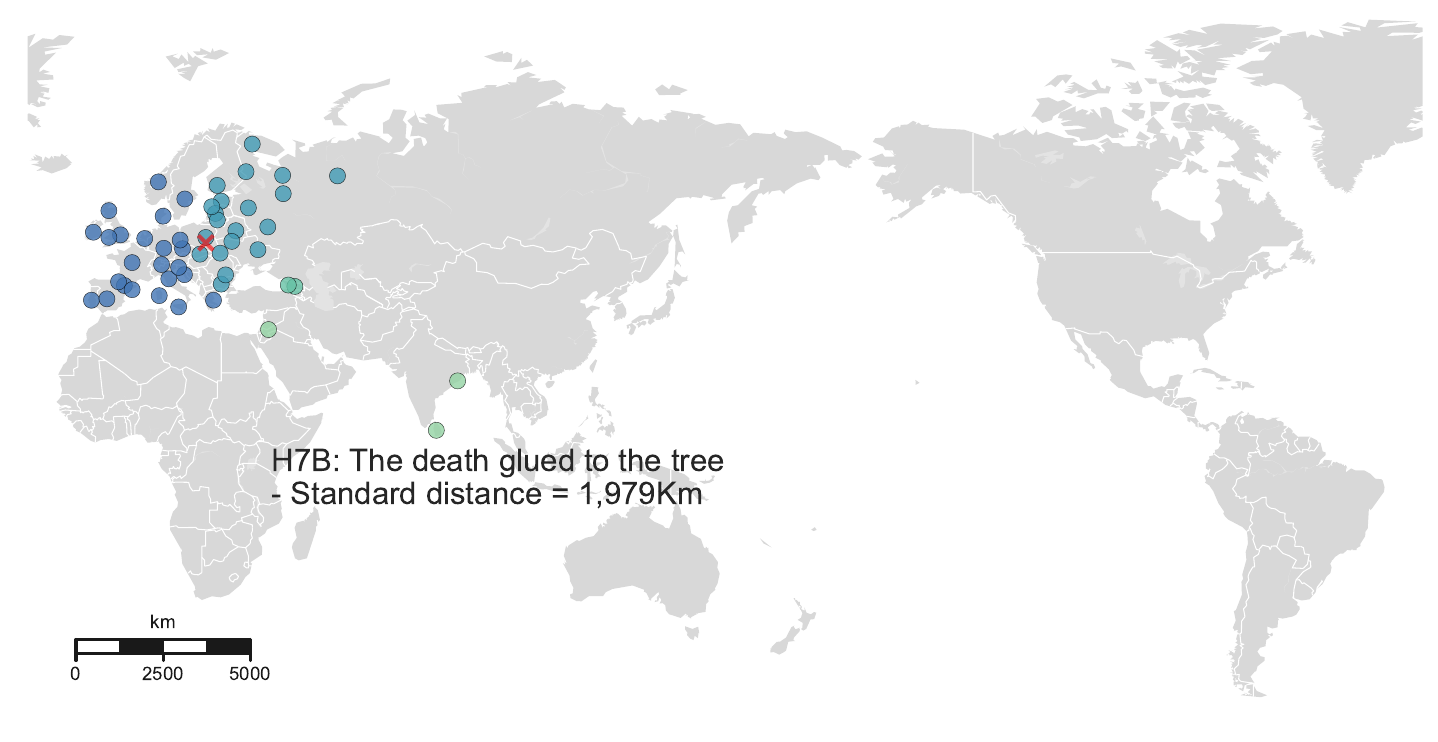}}
\caption{An example of standard distance. The colored points are traditions having the motif `The death glued to the tree' (H7B). The ubiquity index of H7B is 2 as the motif is significant in Western and Eastern Europe. The red cross is the geometric center of traditions having the motif.}
\label{si:fig:sd}
\end{figure}

Here is an example of standard distance calculation. In FIG \ref{si:fig:sd}, the red cross is the geometric center of traditions having `The death glued to the tree' (H7B) which is found in 49 traditions and significant in two clusters ($U=2$) with the standard distance 1,979km. Distances from the center of 49 traditions are first calculated and then put in Equation \ref{si:eq:sd} to compute the standard distance. In this way, we can obtain the standard distance of each motif that characterizes its spatial distribution. 

Then, what are standard distances of other $U=2$ motifs? As mentioned above, standard distances of other $U=2$ motifs would be concentrated in close distances (FIG \ref{si:fig:exp_decay_global}) like the motif H7B if they are diffused by ``isolation-by-distance''. Interestingly, adventure/trickster motifs are distributed under our assumption, while mythological motifs are not.  

\begin{figure}
\centerline{\includegraphics[width=0.5\textwidth]{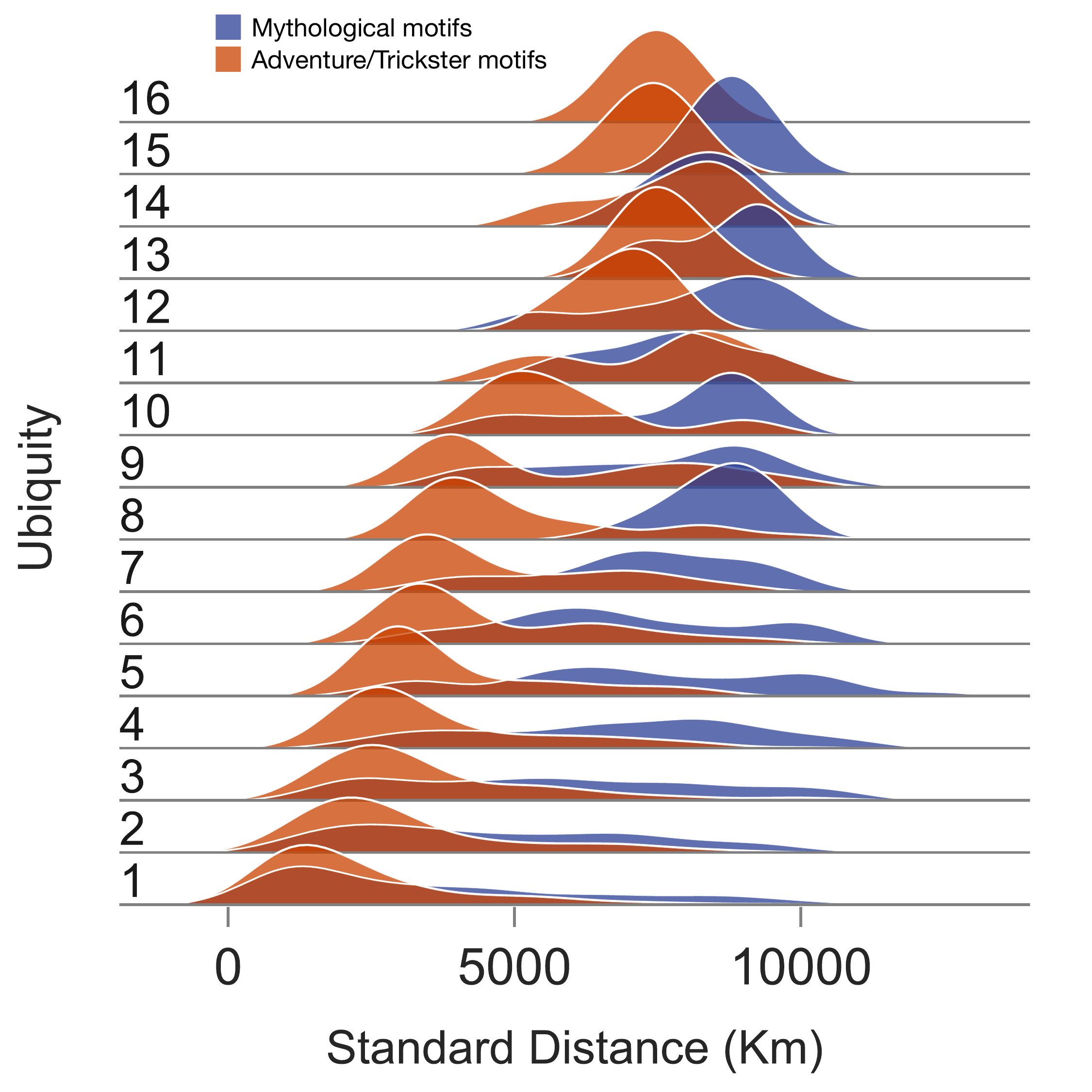}}
\caption{Gaussian smoothed distributions of standard distance by ubiquity. Mythological motifs and adventure/trickster motifs have different spatial patterns. Adventure/trickster motifs tend to be distributed in close distance, while mythological motifs are widespread across the clusters even though their ubiquity is small.}
\label{si:fig:dist_by_u}
\end{figure}

FIG \ref{si:fig:dist_by_u} shows different spatial patterns by motif type. The distributions in the plot are smoothed by Gaussian kernels. Standard distance distributions of adventure/trickster motifs have peaks that tend to be larger by ubiquity. It is reasonable since motifs exist in broader areas as ubiquity increases. On the other hand, mythological motifs behave differently even the ubiquity is two or three. It means that some mythological motifs are found in distant clusters. There are two possible interpretations. First, mythological motifs could be replaced or removed during the history while they were transferred to neighboring traditions in early human migrations. Second, mythological motifs could be developed independently based on the common human cognition. Further studies are needed to examine these hypotheses.

\clearpage

\bibliography{scibib}